\documentclass[twocolumn,pre,showpacs,amsmath]{revtex4-1}
\usepackage{color}
\usepackage[dvipdfmx]{graphicx}
\usepackage{epsfig,amsfonts,amsbsy,color}

\newcommand{\bra}{\left\langle}
\newcommand{\ket}{\right\rangle}

\newcommand{\pder}[2]{\frac{\partial #1}{\partial  #2}}
\newcommand{\pderf}[3]{\left({\frac{\partial #1}{\partial  #2}}\right)_{#3}}

\newcommand{\der}[2]{\frac{d #1}{d  #2}}

\newcommand{\var}[2]{\frac{\delta #1}{\delta  #2}}
\newcommand{\vart}[3]{\frac{\delta^2 #1}{\delta  #2\delta  #3}}

\newcommand{\bv}[1]{{\boldsymbol #1}}

\newcommand{\e}{{\rm e}}
\newcommand{\ep}{\epsilon}

\newcommand{\Tc}{T_c}

\makeatletter
\@addtoreset{equation}{section}
\makeatother

\begin{document}
\title{Stochastic order parameter dynamics for phase coexistence
      in heat conduction}

\author{Shin-ichi Sasa}
\email{sasa@scphys.kyoto-u.ac.jp}
\affiliation {
Department of Physics, Kyoto University, Kyoto 606-8502, Japan}
\author{Naoko Nakagawa}
\email{naoko.nakagawa.phys@vc.ibaraki.ac.jp}
\affiliation {Department of Physics, Ibaraki University,
  Mito 310-8512, Japan}
\author{Masato Itami}
\email{itami@r.phys.nagoya-u.ac.jp}
\affiliation {Department of Physics, Nagoya University,
  Nagoya 464-8602, Japan}

\author{Yohei Nakayama}
\email{r\_nakayama@tohoku.ac.jp}
\affiliation {Department of Applied Physics,
  Tohoku University, Sendai 980-8579, Japan}

\date{\today}

\begin{abstract}
We propose a stochastic order parameter model for describing 
phase coexistence in steady heat conduction near equilibrium.
By analyzing the stochastic dynamics with a non-equilibrium adiabatic
boundary condition, where total energy is conserved over time,
we derive a variational principle that determines 
thermodynamic properties in non-equilibrium steady states.
The resulting variational principle indicates that the temperature of the
interface between the ordered region and the disordered region becomes
greater (less) than  the equilibrium transition temperature in the linear
response regime when the thermal conductivity in the ordered region
is less (greater) than that in the disordered region.
This means that a super-heated ordered (super-cooled disordered)
state appears near the interface, which was predicted by an extended
framework of thermodynamics proposed in [N. Nakagawa and
S.-i. Sasa, Liquid-gas transitions in steady heat conduction,
Phys. Rev. Lett. {\bf 119}, 260602, (2017).]
\end{abstract}


\maketitle

\section{Introduction}


Phase coexistence, such as liquid-gas coexistence, is ubiquitous
in nature. As the most idealized situation, phase coexistence
under equilibrium conditions has been studied. For example, the liquid-gas
coexistence temperature is determined by the  equality of the chemical
potential of liquid and gas at constant pressure. The pressure dependence
of the coexistence temperature is related to the latent heat 
and the volume jump at the transition point, which is known as
the Clausius-Clapeyron equation. These are important
consequences of thermodynamics \cite{Callen}. 


In addition to equilibrium systems, phase coexistence gives
rise to a rich variety of phenomena out of equilibrium such as 
flow boiling heat transfer, pattern formation in crystal growth,
and motility-induced phase separation
\cite{boiling, crystal, Cannell,Zhong,Ahlers,mips}.
Moreover, as an interesting phenomenon, it has been
reported  that heat flows from a colder
side to a hotter side in a transient regime for continuous heating
\cite{Urban}.
One may expect that a deterministic hydrodynamic equation
incorporating interface thermodynamics, 
which is referred to as the Navier-Stokes-Korteweg equation
(See \cite{Anderson98} for a review), generalized hydrodynamics
\cite{Bedeaux03}, or dynamical van der Walls
theory \cite{Onuki}, could describe such dynamical phenomena.

\begin{figure}[tb]
\centering
\includegraphics[width=7cm]{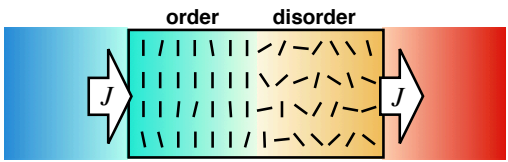}
\caption{Schematic of setup.
  The configuration of a single interface is displayed, where the heat flux
  $J <0$.}
\label{fig:setup}
\end{figure}

However, the situation is not so obvious.  
Because the macroscopic description is obtained by the coarse-graining
of microscopic mechanical systems, the noise inevitably appears.
The noise properties are determined by the fluctuation-dissipation
relation of the second kind at equilibrium, and the relation is also
assumed for systems  out of equilibrium. Such a framework is called
fluctuating hydrodynamics \cite{Schmitz} or macroscopic fluctuation theory
\cite{Bertini-rev}. For
standard cases such as simple homogeneous fluids, the noise effects are so weak
that the thermodynamic behavior is well-approximated by the noiseless limit,
while it has been known that noises substantially modify the macroscopic
behavior for systems in low-dimensions \cite{FNS} or near the critical point
\cite{HH}. 
As  another    example of such strong noise effects, in this paper, we  
study phase coexistence in steady heat conduction. For simplicity,
we assume that the system is divided into two phases by a macroscopic
planer interface across which the heat flows in a simple cuboid
geometry, as shown in Fig. \ref{fig:setup}.


  The  most impressive phenomenon exhibited by the strong noise
effect is that the interface temperature $\theta$ deviates from
the equilibrium transition temperature $T_c$.  
That is, a super-heated  ordered state or a super-cooled disordered
state stably appears locally near the interface. It should be noted that
this phenomenon  was
predicted by an extended framework of thermodynamics for heat conduction
systems \cite{NS},  which we call {\it global thermodynamics} \cite{NS2}.
Remarkably, despite the difference of theoretical frameworks, our result
on $\theta-\Tc$ qualitatively agrees with the prediction of this
thermodynamic framework up to a multiplicative numerical constant.  
The main purpose of this paper is to calculate the interface
temperature based on a stochastic model that exhibits the phenomenon.
See (\ref{temp-1}) for the main result.  

\subsection{Highlight of the paper}

Now, we describe the highlight of the paper.

\paragraph{Model}


Among many first-order transitions, we specifically study
the order-disorder transition associated with the ${\bf Z}_2$ symmetry
breaking. This is the simplest case of symmetry
breaking, and it is easily generalized to other complicated
symmetry breakings, such as the nematic-isotropic transition in
liquid crystals, which may be relevant in experiments \cite{nematic-isotropic}.
Although the liquid-gas transition may be most popular in the first-order
transition, we study this phenomenon in another paper. See
the second paragraph in Sec. \ref{remark} for related discussions.


For the order-disorder transition associated with the ${\bf Z}_2$
symmetry, one may recall a Ginzburg-Landau equation that includes
the interface thermodynamics as a gradient term. However, because
this model describes  the order parameter dynamics
with the isothermal condition,  it cannot be used for
heat conduction systems. We must at least
consider a  coupled equation of the order parameter
density field and the energy density field.
When we consider a stochastic model as a generalization
of the Ginzburg-Landau model, it is  best  to use the
concept of the Onsager theory as follows. First, we specify
a set of dynamical variables. Then, under the assumption of 
local thermodynamics, we consider the minimum form of 
dissipation and noise with the detailed balance condition
at equilibrium.    In Sec. \ref{Sec:det}, following these
concepts, we present deterministic and stochastic order parameter
dynamics.  
See  (\ref{model-s1}),  (\ref{model-s2}), and (\ref{model-s3})
for the final form of stochastic dynamics.  
We next describe an interface between the ordered  region    and the
disordered region within a framework of the deterministic dynamics
in Sec. \ref{interface}. We then discuss how  fluctuations
of the interface position play an inevitable role in 
thermodynamic  behavior.   

\paragraph{Theoretical method}


The theory for stochastic models related to thermodynamics
has developed significantly over the last two decades
\cite{Sekimoto-book,SeifertRPP}. 
This  mainly comes from the discovery of simple and universal
relations: the fluctuation theorem
\cite{Evans-Cohen-Morriss,Gallavotti,Kurchan,LS,Maes,Crooks} 
and Jarzynski equality \cite{JarzynskiPRL}. Even
for the theoretical calculation of quantities,  these formulas
can simplify the derivation of macroscopic evolution such as
the Navier-Stokes equation 
\cite{Sasa-fluid} and the order parameter dynamics of coupled 
oscillators \cite{Sasa-oscillator}. In the present problem,
we start by deriving the stationary
distribution for the system out of equilibrium.  
See (\ref{Zubarev-flux-0}) with (\ref{tildeS-0}) and 
(\ref{Idef}).   
It has been known that the stationary distribution is
formally expressed in terms of the time integration of the
excess entropy production rate \cite{Zubarev,Mclennan,KN,KNST-rep,Maes-rep}.
We attempt to derive a potential function of   thermodynamic
quantities    for the phase coexistence in the heat conduction by
contracting the stationary distribution of configurations.
  Once the potential function is derived, all thermodynamic quantities
are determined as an extremal point of the potential.  
This is nothing but a variational principle for determining
  thermodynamic properties.    We may say that our theoretical challenge is
the derivation of such a variational principle. 

\paragraph{Key concept}


  For a  standard  setup where two heat baths
contact to boundaries of the system, the problem mentioned above is
too difficult to solve because of the following two reasons. 
First, since the expectation value of a thermodynamic 
quantity is determined from the time correlation between this quantity
and the excess entropy production, derivation of the potential function
requires  analysis of such time-dependent statistical quantities.
Second, in the equilibrium limit for this setup, the thermodynamic
quantities are not uniquely determined so that the variational principle
is not formulated. Thus, it is not straightforward to perform a perturbation
approach from the equilibrium case.  In order to overcome   these
two difficulties,    we come up with a key concept of this paper.   
We impose a special boundary condition, where the constant energy flux
is assumed at boundaries so that the energy of the system is conserved.
  See Fig. \ref{fig:setup} as an illustration.  
We refer  to this as the {\it non-equilibrium adiabatic condition}.
In equilibrium cases, this boundary condition is the standard
adiabatic condition, where the total energy is conserved  over time
  without an external operation. The variational
principle for determining thermodynamic properties here
is well-established as the maximal principle of the total entropy.
Thus, for the non-equilibrium adiabatic condition in the linear
response regime, we can develop  a perturbation theory for extending 
this variational principle.

\paragraph{Analysis}

Towards the derivation of the variational principle, 
in   Secs. \ref{stationary distribution}
and \ref{ent-interface},    we derive the stationary
distribution of interface configurations by analyzing the
Zubarev--Mclennan distribution.   
We can calculate the time integration
of excess entropy production rate for the configuration with a single
interface shown in Fig. \ref{fig:setup}. 
Explicitly, we consider the relaxation to the equilibrium
state from this configuration   and we find that the time integration
of excess entropy production rate is decomposed into three parts,
each of which is defined in the ordered region, the disordered region,
and the interface region.    See  (\ref{I-decom-int})
for the decomposition.     In the ordered and disordered
regions, because the process may be well-described by the deterministic
equation, we can explicitly solve it. 
We then  estimate this contribution 
to the excess entropy production    as (\ref{I-bulk-result}).

However, calculating the contribution to the excess entropy
production in the interface  region is not straightforward.
Physically, the latent heat is generated at the moving interface
in the relaxation 
process. This heat diffuses into  both regions, and as the result,
the entropy production is observed. Moreover,
a macroscopic temperature gap appears in the moving
interface, as observed in experiments \cite{gap}. This is another
source of entropy production. We estimate this contribution
with some approximation   as (\ref{Ical-int-result}).  

\paragraph{Result}


By using these results for the particular setup,
in Sec. \ref{Sec:var},  we  derive a potential function
of the interface position in the macroscopic limit.   See
(\ref{Phi-def}) for the final form of the potential function 
defined by (\ref{pot-def}).  
That is, the interface 
position is  uniquely determined by the variational principle
for  the phase coexistence in  heat conduction. The variational
function is a modified  entropy of the steady-state profile
for a given interface position.  
Solving the variational equation, we calculate the interface
temperature $\theta$   as  (\ref{temp-1}), which indicates that
a super-heated  ordered state or a super-cooled disordered
state stably appears locally near the interface.  
It should be noted that the expectation value of a thermodynamic
quantity would be independent of boundary conditions if the energy
flux and  energy are specified. We thus expect that  our result
is available even for  cases where two heat baths contact at boundaries,
which is a standard setup for heat conduction. 


From a theoretical viewpoint, the variational principle
for determining thermodynamic properties out of equilibrium
has never been considered in previous studies. 
For example, it has been known that the minimum entropy production
principle may characterize 
the steady state in the linear response regime \cite{min-ent}.
However, in the most general form, the variational principle is
formulated for determining 
the statistical ensemble in the linear response regime
as that  minimizes the entropy production as a
function of probability density \cite{Klein,Maes-LD}.
Although one may expect that the variational principle for
thermodynamic properties is obtained from the variational
principle for the statistical ensemble, this remains
too formal to calculate thermodynamic values explicitly. 
As another example of recent activities in the variational
principle, we recall those coming from the large deviation theory
\cite{Derrida,Maes-LD, Nemoto,Bertini-rev}. In these theories,
the main concern is  fluctuation properties, while thermodynamic
values are assumed to be obtained immediately.  Thus, our theoretical
framework is regarded as  essentially different from existing
approaches in fluctuation theory.


\paragraph{Note}

The final section is devoted as concluding remarks and several
technical details are separately discussed in Appendices.
The Boltzmann constant is set to unity, and the inverse temperature
$\beta$ is always connected to the temperature $T$ as $\beta=1/T$
without an explicit remark.

\section{Order parameter dynamics}\label{Sec:det}

We consider a system confined in a cuboid 
\begin{equation}
{\cal D}=\{(x,y,z)|0 \le x \le L, 0 \le y \le L_y,  0 \le z \le L_z \}
\end{equation}
with  $L > L_y, L_z$. When we study an equilibrium system,
we assume that the system is enclosed by adiabatic walls.
We also assume that the system exhibits an order-disorder
transition at $T=\Tc$ under the equilibrium condition and
that the transition is  the first-order, that is, the order
parameter shows discontinuous change at $T=T_c$ when decreasing
the temperature from a sufficiently high-temperature state.
In Sec. \ref{entropy-functional},  we first
consider the entropy functional 
of  the internal energy density field and the order parameter
density field.
In Sec. \ref{e-dynamics}, we derive
a deterministic equation for equilibrium cases following the
Onsager theory.
In Sec. \ref{Sec:formal-model},
we study  a stochastic model associated with
the deterministic equation.
We then  present  a dimensionless
form of the equation in Sec. \ref{scaling}.
  The final form of the model we study is given by 
(\ref{model-s1}),  (\ref{model-s2}), and (\ref{model-s3}).  
In Sec. \ref{heat conduction}, we set up the heat
conduction systems.

\subsection{Entropy functional}\label{entropy-functional}


Let $m(\bv{r})$ be an order parameter density field. For simplicity,
we consider the scalar order parameter. The generalization to
other complicated symmetry breakings is straightforward.
  We employ a mesoscopic description by assuming 
that the internal energy density $u(\bv{r})$ and the order
parameter density $m(\bv{r})$ are defined as those averaged over
a mesoscopic region with a length scale $\Lambda$ at each space $\bv{r}$.
Here, the mesoscopic length $\Lambda$ is chosen so as to satisfy 
\begin{equation}
\ell \ll \Lambda \ll L  
\label{L0}
\end{equation}
with a microscopic length scale $\ell$, such as
the size of atoms. A deterministic macroscopic equation
emerges from a microscopic description as a result of the law of
large numbers \cite{lps},  which is applied  to systems with
the separation of two scales: a microscopic length $\ell$ and
the system size $L$. By introducing the ratio of the two scales as
\begin{equation}
\eta \equiv \frac{\ell}{L},
\label{L00}
\end{equation}  
we express the separation of the scales as $\eta \to 0$, which corresponds
to the thermodynamic limit in equilibrium statistical mechanics. 
Note that the condition (\ref{L0}) is necessary for describing spatial
variation of local thermodynamic quantities.   
In the  argument below, we specifically   set
\begin{equation}
\Lambda =L \sqrt{\eta}
\label{Lam-est}
\end{equation}
for small $\eta$.


We assume  an  entropy density  function  $s(u,m)$
for a given material. We then have
\begin{equation}
  s(\bv{r})=s(u(\bv{r}), m(\bv{r})).
\label{t-relation-0}
\end{equation}
All thermodynamic quantities are determined from 
(\ref{t-relation-0}) with the fundamental relation
\begin{equation}
ds=\frac{1}{T} du+\frac{\sigma}{T} dm,   
\label{f-relation}
\end{equation}
where $T$ is the temperature and $\sigma$ corresponds
to the   thermodynamic force     conjugate to $m$. 
The free energy density $f(\bv{r})$ is defined by
\begin{equation}
  f(\bv{r})=u(\bv{r})-T(\bv{r})s(\bv{r}).
\end{equation}

For any field $a(\bv{r})$,  the configuration
$(a(\bv{r}))_{\bv{r} \in {\cal D}}$ is simply denoted by $a$.
The total entropy of the system, which is given as a functional
of configurations  $(m,u)$ , is expressed as 
\begin{equation}
  {\cal S}(m,u) =  \int_{\cal D} d^3\bv{r}  
  \left[ s(u(\bv{r}), m(\bv{r}))
    -\frac{d_s}{2} |\bv{\nabla} m|^2 \right],
\label{e-fun}
\end{equation}
where the gradient term represents an  entropy associated with
the order parameter density gradient which may be most relevant in the
interface.  For mathematical simplicity,  
we impose the boundary condition 
\begin{equation}
(\bv{\nabla}m) \bv{n}=0
\label{m-con}
\end{equation}
at the boundaries with the unit normal vector $\bv{n}$.  
Hereafter, the notation ${\cal D}$ in the space integral
will be omitted. We assume that $d_s$ is
constant, for simplicity.
The inclusion of the gradient term implies that
$s(u(\bv{r}), m(\bv{r}))$ is interpreted as the
mesoscopic entropy density. We assume that the
mesoscopic entropy density is given by the mean-field
entropy density, in which nucleation events are not
taken into account.
Although it seems difficult to justify this picture
from a microscopic description,  (\ref{e-fun}) with $s(u,m)$
may be a good starting hypothesis for a phenomenological mesoscopic
approach. We ignore  an entropy term of the form $|\bv{\nabla} u|^2$
in (\ref{e-fun}), for simplicity.

\begin{figure}[tb]
\centering
\includegraphics[width=6.5cm]{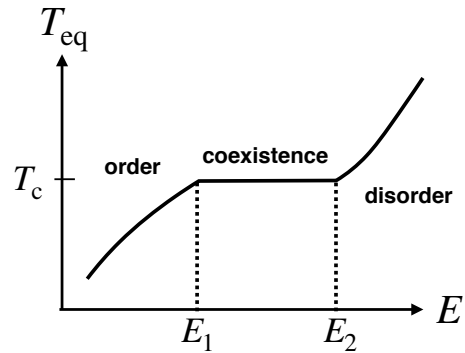}
\caption{ Schematic graph of  
$T_{\rm eq}$ as a function of $E$.
The phase coexistence is
observed at $T_{\rm eq}=T_c$ for $E_1 \le E \le E_2$.}
\label{fig:T-E}
\end{figure}

For a given total energy $E$,
the equilibrium value
\begin{equation}
  (m_{\rm eq}(\bv{r}),u_{\rm eq}(\bv{r}))  
\end{equation}
is determined as that  maximizes ${\cal S}$ under the energy
conservation 
\begin{equation}
  \int d^3\bv{r} u(\bv{r})=E.
\label{e-con-0}
\end{equation}
In the equilibrium state, the temperature $T(\bv{r})$
is uniform in space, which is denoted by $T_{\rm eq}$. In Fig.~\ref{fig:T-E},
we plot this $T_{\rm eq}$ as a function of $E$. We here find a plateau 
\begin{equation}
  T_{\rm eq}=\Tc
\end{equation}
in the region $E_1 \le E \le E_2$, where $E_1$ and $E_2$ are
calculated as
\begin{eqnarray}
  E_1 &=& u^{\rm o}(\Tc)LL_yL_z, \label{53}\\
  E_2 &=& u^{\rm d}(\Tc)LL_yL_z. \label{54}
\end{eqnarray}
  $u^{\rm o}(\Tc)$ and $u^{\rm d}(\Tc)$ are internal energy
densities in the ordered region and disordered region, respectively,
in the coexistence phase.  
Let $m_{\rm loc}(T)$ be the non-trivial value of $m$
for a specific model.   An explicit example of
$m_{\rm loc}(T)$ is shown in Appendix \ref{example}. 
See (\ref{m-loc}) for the model (\ref{GL-ex}).   
We then have 
\begin{eqnarray}
  u^{\rm o}(T) &=&  u(T, m_{\rm loc}(T)), \label{uo-def}  \\
  u^{\rm d}(T) &=&  u(T, m=0) . \label{ud-def}
\end{eqnarray}
These provide the explicit forms of $u^{\rm o}(\Tc)$
and $u^{\rm d}(\Tc)$  in (\ref{53}) and (\ref{54}).
In the plateau region,  $(m_{\rm eq}(\bv{r}),u_{\rm eq}(\bv{r}))$  
is not homogeneous in space; the ordered state $(m=m_{\rm loc}(\Tc))$
and the disordered state $(m=0)$ coexist with the minimum surface
of the interface between the two states.

Now, we define the momentum density field $v(\bv{r})$
conjugate to ${m}(\bv{r})$ as 
\begin{equation}
v \equiv \partial_t m.
\label{vdef}
\end{equation}
The energy density field $\phi(\bv{r})$ consists
of the internal energy density field $u(\bv{r})$,
the kinetic energy of the order parameter density
$v(\bv{r})^2/2$, and the energy contribution
of the order parameter density gradient which is
most relevant in the interface.
Note that
$v(\bv{r})^2/2$ is separated from $u(\bv{r})$,
which is standard in fluid dynamics \cite{Landau-Lifshitz-Fluid}. 
That is, $\phi(\bv{r})$ is expressed as
\begin{equation}
\phi(\bv{r})= u(\bv{r})+\frac{v(\bv{r})^2}{2}
+\frac{d_e}{2} 
|\bv{\nabla} m|^2 ,
\label{phi-wt-2}
\end{equation}
where $d_e$ is assumed to be constant, for simplicity.
The energy conservation is now written as 
\begin{equation}
  \int d^3\bv{r} \phi(\bv{r})=E.
\label{e-con}
\end{equation}

We consider the entropy functional ${\cal S}$
as a functional of $( m, v, \phi)$  with the energy
conservation (\ref{e-con}).
Explicitly, we express
\begin{align}
&  {\cal S}(m,v,\phi) \nonumber  \\
&=  \int d^3\bv{r}  
  \left[ s\left(\phi(\bv{r})
         -\frac{v(\bv{r})^2}{2}
        -\frac{d_e}{2} |\bv{\nabla} m|^2 
    , m(\bv{r})  \right) \right. \nonumber \\
& \quad  \left.  -\frac{d_s}{2} |\bv{\nabla} m|^2 \right].
\label{e-fun2}
\end{align}
The entropy functional including the gradient term
was used in Refs. \cite{Penrose,HHM}.
The same concept naturally appears in the  hydrodynamic equations
with the interface thermodynamics \cite{Onuki,Fujitani}.
The entropy functional in Ref. \cite{Fukuma-Sakatani} also takes
a similar form, but it employs
the gradient expansion around the global equilibrium which
is different from the gradient expansion around the local
equilibrium shown in (\ref{e-fun2}).

  Related to $u^{\rm o}$ and $u^{\rm d}$,
it is useful to introduce the heat capacity $c^{\rm o/d}$
without an external field defined as
\begin{eqnarray}
  c^{\rm o}(T) &\equiv&  \frac{d u(T, m_{\rm loc}(T))}{dT},  \label{zero-cap-o} \\
  c^{\rm d}(T) &\equiv&  \frac{d u(T, m=0)}{dT}  .\label{zero-cap-d}
\end{eqnarray}
We also define   the entropy densities as   
\begin{eqnarray}
  s^{\rm o}(T) &\equiv& s(T, m_{\rm loc}(T)), \label{so-def} \\
  s^{\rm d}(T) &\equiv& s(T, m =0). \label{sd-def} 
\end{eqnarray}
We then have
\begin{eqnarray}
  c^{\rm o}(T) &=&  T\frac{d s^{\rm o}(T)}{dT},  \label{zero-cap-o-2} \\
  c^{\rm d}(T) &=&  T\frac{d s^{\rm d}(T)}{dT} . \label{zero-cap-d-2}
\end{eqnarray}

\subsection{Deterministic dynamics for equilibrium cases}\label{e-dynamics}

For  the entropy functional ${\cal S}$ in (\ref{e-fun2}), we calculate
the functional derivative  as 
\begin{align}
&  \var{\cal S}{m(\bv{r})} =  \pderf{s}{m}{u}+
  d_e (\bv{\nabla}\beta)(\bv{\nabla} m) 
  + \beta d_f\Delta m , \label{37}\\
&  \var{\cal S}{v(\bv{r})} =  - {\beta v}, \label{36}\\
&  \var{\cal S}{\phi(\bv{r})} = \beta. \label{38}
\end{align}  
Here, we have defined the coefficient of the gradient
contribution to the free energy density as 
\begin{equation}
  d_f\equiv d_e+T d_s 
\label{dfdef}
\end{equation}
with constants $d_e$ and $d_s$.
 From (\ref{vdef}) and (\ref{36}), we have 
\begin{equation}
 \partial_t m = -T  \var{\cal S}{v(\bv{r})}. \label{ons-m} 
\end{equation}
      Since
      the right-hand side of (\ref{ons-m}) is 
      a reversible term that yields no entropy production,
      $\partial_t v$ should contain a corresponding reversible term.
      We then assume that the simplest momentum dissipation
      term $-\gamma v$ is contained in $\partial_t v$, where
      $\gamma$ is assumed to be a positive constant.
      That is, using (\ref{36}) and (\ref{ons-m}), we write
\begin{equation}
 \partial_t v = \gamma T  \var{\cal S}{v(\bv{r})}
                 + T \var{\cal S}{m(\bv{r})}.
                 \label{ons-v}  
\end{equation}
Finally, from the energy conservation (\ref{e-con}),
we assume the minimum form of the time evolution of $\phi$:
\begin{equation}
 \partial_t \phi =
- \bv{\nabla} \left(\lambda \bv{\nabla}  \var{\cal S}{\phi(\bv{r})} \right),
   \label{ons-phi}
\end{equation}
where $\lambda$ is a function of $(T,m)$.
The thermal conductivity $\kappa$ is related to
$\lambda$ as 
\begin{equation}
\kappa=  \frac{\lambda}{T^2}.
\label{kappa-def}  
\end{equation}
For the model (\ref{ons-m}), (\ref{ons-v}) and (\ref{ons-phi}),
we confirm the monotonic increment of ${\cal S}$ in time,
which is explicitly calculated as 
\begin{align}
    \der{\cal S}{t}
  &= 
 \int d^3{\bv r}
 \left[  \var{\cal S}{m }\partial_t m+\var{\cal S}{v}\partial_t v
   +\var{\cal S}{\phi}\partial_t \phi
   \right] \nonumber \\
  &= 
 \int d^3{\bv r}
 \left[ \gamma T \left(\var{\cal S}{v(\bv{r}) } \right)^2
   + \lambda   \left| \bv{\nabla}\var{\cal S}{\phi(\bv{r})} \right|^2 \right]
    \nonumber \\
& \qquad
 - \int d^3{\bv r} \bv{\nabla}( \beta \lambda \bv{\nabla} \beta ) \nonumber \\
&=
 \int d^3{\bv r}
 \left[ \gamma T \left(\var{\cal S}{v(\bv{r}) } \right)^2
   + \lambda   \left| \bv{\nabla} \var{\cal S}{\phi(\bv{r})} \right|^2
   \right] \nonumber \\
& \ge  0, 
 \label{ent-dt}
\end{align}
where  we have used the adiabatic condition
\begin{equation}
(\bv{\nabla}\beta) \bv{n}=0
\label{ad-con}
\end{equation}
at the boundaries with the unit normal vector $\bv{n}$.
The expression (\ref{ent-dt}) shows that the right-hand side of
  (\ref{ons-m}) and the second term in the right-hand side of
  (\ref{ons-v}) yield no entropy production.

By substituting (\ref{37}), (\ref{36}) and (\ref{38}) into
the equations (\ref{ons-m}), (\ref{ons-v}), and (\ref{ons-phi}),
we obtain the explicit form of the equations as 
\begin{align}
&  \partial_t m = {v},  \label{ons-m-d} \\
&  \partial_t v = -\gamma v +\sigma
  +T d_e (\bv{\nabla}\beta)(\bv{\nabla} m)
  + d_f \Delta m, 
  \label{ons-v-d} \\
&  \partial_t \phi =
  - \bv{\nabla}  \left(\lambda \bv{\nabla} \beta \right),
   \label{ons-phi-d}
\end{align}
where the thermodynamic force $\sigma$ is given by
\begin{equation}
  \sigma= T\pderf{s}{m}{u}.
\label{sigma}
\end{equation}
  See (\ref{f-relation}).   
From the thermodynamic relation   
\begin{equation}
  -\pderf{f}{m}{T}=T \pderf{s}{m}{u},
\label{fmt}
\end{equation}  
one can rewrite the thermodynamic force $\sigma$ as
\begin{equation}
  \sigma= -\pderf{f}{m}{T}.
\label{sigma2}
\end{equation}
By using (\ref{phi-wt-2}) and (\ref{ons-v-d}),
      we can express the last equation
(\ref{ons-phi-d})  for the case that  $d_e=d_s=0$ as 
\begin{equation}
  \partial_t u =
  {\gamma v^2} -\sigma \partial_t m
    - \bv{\nabla} (\lambda \bv{\nabla} \beta) .
\label{u-evol}
\end{equation}
The first term of the right-hand side represents the generating heat
caused by the momentum dissipation,  the second term is associated
with the work done by the thermodynamic force, 
and the third term the heat conduction.

\begin{figure}[tb]
\centering
\includegraphics[width=7cm]{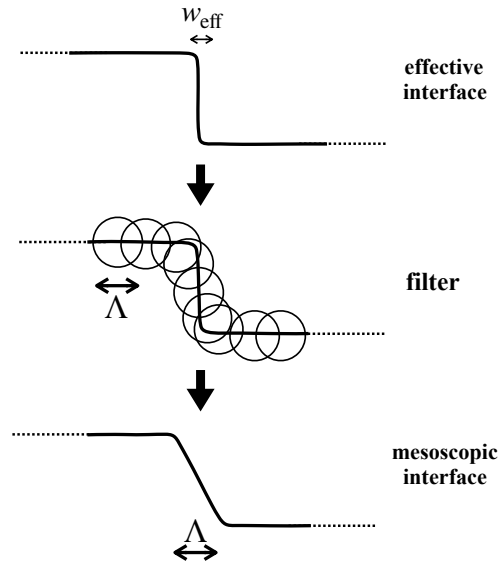}
\caption{
  The statistical average of a single interface is represented by
  an effective interface whose width remains microscopic. By the spatial
  average over a region of length $\Lambda$,  the interface in the
  mesoscopic description is defined.  }
\label{fig:meso-int}
\end{figure}

The parameters $d_e$ and $d_f$ characterize the interface energy
and the interface free energy, respectively.  
Let us estimate the magnitude of $d_e$ and $d_f$. We first discuss
the interface width in the mesoscopic description.
Physically,  
the interface is identified as a deformed
surface of an intrinsic width $w$ which is at most $10^{-7}$ cm \cite{width}.
This width $w$ is of the same order as the  microscopic length $\ell$, and
the deformation of the surface is described by a capillary wave theory
or  fluctuation theory \cite{Triez-Zwanzig}. 
By averaging density profiles in the equilibrium ensemble, 
one has an effective interface of the width $w_{\rm eff}$
which is estimated as 
$w_{\rm eff}\simeq \ell \sqrt{\log (L/\ell)}$ for three-dimensional
systems \cite{Weeks}. 
We note here  that $w_{\rm eff}/L \to 0$ in the limit $\eta \to 0$.
That is, the interface in the deterministic hydrodynamic equation
is a singular surface  whose motion   has been formulated as
a free boundary problem \cite{Anderson98},  
but it should be noted that  when we keep the finiteness of the interface
width in the dynamics, the noise intensity also remains finite.
In the mesoscopic description we employ, all thermodynamic
quantities are spatially averaged over a region with the mesoscopic length 
$\Lambda$.  Thus, the interface width of the spatially averaged configuration
is given by   the mesoscopic length  $\Lambda$
up to a multiplicative numerical constant, as shown in Fig. \ref{fig:meso-int}.
Then,   since a typical value of  $d_f |\bv{\nabla} m|^2$ in the interface
region is estimated as $T_c \ell^{-3}$,     we have
\begin{equation}
  d_f \frac{m_*^2}{\Lambda^2} \simeq T_c \ell^{-3},
\label{df-est}
\end{equation}
where $m_*$ is the characteristic value of $m$ in the ordered
state and $\ell$ represents the microscopic length scale
mentioned in the first paragraph of Sec. \ref{entropy-functional}.   

\subsection{Stochastic dynamics for equilibrium cases}
\label{Sec:formal-model}

A collection of the configurations 
$m$, $v$, and $\phi$ is denoted by
\begin{equation}
\alpha=(m, v, \phi).
\label{alpha:def}
\end{equation}  
 
Recalling that the system is enclosed by
the adiabatic wall, we construct   a  
stochastic  model that 
yields the stationary distribution 
\begin{equation}
{\cal P}_{\rm eq}(\alpha)
= {\cal N}\exp({\cal S}(\alpha))
\delta \left( \int d^3\bv{r} \phi(\bv{r})-E \right)
\label{micro-canonical}
\end{equation}
for  the equilibrium case,  
where ${\cal N}$ is the normalization constant. It should
be noted that the energy conservation (\ref{e-con}) holds
for the stochastic systems. 
We  add Gaussian
white noises to  (\ref{ons-m-d}), (\ref{ons-v-d}),
and (\ref{ons-phi-d}) that  satisfy the detailed balance
condition. The noise intensity is related to the
dissipation intensity, which is  called the  fluctuation-dissipation
relation of the second kind.  We  then write
\begin{align}
&  \partial_t m = { v},
  \label{ons-m-n} \\
&  \partial_t v = -\gamma  v   +\sigma  +d_e  T(\bv{\nabla}\beta)(\bv{\nabla} m) \nonumber \\
&  \qquad\qquad
+ d_f \Delta m  
  +\sqrt{2\gamma   T} \xi^v ,    \label{ons-v-n} \\
&  \partial_t \phi =
  - \bv{\nabla} \left( \lambda  \bv{\nabla} \beta 
 +  \sqrt{2\lambda  } \bv{\xi}^\phi \right)
   \label{ons-phi-n},
\end{align}
where $\xi^v$ and $\bv{\xi}^\phi$ are Gaussian white noise.
For later convenience, we set 
\begin{eqnarray}
\xi^1 &=& 0 , \\  
\xi^2 &=& \xi^v ,  \\
(\xi^3,\xi^4,\xi^5)&=&{\bv \xi}^{\phi} .
\end{eqnarray}
The property of the Gaussian white noise is formally expressed as
\begin{eqnarray}
  \bra \xi^{a}(\bv{r},t) \xi^b(\bv{r}',t') \ket&=& 
  \delta^{ ab}\delta(\bv{r}-\bv{r'})\delta(t-t'),
    \label{ons-naive-noise}
\end{eqnarray}
where $2 \le a,b \le 5$.   It should be noted that the argument
so far is too formal. Indeed,  due to the multiplicative nature
of the noise, the formal model exhibits a singular behavior.
In Appendix \ref{s-model:detail}, we perform a  careful analysis
of the stochastic process.   

Historically, a deterministic
order parameter model with energetics was derived from
an entropy functional   as a phase field  model  that describes
crystal growth \cite{Penrose}. From this direction of research,
one may interpret the model we study as  a phase field model with noise.
  The equations in this
previous study correspond to the over-damped version of (\ref{ons-m-d}),
(\ref{ons-v-d}), and  (\ref{ons-phi-d}) with $d_e=0$. Similar
equations  were also considered in the context of critical phenomena
\cite{HHM}, where   another simple entropy
functional is assumed differently from our case.    
The model in this previous study \cite{HHM}, where the noise
was taken into account, 
was called {\it Model C} \cite{HH}.

\subsection{Scaling}\label{scaling}

We consider a dimensionless form of the equations 
(\ref{ons-m-n}), (\ref{ons-v-n}), and  (\ref{ons-phi-n}).
First, we define the dimensionless quantity $\breve Q$ for
any quantity $Q$  by
\begin{equation}
  Q=\breve Q Q_* ,
\label{Q-trans}
\end{equation}  
where $Q_*$, which is a characteristic value with the dimension, is
estimated below. We then introduce dimensionless space
coordinate $\breve{\bv{r}}$ and dimensionless time $\breve t$ 
so that the relaxation time of thermodynamic quantities,  
which is denoted by $\tau$, becomes the unity  in this dimensionless
time $\breve t$.  That is, we set  
\begin{equation}
(\bv{r},t)=  (L \breve{ \bv{r}}, \tau \breve t).
\label{c-trans}
\end{equation}
Note that the choice of dimensionless coordinates
$( \breve{\bv{r}}, \breve t)$ is arbitrary, and we choose this
macroscopic unit for later convenience.  This is in contrast with
$\breve Q$, which is determined by  the physical properties of
natural phenomena. 

By substituting (\ref{Q-trans}) and (\ref{c-trans}) into
(\ref{ons-m-n}), (\ref{ons-v-n}) and  (\ref{ons-phi-n}),
we have
\begin{align}
&  \partial_{\breve t} \breve m = \Gamma_1 \breve v ,
  \label{ons-m-n2} \\
&  \partial_{\breve t} \breve v = -  \Gamma_2 \breve v
  +\Gamma_3 \breve \sigma
  + \Gamma_4 \left[ \breve T \breve d_e( \breve{ \bv{\nabla}} \breve\beta)
     (\breve{ \bv{\nabla}} \breve m) 
  +   \breve d_f \breve \Delta \breve m \right]
 \nonumber \\
& \qquad\qquad +\sqrt{2 \Gamma_5 \breve T} \breve \xi^v ,
  \label{ons-v-n2} \\
&  \partial_{\breve t} \breve \phi =
  - \breve{ \bv{\nabla}}
  \left( \Gamma_6 \breve \lambda  \breve{\bv{ \nabla}}\breve \beta 
 +  \sqrt{2\breve \lambda \Gamma_7 } \breve{ \bv{\xi}}^\phi \right)
   \label{ons-phi-n2},
\end{align}
where we have introduced dimensionless parameters
\begin{align}
  &  (\Gamma_1, \Gamma_2, \Gamma_3, \Gamma_4, \Gamma_5,\Gamma_6, \Gamma_7)
  \nonumber \\
&=
  \left(
  \frac{v_* \tau}{m_* }, \gamma \tau, \frac{\sigma_* \tau}{v_*},
  \frac{(d_f)_* m_* \tau}{L^2 v_*}, 
  \frac{\gamma  T_* \tau}{L^3 v_*^2},
  \frac{ \lambda_* \beta_* \tau}{L^2 \phi_*},
  \frac{\lambda_* \tau}{L^5 \phi_*^2}
  \right).
\end{align}
Here, we have assumed $(d_e)_*=(d_f)_*$ from (\ref{dfdef}). 
The characteristic values of the quantities are estimated
by using $T_c$, $\tau$, $L$, and the microscopic length $\ell$.
Concretely, first, it is obvious $T_*=T_c$. Second, from the
equipartition law, $\phi_*$ is estimated as $T_c\ell^{-3}$
up to a multiplicative numerical constant.
From (\ref{vdef}) and (\ref{phi-wt-2}),
we find that $v_*^2 =  \phi_*$ and $m_*=\tau v_*$;
and from  (\ref{sigma2}), we have $\sigma_*= \phi_*/m_*$. Finally,
since $\lambda$ determines the diffusion time
scale of the energy, we obtain 
\begin{equation}
  \lambda_* = \Tc \phi_*\frac{L^2}{\tau}.
\end{equation}
From (\ref{df-est}), we also have 
\begin{equation}
(d_f)_*= \frac{\Lambda^2}{m_*^2}T_c\ell^{-3}.  
\end{equation}
By substituting these results, we obtain
\begin{align}
 (\Gamma_1, \Gamma_2, \Gamma_3, \Gamma_4, \Gamma_5,\Gamma_6, \Gamma_7)
=
  \left(
  1, \breve \gamma, 1, \eta, \breve \gamma\eta^3,
  1, \eta^3
  \right),
\end{align}
where we set $ \breve \gamma = \gamma \tau$ and
  we have used $\eta$ defined by (\ref{L00}),
which is assumed to be sufficiently small.  
Moreover, we consider the dimensionless energy
$\breve E$   defined    by
\begin{equation}
  E = \breve E  T_c \left( \frac{L}{\ell} \right)^3 .
\end{equation}
The   mesoscopic length    $\Lambda$ is also expressed
as $\Lambda = \breve \Lambda  L$, where $\breve \Lambda$
is written as  
\begin{equation}
  \breve \Lambda = \sqrt{\eta}
\label{Lambda-L}
\end{equation}
from (\ref{Lam-est}).

Here, in order to simplify the notation, we remove all breve
symbols. The final expression then becomes
\begin{align}
&  \partial_{ t}  m =   v ,
  \label{model-s1} \\
&  \partial_{ t}  v = -   \gamma  v   +\sigma
  +  \eta \left[  T d_e ( \bv{\nabla}\beta)
  ( \bv{\nabla}  m)  +  d_f   \Delta  m \right]
 \nonumber \\
& \qquad\qquad +\sqrt{2  \gamma  \eta^3 T} \xi^v ,
  \label{model-s2} \\
&  \partial_{ t}  \phi =
  -\bv{\nabla} \left(   \lambda   \bv{\nabla}  \beta 
 +  \sqrt{2 \lambda \eta^3 }  \bv{\xi}^\phi \right)
   \label{model-s3}
\end{align}
with the small parameter  $\eta \ll 1$ that represents
the separation of scales.

  Furthermore, when we study the
deterministic systems, we analyze the noiseless limit
of (\ref{model-s1}), (\ref{model-s2}), (\ref{model-s3}):
\begin{align}
&  \partial_{ t}  m =   v ,
  \label{ons-m-d-s} \\
&  \partial_{ t}  v = -   \gamma  v   +\sigma
  +  \eta \left[  T d_e ( \bv{\nabla}\beta)
  ( \bv{\nabla}  m)  +  d_f   \Delta  m \right]
  \label{ons-v-d-s} \\
&  \partial_{ t}  \phi =
  -\bv{\nabla} \left(   \lambda   \bv{\nabla}  \beta  \right)
   \label{ons-phi-d-s}
\end{align}
instead of (\ref{ons-m-d}), (\ref{ons-v-d}), and (\ref{ons-phi-d}).
It should be noted that the dimensionless space coordinate $(x,y,z)$ satisfies
$0 \le x \le 1$, $0 \le y \le L_y/L$, and $0 \le z \le L_z/L$. Hereafter,
we set
\begin{equation}
  A=\frac{L_yL_z}{L^2} ,
\label{Adef}
\end{equation}  
which is the dimensionless area of the cross-section
of the system.


When we consider a symmetry-breaking phase, the long time
behavior of the system for finite $\eta$ is different from
that for the system in the limit $\eta \to 0$. In order to
avoid such a singular behavior, we add a small symmetry-breaking field
$\sigma^{\rm ex}(x)$ to the right-hand side of (\ref{model-s2}), and
consider the limit $\sigma^{\rm ex }(x) \to 0$ in the last step.
  Here, $\sigma^{\rm ex }(x)$ is spatially inhomogeneous  so as
to break the left-right symmetry. Specifically, we set 
$\sigma^{\rm ex}(x) > 0 $ for $x \in [0, 1/2]$ and 
$\sigma^{\rm ex}(x) = 0 $ for $x \in [1/2, 1]$ such that
the equilibrium configuration is continuously deformed to that
in the heat conduction state with $J <0$.  
In the argument below, we do not write this term explicitly but
we always keep this process in mind.

\subsection{Non-equilibrium adiabatic conditions}\label{heat conduction}

\subsubsection{Deterministic cases}

We study the heat conduction by using the equations
(\ref{ons-m-d-s}), (\ref{ons-v-d-s}), and (\ref{ons-phi-d-s}) 
with the boundary condition
\begin{equation}
  \lambda \partial_x\beta(0,y,z)=  \lambda\partial_x\beta(1,y,z)=J
\label{flux-con-det}
\end{equation}
at the boundaries $x=0$ and $x=1$ instead of (\ref{ad-con}),
while (\ref{ad-con}) holds at the other boundaries.
Without loss of  generality, we assume $J \le 0$.
The condition (\ref{flux-con-det}) implies that the energy
flux is kept constant at the boundaries. A remarkable
property of the boundary condition is that the total energy
of the system is conserved. From this property, we call
(\ref{flux-con-det}) with $J\not =0$ a {\it non-equilibrium
adiabatic condition}, which is contrasted with more standard boundary
conditions $T(0,y,z)=T_{\rm L}$ and $T(1,y,z)=T_{\rm R}$.
We impose the special boundary condition (\ref{flux-con-det})
for a technical reason to analyze stochastic systems.

\subsubsection{Stochastic cases}

  We attempt to extend (\ref{flux-con-det}) to the stochastic systems.
We expect the following two conditions. The first condition is that 
the stationary distribution is given by (\ref{micro-canonical})
when $J =0$. The second condition is that when $J \not = 0 $, similarly
to the deterministic description, non-equilibrium nature is brought
only by the boundary condition with keeping the energy conservation.
Concretely, we impose the boundary condition
\begin{eqnarray}
 j_x(x=0,y,z,t) &=& J,  \label{noise-21}\\
 j_x(x=1,y,z,t) &=& J ,
 \label{noise-22}
\end{eqnarray}
and $\bv{j}\bv{n}=0$ at the other boundaries, 
where $\bv{j}$ is defined as
\begin{equation}
\bv{j}\equiv   \lambda   \bv{\nabla}  \beta 
 +  \sqrt{2 \lambda \eta^3 }  \bv{\xi}^\phi .
\label{j-def}
\end{equation}
We  easily confirm  that the two conditions are satisfied by
this boundary condition.   

\subsubsection{Linear response regime}

In order to represent the extent of the non-equilibrium,
we introduce a  dimensionless small parameter
\begin{equation}
  \epsilon \equiv \frac{|J |L T_c}{ \lambda_{*}}
\end{equation}  
using the original dimensional quantities. 
By introducing the dimensionless heat flux $\breve J$  as 
\begin{equation}
J = \breve J  \frac{\lambda_*}{T_c L} , \label{63}
\end{equation}
we find that $  |\breve J|=\epsilon $. Therefore,
\begin{equation}
  |J|=\epsilon
\label{J-ep}  
\end{equation}
in this dimensionless form.  
In the argument below, we focus on the linear response regime 
by studying  only the contribution of $O(\ep)$.

\section{Interface in the deterministic system}\label{interface}

In this section, we study the properties of the interface in the
deterministic system.   
In Sec. \ref{s-interface}, we analyze
the stationary interface in the equilibrium state.
In Sec. \ref{interface in heat conduction}, we analyze the interface
in the heat conduction. In Sec. \ref{problem}, we summarize 
the result for the deterministic system, and we show our
motivation of studying the stochastic system. 

\subsection{Equilibrium interface}\label{s-interface}

We study the deterministic system described by 
(\ref{ons-m-d-s}), (\ref{ons-v-d-s}), and (\ref{ons-phi-d-s}).
For any initial value of  $(m(\bv{r}), v(\bv{r}),\phi(\bv{r}))$,
the energy  $E$ is conserved over time
and $d {\cal S}/dt \ge 0$ for any $t$ as shown in
(\ref{ent-dt}). This means that
$(m(\bv{r},t), v(\bv{r},t),\phi(\bv{r},t))$ goes to
the equilibrium value
\begin{equation}
   \alpha_{\rm eq}(\bv{r})=  
     (m_{\rm eq}(\bv{r}), v_{\rm eq}(\bv{r})=0,\phi_{\rm eq}(\bv{r})),
\end{equation}
which maximizes ${\cal S} (\alpha) $
under the energy conservation.
In particular, when $E_1 \le E \le E_2$,
where $E_1$ and $E_2$ are given by (\ref{53}) and (\ref{54}),
the equilibrium temperature takes the constant value $\Tc$ 
as shown in Fig.~\ref{fig:T-E}.
In this equilibrium state,
the temperature  is homogeneous in space  
such that $T(\bv{r})=\Tc$,   while
$(u_{\rm eq}(\bv{r}),m_{\rm eq}(\bv{r}))$ is not homogeneous in space;
the ordered state $(m=m_{\rm loc}(\Tc))$ and the disordered state $(m=0)$
coexist with the minimum surface of the interface between the
two states. 

\begin{figure}[tb]
\centering
\includegraphics[width=8cm]{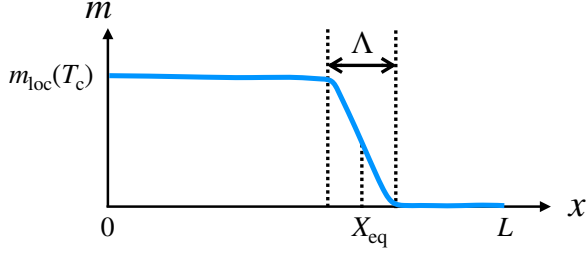}
\caption{Schematic figure of a stationary interface   in equilibrium.
$\Lambda = \sqrt{\eta}$ and $L=1$ in the dimensionless form.  }
\label{fig:s-interface}
\end{figure}

We derive an expression of $m_{\rm eq}(\bv{r})$ for $E_1 \le E \le E_2$. 
Since  the horizontal length $L$, which is now normalized as unity,
is larger than the lengths 
of other directions $L_y$ and $L_z$,  the stationary
interface  is perpendicular to the $x$-axis.
Furthermore, from (\ref{ons-v-d-s}) with (\ref{sigma2}),
we find that the stationary interface described by
$m=m_{\rm eq}(x)$ satisfies
\begin{equation}
-\pder{f(\Tc,m)}{m} +  \eta d_f\partial_x^2 m =0
\label{eq-eq}
\end{equation}
with the boundary conditions (\ref{m-con}).
Let $X_{\rm eq}$ be the stationary
interface position for a given value
of the total energy $E$, as shown in Fig. \ref{fig:s-interface}.
We consider the case that the ordered
state appears on the left-side. Then, $X_{\rm eq}$ is determined by
\begin{equation}
X_{\rm eq}  u^{\rm o}(T_c) +(1-X_{\rm eq}) u^{\rm d}(T_c)=\frac{E}{A}
\label{X-eq}
\end{equation}
in the limit $\eta \to 0$. Looking at (\ref{eq-eq}), 
we express the solution of (\ref{eq-eq}) with $ \eta \ll 1$ as 
\begin{equation}
  m_{\rm eq}(x)=\bar{\bar m}\left( \frac{x-X_{\rm  eq}}{\sqrt{\eta}}
  \right) m_{\rm loc}(T_c).
\label{m-st}
\end{equation}
The quantity $\bar{\bar m}(\xi)$, which 
describes an internal structure of the interface,
then satisfies
\begin{equation}
  -\pder{f(T_c,\bar{\bar m} m_{\rm loc}(T_c))}{\bar{\bar m}} +
  {m_{\rm loc}^2(T_c) d_f}
  \partial_{\xi}^2 \bar{\bar m} =0
\end{equation}
with $\xi = (x-X_{\rm eq})/\sqrt{\eta}$,
$\bar{\bar m}(0)=1/2$, $\bar{\bar m}(-\infty)=1$, 
$\bar{\bar m}(\infty)=0$.

\subsection{Interface in the heat conduction steady state}
\label{interface in heat conduction}

In this section, we derive the stationary interface
in the heat conduction based on the deterministic
description. That is, 
from (\ref{ons-m-d-s}), (\ref{ons-v-d-s}), and (\ref{ons-phi-d-s})
with (\ref{sigma2}), we find that 
the stationary solution satisfies
\begin{align}
-\pderf{f}{m}{T} +  \eta Td_e (\partial_x\beta)(\partial_x m)
+  \eta d_f \partial_x^2 m  &= 0 , \label{st-1} \\
\lambda \partial_x \beta &= J, \label{st-2} 
\end{align}
which are interpreted as the non-equilibrium extension of (\ref{eq-eq}). 

We analyze the equations (\ref{st-1}) and (\ref{st-2}).
Let $X_{\rm ss}$ be the position of the stationary interface
for $(E,J)$. We then determine the temperature of the interface
$\theta$ from (\ref{st-1}) and (\ref{st-2}) with $X_{\rm ss}$.
Multiplying $(\partial_x m)$ to (\ref{st-1}) and integrating
it over $I \equiv [X_{\rm ss}-M \sqrt{\eta},  X_{\rm ss}+ M\sqrt{\eta}]$
with a large $M$ independent of $\eta$, we obtain
\begin{align}
&  - \int_I  dx (\partial_x  m) \pderf{f}{m}{T}
  +  \eta\int_I dx Td_e (\partial_x\beta)(\partial_x m)^2 \nonumber \\
 &\qquad +  \eta \int_I dx d_f (\partial_x  m) (\partial_x^2  m)  = 0.
\label{82}  
\end{align}
Here, we note
\begin{equation}
\partial_x f= \pderf{f}{T}{m} \partial_x T+\pderf{f}{m}{T} \partial_x m
\end{equation}
and
\begin{align}
&  d_f (\partial_x m) (\partial_x^2 m)
  = \frac{d_f}{2}
   \partial_x ( (\partial_x m)^2 )  \nonumber \\
 &\qquad\qquad = \frac{1}{2}
   \partial_x ( d_f(\partial_x m)^2 )
         - \frac{d_s}{2} (\partial_x T) (\partial_x m)^2 .
\end{align}
By using these results, we further rewrite (\ref{82}) as 
\begin{align}
& f(X_{\rm ss}+M \sqrt{\eta})-f(X_{\rm ss}-M \sqrt{\eta})  \nonumber \\
  &~~ =  J
  \int_{X_{\rm ss}-M \sqrt{\eta}}^{X_{\rm ss}+M \sqrt{\eta}} dx
  \left[
    \frac{T \eta d_e (\partial_x  m)^2}{\lambda}
    + \frac{T^2 \eta d_s  (\partial_x  m)^2}{2\lambda} 
    \right]
\nonumber \\
&~~~~   -   \int_{X_{\rm ss}-M \sqrt{\eta}}^{X_{\rm ss}+M \sqrt{\eta}} dx
           s \partial_x T  
\nonumber \\
 &~~~~  + \frac{d_f}{2}
   \left. (\partial_x  m)^2 \right|_{X_{\rm ss}-M \sqrt{\eta} L}^{X_{\rm ss}+M \sqrt{\eta} L} .
\label{ba}
\end{align}  
The last term is proportional to $\epsilon^2$, because
\begin{equation}
  \partial_x m \simeq  -\der{m_{\rm loc}(T)}{T}T^2\frac{J}{\lambda}
\end{equation}  
at $x=X_{\rm ss}-M \sqrt{\eta}$, and
$\partial_x m=0$ at $x=X_{\rm ss}+M \sqrt{\eta}$.
The first  line   of the right-hand side
of (\ref{ba})  is rewritten   as
\begin{equation}
J \int_{X_{\rm ss}-M \sqrt{\eta}}^{X_{\rm ss}+M \sqrt{\eta}} dx
\eta \frac{T   (d_e+T d_s/2)  (\partial_x  m)^2}{\lambda}.
\end{equation}
 This   is estimated as 
\begin{equation}
\epsilon  {\eta}^{\frac{1}{2}}  d_f m_{\rm loc}(\Tc)^2 
\end{equation}
up to a numerical factor  when $\lambda$ in the interface
region is estimated as $\lambda_*$.  Thus, the first line 
of (\ref{ba}) is $O(\eta^{1/2})$. 
From these, the leading term of (\ref{ba}) becomes  
\begin{align}
f(X_{\rm ss}+M \sqrt{\eta})-f(X_{\rm ss}-M \sqrt{\eta}) 
=  - \int_{X_{\rm ss}-M \sqrt{\eta}}^{X_{\rm ss}+M \sqrt{\eta}}
dx s (\partial_x T). 
\label{ba22}
\end{align}
Furthermore, recalling $f=u-Ts$, we have
\begin{align}
u(X_{\rm ss}+M \sqrt{\eta})-u(X_{\rm ss}-M \sqrt{\eta}) 
=   \int_{X_{\rm ss}-M \sqrt{\eta}}^{X_{\rm ss}+M \sqrt{\eta}}
dx T (\partial_x s) . 
\label{ba3}
\end{align}
Let $\theta $ be the temperature of the interface,
defined by
\begin{equation}
  \theta=T(X_{\rm ss}).
\end{equation}
  Noting the continuity of $T(x)$ and ignoring $O(\eta^{1/2})$ terms,
we find that    (\ref{ba3}) becomes 
\begin{align}
  u^{\rm o}(\theta)-u^{\rm d}(\theta)
  = \theta [s^{\rm o}(\theta)-s^{\rm d}(\theta)],
\label{ba4}
\end{align}
where  $u^{\rm o}$ and $u^{\rm d}$ are defined by (\ref{uo-def})
and (\ref{ud-def}). $s^{\rm o}$ and $s^{\rm d}$ are also
defined by   (\ref{so-def}) and (\ref{sd-def}).  
We thus  obtain
\begin{equation}
\theta=T_c + O({\eta}^{\frac{1}{2}}).
  \label{theta-det-0}
\end{equation}
This estimate indicates that, in the limit $\eta \to 0$ with $\epsilon$ fixed, 
the stationary interface temperature in the heat
conduction state remains $\Tc$.

\subsection{Role of fluctuation}\label{problem}

If the deterministic equation correctly describes the
thermodynamic behavior, all thermodynamic quantities
are determined from  the stationary solution of the
equation. In particular, the interface temperature in
the heat conduction systems is equal to the equilibrium
transition temperature in the limit $\eta \to 0$. Now,
the question is whether or not the deterministic equation
is valid for the phase coexistence under heat conduction.

As a related example, let us recall the understanding of
a fluid consisting of many particles in two dimensions.
One may write the standard two-dimensional hydrodynamic
equation as a deterministic model describing the hydrodynamic
behavior. However, it has been known that the parameters in the equation,
the transportation coefficients, do not have a definite
value measured in experiments. Theoretically, this result
is understood as a singular (divergent) behavior of the
parameter values in the macroscopic limit on the  basis of
microscopic dynamics. In this sense, deterministic hydrodynamic
equations  are not valid for describing the dynamical behaviors of
a fluid consisting of many particles in two dimensions.
Even for this case, it is  expected that stochastic hydrodynamic
equations with well-defined parameters can describe the behavior
quantitatively. The consistency between the two models has
been understood from the renormalization group analysis \cite{FNS}.

In the phase coexistence under heat conduction, the interface
region is singular   because the interface width is
$O(\sqrt{\eta})$.    Thermodynamic quantities
in this thin interface region may be described  by equilibrium
statistical mechanics.   
Here we discuss how an energy fluctuation of $O(\sqrt{\eta}) $
in the ordered region evolves over time under the equilibrium condition.
The corresponding temperature fluctuation in the ordered region is
$O(\sqrt{\eta})$ because the heat capacity is $O(1)$. 
Then the energy flows into the interface region and  this leads to
the temperature change of $O(\sqrt{\eta})$ in the interface region,
which is achieved by the change in the interface position of
$O(\sqrt{\eta})$. Since the temperature difference over the interface
region is estimated as $O(\sqrt{\eta})$, $\partial_x \beta$ in the
interface region is $O(1)$.  Thus, the energy flux in the
interface region is expressed as  $\lambda_{\rm int} \times O(1)$,
where $\lambda_{\rm int}$ is the thermal conductivity in the interface
region. Since the energy flux of $O(\sqrt{\eta})$ in the bulk is
balanced with the energy flux in the interface region, it is expected
$\lambda_{\rm int} =O(\sqrt{\eta})$. That is, the singularity appears
in the limit $\eta \to 0$.

When we study the deterministic system (\ref{ons-m-d-s}), (\ref{ons-v-d-s}),
and (\ref{ons-phi-d-s}) with $\lambda_{\rm int} =O(\sqrt{\eta})$, the behavior
depends on the detail of $\lambda_{\rm int}$ even in the limit $\eta \to 0$.
For example, for heat conduction steady state, the temperature gap of $O(1)$
appears and the amount of the gap depends on $\lambda_{\rm int}/\sqrt{\eta}$
in the limit $\eta \to 0$.   
Here, let us recall that the energy transfer from/to the interface
region to/from the bulk is basically induced by fluctuations of the interface.
Therefore, the stochastic noise  is inevitable for the description of the
 energy transfer.  
Even if we assume that the ``bare conductivity'' in the interface region,
which is a parameter of the stochastic model, is $O(\sqrt{\eta})$,
the ``measured conductivity'' in the interface region may be $O(1)$
as the result of
the renormalization of fluctuations. This leads to no  temperature
gap in the limit $\eta \to 0$, but this is not described as the
limit $\eta \to 0$ of a deterministic equation. It should be
noted that the energy transfer occurs as the result of fluctuations
of the interface position is similar to the so-called adiabatic piston problem
\cite{Callen,Feynman,Lieb,Gruber,Gruber2}.


\section{Stationary distribution for interface configurations}
\label{stationary distribution} 


We start this section with the Zubarev-Mclennan representation
of the stationary distribution for heat conduction systems
in Sec. \ref{ZM}. The probability density is an extension
of the micro-canonical ensemble and we naturally define a
modified entropy which contains a correction term ${\cal I}$
in addition to the entropy ${\cal S}$.  Note that ${\cal I}$ is the
time integration of the entropy production. Then, 
for a single interface configuration $\alpha_X$ defined
in Sec. \ref{int-def}, we attempt to express ${\cal I}$ as a form 
without the time integration. If it is done successfully, we
can formulate the variational principle so that all thermodynamic
quantities can be determined as that maximizing the modified entropy.
  We thus attempt to evaluate ${\cal I}$.    
Concretely,   in Sec. \ref{correction}, we decompose 
${\cal I}$ into the bulk contribution and the interface contribution.
Then,  in Sec. \ref{bulk},
we estimate the bulk contribution to ${\cal I}$. This will be done quite
easily thanks to the boundary condition we impose. This calculation also
gives the correction term ${\cal I}$ for configurations without interfaces.
In Sec. \ref{int}, we argue that the temperature gap of $O(\sqrt{\eta})$
gives a contribution to ${\cal I}$.

\subsection{Zubarev-Mclennan representation}\label{ZM}

Let ${\cal P}_{\rm ss}(\alpha;E,J)$
be the stationary distribution of $\alpha$ for a system with $(E, J)$,
where $E$ and $J$ are  values of the dimensionless total energy
and the dimensionless boundary current, respectively.
In this subsection,  we derive an expression of 
${\cal P}_{\rm ss}(\alpha;E,J)$, which is
called the Zubarev-Mclennan representation
\cite{Zubarev,Mclennan,KN,KNST-rep,Maes-rep},
in the linear response regime around the 
equilibrium state.

Let $\hat \alpha$ denote the trajectory of $\alpha$ from $t=0$
to $t=t_f$. That is, $\hat \alpha=(\alpha(t))_{t=0}^{t_f}$. 
The probability density (measure) of trajectory $\hat \alpha$
with $\alpha(0)$ fixed at $t=0$ is denoted by
$ \hat {\cal P}(\hat \alpha|\alpha(0); E, J)$.
From (\ref{model-s1}), (\ref{model-s2}), and (\ref{model-s3}),
we obtain 
\begin{equation}
  \log \hat {\cal P}(\hat \alpha|\alpha(0);E,J) =
  - \frac{1}{\eta^3}  \hat {\cal I}(\hat \alpha|\alpha(0);E,J)
  +{\rm const}
\label{III-64}
\end{equation}
with
\begin{align}
&   \hat {\cal I}(\hat \alpha|\alpha(0); E,J) 
=    \int_0^{t_f}dt\int d^3\bv{r}
\left\{      
  \frac{1}{4\lambda}
  \left|\bv{j}-\lambda \bv{\nabla} \beta \right|^2
   \right. \nonumber \\
&  +   \left.   
\frac{1}{4 \gamma T}
    \left[ \partial_t v+\gamma v -\sigma
-\eta  \left(   T
   d_e (  \bv{\nabla}  \beta)(  \bv{\nabla}  m) 
   +   d_f   \Delta   m \right) 
   \right]^2 
   \right\},
\label{III-65}
\end{align}
where $\partial_t m$ and $\partial_t \phi$
are connected to $v$ and $ \bv{j}$ as 
\begin{eqnarray}
  \partial_t m-{v}  &=& 0, \\
  \partial_t \phi+\bv{\nabla}\bv{j} &=&  0.
\label{phi-j}  
\end{eqnarray}
By a standard technique related to the local
detailed balance condition, which is reviewed
in Appendix \ref{der-zubarev}, we can derive 
\begin{align}
&{\cal P}(\alpha,t_f;E,J)=  {\cal N} \e^{
     {\cal S}(\alpha)/\eta^3
   } \nonumber \\
& \times    
\bra  \e^{
  J/\eta^3\int d^2\bv{r}_\perp
  \int_0^{t_f} dt  (\beta(1,\bv{r}_\perp,t)-\beta(0,\bv{r}_\perp,t))
          }
\ket_{\alpha^\dagger \to *}^{-J} 
\nonumber \\
& \times 
\delta\left( \int d^3\bv{r} \phi(\bv{r})-E \right) 
\label{zubarev}
\end{align}
with $\bv{r}_\perp=(y,z)$, 
where $ \bra \ \ket_{\alpha^\dagger \to *}^{-J}$ represents
the expectation value over trajectories $\alpha(t)$ starting from
$\alpha(0)=\alpha^\dagger=(m,-v,\phi)$ for $\alpha=(m,v,\phi)$
with respect to the path probability density 
in the system with $-J$.

Here, we consider the steady state obtained in the long
time limit $t_f \to \infty$ for the system with the separation
of scales $\eta \to 0$, with focusing on the linear response
regime in $J$. That is, precisely speaking, three limits
$t_f \to \infty$, $\eta \to 0$, and $\ep=|J| \to 0$, should be
taken into account. (In addition to those, 
the symmetry breaking external field $\sigma^{\rm ex}(x)$ 
should be taken to be zero in the last step, as discussed in
the previous section.) Now, if  we first took the limit
$t_f \to \infty$ for fixed $\eta$, we could not observe the symmetry
breaking in the limit $\sigma^{\rm ex}(x) \to 0$.  On the other hand, if we
first took $\eta \to 0$, the interface motion could not be
observed even in the equilibrium system,   as reviewed in Appendix 
\ref{i-motion}. More explicitly, let $\tau_{\rm int}$ be the time scale
of the interface motion. We then  confirm that $\tau_{\rm int} \to \infty$
for $\eta \to 0$. See (\ref{tau-int}).  The proper limit may 
be that we first set  $t_f=K\tau_{\rm int}$ in the limit $\eta \to 0$
with fixed $K$, and take the limit $K \to \infty$. We then consider
the limit $\ep \to 0$. 

Keeping this remark in mind, we define a modified entropy
$\tilde {\cal S}$ as
\begin{equation}
  \tilde {\cal S}(\alpha  ;E,J)
  \equiv \lim_{K \to \infty}
  \lim_{\eta \to 0} \eta^3
  \log \frac{{\cal P}(\alpha,K\tau_{\rm int};E,J)}
 {{\cal N}\delta\left( \int d^3\bv{r} \phi(\bv{r})-E \right) } .
\end{equation}
We then assume that the stationary probability distribution
in our problem is expressed as
\begin{equation}
 {\cal P}_{\rm ss}(\alpha;E,J)=
       {\cal N}\e^{
         \frac{1}{\eta^3} \tilde {\cal S}(\alpha;E,J)
          }
  \delta\left( \int d^3\bv{r} \phi(\bv{r})-E \right).
\label{Zubarev-flux-0}
\end{equation}
Now, recalling (\ref{J-ep}), we expand $\tilde S$ in $J$ as
\begin{equation}
  \tilde {\cal S}(\alpha  ;E,J)=
   {\cal S}_0(\alpha)+ J{\cal I}(\alpha;E)+O(\ep^2)
\label{tildeS-0}
\end{equation}
with
\begin{equation}
  {\cal S}_0(\alpha)=   \lim_{\eta \to 0}  {\cal S}(\alpha).
\end{equation}
Here, the functional ${\cal I}$ is calculated as 
\begin{eqnarray}
{\cal I}(\alpha;E) = &&\lim_{K \to \infty}
  \lim_{\eta \to 0}  \int d^2\bv{r}_\perp \int_0^{K\tau_{\rm int}} dt \nonumber \\
&& \times  \bra (\beta(1,\bv{r}_\perp,t)
  -\beta(0,\bv{r}_\perp, t)) \ket_{\alpha^\dagger \to *}^{\rm eq} ,
\label{Idef}
\end{eqnarray}
where $ \bra \ \ket_{\alpha \to *}^{\rm eq}$ is defined as
\begin{equation}
  \bra \ \ket_{\alpha \to *}^{\rm eq}
  = \lim_{J' \to 0} \bra \ \ket_{\alpha \to *}^{-J'}.
\end{equation}
Note that the right-hand side is uniquely determined in the limit
$J' \to 0$ for $\sigma^{\rm ex}(x)$ fixed.  
(\ref{Zubarev-flux-0}) may be referred to as the Zubarev-Mclennan
representation of the probability density for the system
with the flux control.  
When $J=0$, $ {\cal P}_{\rm ss}(\alpha;E,J=0)$  is the
micro-canonical distribution. The second term of (\ref{tildeS-0})
is the non-equilibrium correction to the entropy, which represents
the entropy production in the relaxation process to the equilibrium
state    from $\alpha^\dagger$ for
the configuration $\alpha$.   
This entropy production
is called {\it excess entropy production}.

\subsection{Interface configuration}\label{int-def}


In this section, we define a single interface configuration
$\alpha_X$ whose interface position  is given by $X$.  

First, we introduce the over-bar to represent
the average over vertical directions to the heat flux. For example, 
\begin{equation}
  \bar \beta(x,t)\equiv  \frac{1}{A}
  \int d^2\bv{r}_\perp \beta(x,\bv{r}_\perp,t ),
\end{equation}
where $A$ is the dimensionless cross-section defined by (\ref{Adef}).  
Let $\alpha_X$ denote  a single interface configuration  with
the interface position $X$.    Precisely, the interface position
is specified by 
\begin{equation}
{\bar m}(X) = \frac{m_{\rm loc}(\bar T(X))}{2}.
\end{equation}
We then define the interface region $[X_-, X_+]$ by
\begin{eqnarray}
  X_- &\equiv&  X-r  \sqrt{\eta},  \\
  X_+ &\equiv&  X+r  \sqrt{\eta},
\end{eqnarray}
where $r$ is a positive constant such that $e^{-r}$ is much smaller than 1,
say $e^{-r}=0.01$.  A single interface configuration $\alpha_X$
with the
interface position $X$   is defined as that satisfying
\begin{eqnarray}
|{\bar m}(x)-m_{\rm loc}(\bar T(x))| &\le&  \delta_m m_{\rm loc}(\bar T(x))
\end{eqnarray}  
for $ x \le X_- $, and
\begin{equation}
|{\bar m}(x)| \le  \delta_m m_{\rm loc}(\bar T(x))
\end{equation}
for  $ x \ge X_+$,  
where the constant $\delta_m$ is  much smaller  than $1$.  
We also impose that the interface configuration
satisfies
\begin{equation}
|\bar v(x)| \le  \delta_v , 
\end{equation}
where the constant $\delta_v$ is  much smaller  than $1$. 
Since we consider the limit $\eta \to 0$, the final result is
independent of the parameters $(\delta_m, \delta_v, r)$.

For a
given single interface configuration $\alpha_X$, we study the
time evolution from $\alpha_X$. We assume that a configuration at
any time $t$ in the time interval $[0, K\tau^{\rm int}]$ still possesses
a single interface at the interface position $X(t)$ which  depends
on the noise realization. Note that $X(0)$ equals to $X$ in $\alpha_X$.

Hereafter, for simplicity, we assume 
\begin{equation}
  \lambda(T,m)=\lambda^{\rm o}  
\label{kappa-o}
\end{equation}
in the ordered region $[0, X_-]$ and 
\begin{equation}
  \lambda(T,m)=\lambda^{\rm d}  
\label{kappa-d}
\end{equation}  
in the disordered region $[X_+,1]$,
where $\lambda^{\rm o}$ and $\lambda^{\rm d}$ are constants, and
$\lambda(T,m)$ in the region $[X_-,X_+]$ is $O(\sqrt{\eta})$,
while  its functional form is not specified. 
See Sec. \ref{problem} for the argument.

\subsection{Correction term}
\label{correction}

We first re-write ${\cal I}$ as 
\begin{equation}
{\cal I}(\alpha_X) = \lim_{K \to \infty}
  \lim_{\eta \to 0}  A I(\alpha_X) ,
\label{caI-start}
\end{equation}
where $I(\alpha_X)$ is expressed as
\begin{eqnarray}
I (\alpha_X) &=&
\int_0^{K\tau_{\rm int}} dt
  \bra \bar \beta(1,t)
  -\bar \beta(0, t) \ket_{\alpha_X^\dagger \to *}^{\rm eq} . \label{Idef-2} 
\label{Ical}
\end{eqnarray}
We consider the decomposition of $I(\alpha_X) $:
\begin{equation}
I(\alpha_X)= I^{\rm o} (\alpha_X) +I^{\rm d} (\alpha_X) + I^{\rm int} (\alpha_X),
\label{I-decom-int}
\end{equation}
where 
\begin{eqnarray}
I^{\rm o} (\alpha_X) 
&\equiv&
\int_0^{K\tau_{\rm int}} dt  \bra
\bar \beta(X_{-},t) -\bar \beta(0,t)
\ket_{\alpha_X^\dagger  \to *}^{\rm eq},
\label{Ical-o} \\
I^{\rm d} (\alpha_X) 
&\equiv&  \int_0^{K\tau_{\rm int}} dt \bra
\bar \beta(1,t)-\bar \beta(X_{+},t)
\ket_{\alpha_X^\dagger  \to *}^{\rm eq},
\label{Ical-d} 
\end{eqnarray}
and 
\begin{equation}
  I^{\rm int} (\alpha_X)
 \equiv  \int_0^{K\tau_{\rm int}} dt \bra
\bar \beta(X_{+},t)-\bar \beta(X_{-},t)
 \ket_{\alpha_X^\dagger  \to *}^{\rm eq} . 
\label{Ical-int}
\end{equation}

In the evaluation of $I^{\rm o/d}(\alpha_X) $ and $I^{\rm int} (\alpha_X)$,
we take account of only the contribution from the most probable process
by ignoring fluctuations, because we consider the weak noise cases of
small $\eta$. Note that, in the bulk region,  $\alpha(t)$ is replaced
by the solution of the deterministic equation with  $\eta \to 0 $, while
the deterministic equation of $\bar \beta(X_{+\-},t)$ is not obtained by
the noiseless limit of the stochastic model. 
In the argument below,   for any fluctuating thermodynamic
quantity $Q(t)$, we use the same notation $Q(t)$ to represent the most
probable value with the initial condition $\alpha(0) =\alpha_X^\dagger$
under the equilibrium condition.  
That is, (\ref{Ical-o}), (\ref{Ical-d}),
and (\ref{Ical-int}) are rewritten as 
\begin{eqnarray}
I^{\rm o} (\alpha_X) 
&\equiv&
\int_0^{K\tau_{\rm int}} dt  
[\bar \beta(X_{-},t) -\bar \beta(0,t)],
\label{Ical-o-1} \\
I^{\rm d} (\alpha_X) 
&\equiv&  \int_0^{K\tau_{\rm int}} dt 
[\bar \beta(1,t)-\bar \beta(X_{+},t)],
\label{Ical-d-1} 
\end{eqnarray}
and 
\begin{equation}
  I^{\rm int} (\alpha_X)
 \equiv  \int_0^{K\tau_{\rm int}} dt 
[\bar \beta(X_{+},t)-\bar \beta(X_{-},t)]. 
\label{Ical-int-1}
\end{equation}
Below we  evaluate $I^{\rm o/d}(\alpha_X) $ and $I^{\rm int} (\alpha_X)$
for small $\eta$ and large $K$.

\subsection{Bulk contribution}\label{bulk}


First,  we express (\ref{Ical-o-1}) and (\ref{Ical-d-1}) as
\begin{eqnarray}
I^{\rm o} (\alpha_X) 
&=&  \int_0^{K\tau_{\rm int}} dt  \int_0^{X_-(t)} dx
\partial_x \bar \beta(x,t), \label{Ical-o-2} \\
I^{\rm d} (\alpha_X) 
&=&  \int_0^{K\tau_{\rm int}} dt  \int_{X_+(t)}^1 dx
\partial_x \bar \beta(x,t). \label{Ical-d-2} 
\end{eqnarray}

Here, we  find a neat idea to use a variable
$\psi(x,t)$ defined by 
\begin{equation}
\bar \phi(x,t)=\frac{E}{A}+\partial_x \psi(x,t)
\label{psi-def0}
\end{equation}  
with the boundary conditions  $\psi(0,t)=\psi(1,t)=0$.
For a given $\bar \phi(x,t)$, $\psi(x,t)$ can be   uniquely
determined   because of the energy conservation: 
\begin{equation}
  A \int_0^1 dx \bar \phi(x,t)= E.
\label{E-con-av}
\end{equation}
We substitute (\ref{psi-def0}) into (\ref{ons-phi-d-s})
and take the boundary condition (\ref{ad-con}) into account.  
We then obtain the deterministic equation of $\psi$ 
\begin{equation}
  \partial_t\psi+ \lambda^{\rm o} \partial_x \bar \beta =0
\label{beta-psi-o}
\end{equation}  
for $x \in [0, X_-(t)]$ and
\begin{equation}
  \partial_t\psi+ \lambda^{\rm d} \partial_x \bar \beta =0
\label{beta-psi-d}
\end{equation}  
for $x \in [X_+(t),1]$.  
Now, by using (\ref{beta-psi-o}), (\ref{Ical-o-2}) is expressed as
\begin{align}
& I^{\rm o}(\alpha_X) =
 - \int_0^{K \tau_{\rm int}} dt
 \int_0^{X_-(t)} dx   \frac{\partial_t \psi}{\lambda^{\rm o}}    \nonumber \\
&\qquad 
=
- \frac{1}{\lambda^{\rm o}}
 \int_{0}^{K\tau_{\rm int}} dt\int_0^{1} dx  H(X_-(t)-x) \partial_t \psi,
\label{Ical-o-3}
\end{align}
where $H(x) =1$ for $x >0$ and $H(x) =0$ for $x <0$.
Since 
\begin{align}
&H(X_-(t)-x) \partial_t \psi\nonumber\\
& =
\partial_t(H(X_-(t)-x)  \psi)- \der{X}{t}\delta(X_-(t)-x) \psi,
\end{align}
we have
\begin{align}
I^{\rm o}(\alpha_X) &=
- \frac{1}{\lambda^{\rm o}}
  \int_{0}^{K\tau_{\rm int}} dt \int_0^{1} dx
   \partial_t(H(X_-(t)-x)  \psi) \nonumber \\
& 
+ \frac{1}{\lambda^{\rm o}}
   \int_{0}^{K\tau_{\rm int}} dt \int_0^{1} dx
   \der{X}{t}\delta(X_-(t)-x) \psi
\label{Ical-o-4}.
\end{align}
We rewrite it as
\begin{align}
  I^{\rm o}(\alpha_X)
&= \frac{1}{\lambda^{\rm o}}
  \left[ \int_0^{X_-(0)} \!\!\!\! dx \psi_X(x) - \int_0^{X_-(K\tau_{\rm int})}\!\!\!\!\!\!\!\!
       dx \psi(x,K\tau_{\rm int})
    \right]\nonumber \\
& 
\quad + \frac{1}{\lambda^{\rm o}}
   \int_{0}^{K\tau_{\rm int}} dt   \der{X}{t} \psi(X_-(t),t)
\label{Ical-o-5},
\end{align}
where  $\psi_X(x)$ is determined from $\alpha_X$  in the
argument of $I^{\rm o}$. Similarly, we obtain
\begin{align}
  I^{\rm d}(\alpha_X)
&= \frac{1}{\lambda^{\rm d}}
  \left[ \int_{X_+(0)}^1 \!\!\!\!  dx \psi_X(x)
    - \int_{X_+(K\tau_{\rm int})}^1 \!\!\!\! \!\!\!\!  dx \psi(x,K\tau_{\rm int})
    \right]\nonumber \\
& 
\quad
- \frac{1}{\lambda^{\rm d}}
   \int_{0}^{K\tau_{\rm int}} dt   \der{X}{t} \psi(X_+(t),t)
\label{Ical-d-5}.
\end{align}
Now, we consider the limit $\eta \to 0$ with large $K$ fixed.
The interface motion is observed with the time scale
$\tau_{\rm int} = O(\eta^{-1/2})$ which is much larger than
the relaxation time of thermodynamic quantities. Thus, $\alpha(x,t)$
is close to the quasi-equilibrium configuration $\alpha_{X(t)}^{\rm qeq}(x)$
with the interface position $X(t)$, where the quasi-equilibrium configuration
$\alpha_{X}^{\rm qeq}(x)$ is characterized by the uniform temperature
$T_X^{\rm qeq}$ satisfying 
\begin{equation}
  Xu^{\rm o}(T_X^{\rm qeq})+ (1-X)u^{\rm d}(T_X^{\rm qeq})=\frac{E}{A}.
\label{t-qeq-def}
\end{equation}    
All thermodynamic quantities in the quasi-equilibrium
state are calculated from   $\alpha_{X}^{\rm qeq}(x)$.

\begin{figure}[tb]
\centering
\includegraphics[width=8cm]{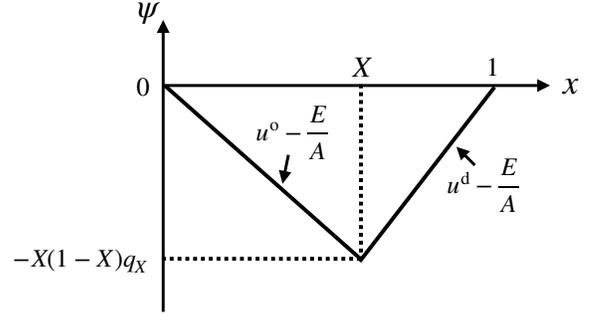}
\caption{Example of the graph $\psi_X^{\rm qeq}(x)$.}
\label{fig:psi}
\end{figure}
As one example,   
the quasi-equilibrium configuration $\psi_X^{\rm qeq}(x)$ is given by 
\begin{equation}
  \psi_X^{\rm qeq}(x) = \left( u^{\rm o}(T_X^{\rm qeq})-\frac{E}{A} \right) x
\label{psi-o-def}
\end{equation}
for $0 \le x \le X$, and
\begin{equation}
  \psi_X^{\rm qeq}(x) = -\left( u^{\rm d}(T_X^{\rm qeq})-\frac{E}{A} \right) (1-x)
\label{psi-d-def}
\end{equation}
for $X \le x \le 1$. Here, we define the latent heat $q_X$ by 
\begin{equation}
q_X \equiv  u^{\rm d}(T_X^{\rm qeq})-  u^{\rm o}(T_X^{\rm qeq}).
\end{equation}
By combining it with the relation (\ref{t-qeq-def}),
we find
\begin{equation}
 u^{\rm o}(T_X^{\rm qeq}) - \frac{E}{A}= -(1-X) q_X .
\label{q-X}
\end{equation}
We thus have
\begin{equation}
  \psi_X^{\rm qeq}(X)=-X(1-X)q_X.
\end{equation}  
Summarizing these results, we show 
an example of quasi-equilibrium configuration
$\psi_X^{\rm qeq}(x)$ in  Fig. \ref{fig:psi}.
By taking the limit $K \to \infty$ and $\eta \to 0$, we have
arrived at 
\begin{align}
& \lim_{K \to \infty} \lim_{\eta \to 0}
  [I^{\rm o}(\alpha_X)+I^{\rm d}(\alpha_X)] \nonumber \\
&=  \frac{1}{\lambda^{\rm o}}\int_0^X dx  \psi_X(x)
   + \frac{1}{\lambda^{\rm d}} \int_X^1 dx  \psi_X(x) \nonumber \\
&\quad - \frac{1}{\lambda^{\rm o}}\int_0^{X_{\rm eq}} dx  \psi_{X_{\rm eq}}^{\rm qeq}(x)
   - \frac{1}{\lambda^{\rm d}} \int_{X_{\rm eq}}^1 dx  \psi_{X_{\rm eq}}^{\rm qeq}(x)
   \nonumber \\
&\quad  + \int_X^{X_{\rm eq}}dY \psi_{Y}^{\rm qeq}(Y)
   \left(
   \frac{1}{\lambda^{\rm o}} - \frac{1}{\lambda^{\rm d}}
   \right) .
\label{I-bulk-result}
\end{align}

\subsection{Interface contribution}\label{int}

We next study the interface contribution (\ref{Ical-int-1}). 
By defining
\begin{equation}
  \beta_{\pm}^{\rm int}(t) \equiv \bar \beta(X_{\pm}(t), t),  
\end{equation}  
we replace (\ref{Ical-int-1}) by  
\begin{equation}
  I^{\rm int} (\alpha_X)=\int_0^{K \tau_{\rm int}} dt 
  \left[\beta_+^{\rm int}(t)- \beta_-^{\rm int}(t) \right].  
\label{Ical-int-2}
\end{equation}
We call  
$\beta^{\rm int}_+(t)-\beta^{\rm int}_-(t)$ ``inverse-temperature
gap''.    We estimate  $ I^{\rm int} (\alpha_X)$ by
dividing the interval $[0, K\tau_{\rm int}]$ into two intervals
$[0, t_c]$ and $[t_c, K\tau_{\rm int}]$, where we take $t_c$
satisfying
\begin{equation}
1 \ll t_c \ll \tau_{\rm int}
\end{equation}
for small $\eta$.  The contribution to $I^{\rm int}$
in the time interval $[0, t_c]$ is expressed as  
\begin{equation}
I_1^{\rm int} (\alpha_X )
=  \int_0^{t_c} dt 
  [  \beta^{\rm int}_+(t) -  \beta^{\rm int}_-(t) ].  
\label{Ical-int-11}
\end{equation}
The initial configuration $\alpha^\dagger_X$ rapidly relaxes in
$t\in [0,t_c]$ to the quasi-equilibrium configuration $\alpha_X^{\rm qeq}(x)$
with keeping the interface position $X$. Using the equilibrium statistical
mechanics, we find that the probability of observing the inverse-temperature
gap of $O(1)$ is extremely small. Thus, considering  cases where
$\beta^{\rm int}_+(t) -  \beta^{\rm int}_-(t) =O(\sqrt{\eta})$, 
we estimate $|I_1^{\rm int}|$ as $O(t_c \sqrt{\eta})$. Since
$t_c \ll O(\eta^{-{1/2}})$,  $|I_1^{\rm int}|$ can be negligible
for small $\eta$.  More precisely, $I_1^{\rm int} \to 0$ in the
limit $\eta \to 0$.

\begin{figure}[tb]
\centering
\includegraphics[width=7cm]{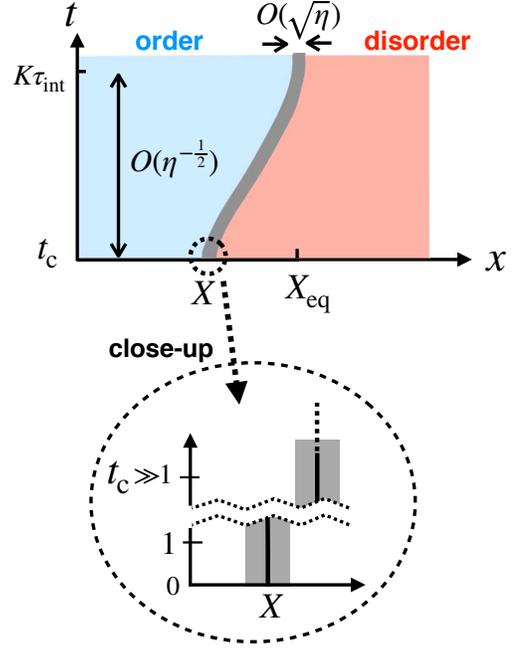}
\caption{Space-time plot associated with interface motion
  whose time scale is $O(\eta^{-1/2})$. Its close-up at a time scale
  of $O(1)$ is also shown.}
\label{fig:space-time}
\end{figure}

In the time interval $t \in [t_c, K\tau_{\rm int}]$,
the slow interface motion with $dX/dt \simeq \sqrt{\eta}$ is observed,
which we call a {\it late stage}. 
Since all quantities in the late stage are assumed to be independent
of $(y,z)$, we, hereafter, describe the configuration as $\alpha(x,t)$
without over-bar. Such a space-time configuration is illustrated
in Fig.~\ref{fig:space-time}. $\alpha(x,t)$ is close to the
quasi-equilibrium configuration $\alpha_{X(t)}^{\rm qeq}(x)$ with
the interface position $X(t)$. We then define 
\begin{equation}
I_2^{\rm int} (\alpha_X)
\equiv  \int_{t_c}^{K \tau_{\rm int}} dt
  [  \beta^{\rm int}_+(t)-   \beta^{\rm int}_-(t)].  
\label{Ical-int-22}
\end{equation} 
Since  $\tau^{\rm int}= O(1/\sqrt{\eta})$, $I_2^{\rm int}$ becomes
finite when the inverse-temperature gap $ \beta^{\rm int}_+(t)-
\beta^{\rm int}_-(t)$ is estimated as $O(\sqrt{\eta})$. This estimation
means  $\partial_x \beta = O(1)$ in the interface region
$x \in [X_-, X_+]$. Note that this is much larger than
$\partial_x \beta = O(\sqrt{\eta})$ expected in the bulk regions,
while it is consistent with the estimation $\kappa^{\rm int}=O(\sqrt{\eta})$
in Sec. \ref{problem}.  Because of this singularity, the description of
the inverse-temperature gap $ \beta^{\rm int}_+(t)-\beta^{\rm int}_-(t)$
cannot be obtained from  the noiseless limit of the stochastic model.  In order to calculate $ \beta^{\rm int}_+(t)-\beta^{\rm int}_-(t)$
quantitatively, one may formulate the renormalization of noise
effects in the interface region.   
Although the study in this direction is interesting, it is
beyond the scope of the present paper. In the next section, we
attempt to estimate the inverse temperature gap
$\beta_+^{\rm int} -\beta_-^{\rm int}$ without
analyzing the stochastic model, but  using a phenomenological
argument.

\section{Entropy production in the interface region}
\label{ent-interface}

In this section, we estimate the entropy production in the
interface region and obtain the final form of the stationary
distribution for interface configurations. Concretely, 
in Sec. \ref{t-gap}, we explain a phenomenological method to
obtain the temperature gap over the interface region.
In Sec. \ref{t-profile}, we derive the temperature profile in the
bulk when the interface slowly moves to the equilibrium position.
By using this result, in Sec. \ref{t-gap-cal}, we  estimate the temperature
gap at the interface. At last, in Sec. \ref{res-correction}, we show the
result of ${\cal I}$ for a single interface configuration $\alpha_X$.


\subsection{Phenomenological argument}\label{t-gap}

\begin{figure}[tb]
\centering
\includegraphics[width=7cm]{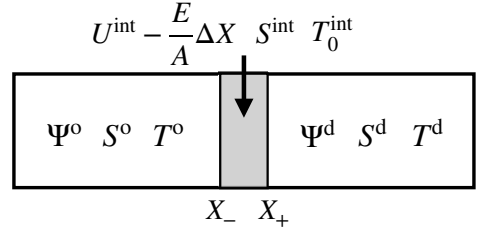}
\caption{Coarse-grained description for determining the
  temperature gap at the interface. See (\ref{T0-def}) for $T_0^{\rm int}$.}
\label{fig:ons-config}
\end{figure}

Let us recall that the interface velocity $dX/dt$ would be determined
by the free energy difference if the temperature of the system were uniform.
  See (\ref{x-m-q-int}) in Appendix \ref{i-motion}.
In the present problem, for a given small $dX/dt$, an inhomogeneous
temperature profiles in the bulk regions are calculated, as shown in
Sec. \ref{t-profile}.    The average temperatures in the
 ordered and disordered   regions
are  determined by two conditions. The first is clearly the
energy conservation, while the second condition should be considered
seriously.  Since  $\eta$ is finite, we consider the interface region
as a thermodynamic subsystem. That is, the system consists of the
three local equilibrium subsystems,  corresponding to the ordered
region, disordered region and the interface region, respectively.
We then describe the energy exchange between each bulk region and the
interface region. This description provides the second condition
for determining the average temperatures for the given $dX/dt$.
Macroscopic variables  corresponding to the energy exchange  are
defined by 
\begin{eqnarray}
  \Psi^{\rm o} &\equiv&  \int_0^{X_-} dx
  \left(u^{\rm o}(T(x,t)) -\frac{E}{A} \right), \label{psio-def} \\
  \Psi^{\rm d} &\equiv&  \int_{X_+}^1 dx
  \left( u^{\rm d} (T(x,t))-\frac{E}{A} \right)  
\end{eqnarray}
with  (\ref{uo-def}) and (\ref{ud-def}) 
for the definition of $u^{\rm o}$ and $u^{\rm d}$.
Note that $\Psi^{\rm o}(t)$  and  $\Psi^{\rm d}(t) $ 
satisfy the energy conservation 
\begin{equation}
  \Psi^{\rm o}+  \Psi^{\rm d}
  + \left(U^{\rm int}-\frac{E}{A} \Delta X \right)
  + O(\eta) =0,
\label{E-con}
\end{equation}
where $\Delta X \equiv X_+-X_-$,  $O(\eta)$ includes the
term proportional to $(dX/dt)^2$, and $U^{\rm int} $
is the internal energy of the interface region, which 
includes the surface energy.
Accordingly, the entropy of the system $S$ is expressed as
\begin{equation}
S=S^{\rm o}+S^{\rm d}+S^{\rm int},
\end{equation}
where $S^{\rm o}$ and  $S^{\rm d}$ are defined as
\begin{align}
&  S^{\rm o}\equiv \int_0^{X_-} dx s^{\rm o}(T(x)), \\
&  S^{\rm d}\equiv \int_{X_+}^1 dx s^{\rm d}(T(x)),
\end{align}
with (\ref{so-def}) and (\ref{sd-def}),
and $S^{\rm int}$ is assumed as a function of $U^{\rm int}$. 
Assuming that $\Psi^{\rm o}$ and $\Psi^{\rm d}$ are slow variables for
the given interface motion $X(t)$, we write the  Onsager form of their
time evolution as 
\begin{eqnarray}
  \der{\Psi^{\rm o}}{t} &=& L^{\rm o}
  \pderf{S}{ \Psi^{\rm o}}{\Psi^{\rm d}},
\label{ons-psi-o}  \\
\der{\Psi^{\rm d}}{t} &=&   L^{\rm d} 
\pderf{S}{ \Psi^{\rm d}}{\Psi^{\rm o}},
\label{ons-psi-d}
\end{eqnarray}
where ${L}^{\rm o}$ and ${L}^{\rm d}$ are new Onsager coefficients in
this projected dynamics.   Note that we  do not take account of 
  off-diagonal components of Onsager coefficients.  
See Fig. \ref{fig:ons-config} for a schematic figure of the setup.  

Since  $\Psi^{\rm o/d}$ is related to $\beta_{-/+}^{\rm int}$
  as shown in Sec. \ref{t-gap-cal},  
(\ref{ons-psi-o}) and (\ref{ons-psi-d}) give an expression of 
$\beta_+^{\rm int} -\beta_-^{\rm int}$ in terms of
$X$, $dX/dt$, $\lambda^{\rm o/d}$, and ${L}^{\rm o/d}$.
  Here, fluctuations are renormalized into ${L}^{\rm o/d}$ 
  so that  
${L}^{\rm o/d}$ is determined by finite time fluctuations of
the energy transfer into the ordered/disordered region from
the interface. Moreover,
we assume that fluctuations of $\Psi^{\rm o}$ and $\Psi^{\rm d}$
are not correlated, because the main contribution to the energy
transfer comes from the latent heat generated at the interface.
Therefore, ${L}^{\rm o/d}$ is given by quantities defined in the
ordered/disordered region.  Recalling that the dimension of 
${L}^{\rm o/d}$ is that of $\lambda^{\rm o/d}$ divided by the
length dimension,  we set ${L}^{\rm o}$ and ${L}^{\rm d}$ as
\begin{eqnarray}
{L}^{\rm o} &=&  \frac{\lambda^{\rm o}}{g X_-},  \label{ons-cof-o}\\
{L}^{\rm d} &=&  \frac{\lambda^{\rm d}}{g(1-X_+)}, \label{ons-cof-d} 
\end{eqnarray}
where $g$ is a dimensionless factor, which is assumed to
be independent of $X$.  When we impose the condition
that the inverse temperature gap  $\beta_+^{\rm int} -\beta_-^{\rm int}$
 vanishes in the limit $X_- \to 0$ and $X_+ \to 1$,   
we can determine the value of $g$  uniquely, 
as shown in the next section.  

Precisely writing, (\ref{ons-psi-o}) and
(\ref{ons-psi-d}) with (\ref{ons-cof-o}) and (\ref{ons-cof-d})
are not yet derived from the stochastic model we study. Rather,
this description involves uncontrolled approximations. For example,
the dynamics of
$\Psi^{\rm o}$
may influence the interface motion and ${L}^{\rm o}$ may depend
on $\lambda^{\rm d}$. We do not find clear reasons to ignore these
effects. Nevertheless, we expect that  (\ref{ons-psi-o}) and
(\ref{ons-psi-d}) with (\ref{ons-cof-o}) and (\ref{ons-cof-d})
  describe qualitative behaviors.   
In the   subsequent subsections, we calculate the temperature
profiles in the bulk regions and determine    the
temperature gap by explicitly expressing (\ref{ons-psi-o})
and (\ref{ons-psi-d}) in terms of $\beta_{+/-}^{\rm int}$.

\subsection{Temperature profile in the bulk}\label{t-profile}

In the bulk regions $[0, X_-(t)]$ and $[X_+(t),1]$ for the
time interval $[t_c,K\tau_{\rm int}]$, the time evolution is described by
the deterministic equation. We ignore the terms associated with interface
thermodynamics by setting $d_e=d_f=0$.   We then study the
behavior in the two bulk regions separately.  Specifically,    
we study  the entropy density $s(x,t)$.
By substituting the thermodynamic relation
\begin{equation}
  \partial_t s= \beta \partial_t u+\pderf{s}{m}{u}\partial_t m
\end{equation}
into (\ref{u-evol}), we obtain
\begin{equation}
  T \partial_t  s =  \partial_x(\kappa \partial_x T)
  + \gamma  (\partial_tm)^2 .
  \label{t-evol}
\end{equation}
In  the ordered region $[0, X_-(t)]$, we may assume $m(x,t)=m_{\rm loc}(T(x,t))$,
because $m(x,t)$ quickly relaxes to the local stable state for
a given temperature $T(x,t)$. Then, since $s(x,t)=s^{\rm o}(T(x,t))$,
we have 
\begin{eqnarray}
T  \partial_t s &=&  T \partial_t s^{\rm o} \\
&=&  c^{\rm o} \partial_t T,
\end{eqnarray}  
where we have used (\ref{zero-cap-o-2}).
By using this relation and noting
$(\partial_tm)^2 = O(\eta)$, we obtain
\begin{equation}
c^{\rm o} \partial_t T = \partial_x(\kappa\partial_x T) +O(\eta).
\label{t-evol-o}
\end{equation}
Let us recall $\kappa=\lambda/T^2$ and we set
\begin{equation}
  \kappa_X^{\rm o}= \frac{\lambda^{\rm o}}{(T_X^{\rm qeq})^2}.
\end{equation}  
Since the time derivative of $T_{X(t)}^{\rm qeq}$ is given by
\begin{equation}
  \der{T_{X(t)}^{\rm qeq}}{t}=
  \left. \der{T_X^{\rm qeq}}{X}\right|_{X=X(t)}\der{X}{t}=O(\sqrt{\eta}),
\end{equation}
the solution for small $\eta$ can be expanded as
\begin{equation}
  T(x,t)=T^{(0)}(x,t)+\sqrt{\eta}T^{(1)}(x,t)+O(\eta).
\label{T-exp}
\end{equation}
By substituting (\ref{T-exp}) into   (\ref{t-evol-o}),
we first have
\begin{equation}
\partial_x  T^{(0)} +O(\sqrt{\eta})=0
\label{t-evol-o-0}
\end{equation}
as the lowest order equation.   The solution $T^{(0)}$ is constant
in $x$. Since we study an interface configuration with the interface
position $X$,  the solution is   the quasi-equilibrium profile 
\begin{equation}
T^{(0)}(x,t)= T_{X(t)}^{\rm qeq} ,
\end{equation}
which slowly evolves through the interface position $X(t)$. 
Next, by substituting 
\begin{equation}
  T(x,t)=T_{X(t)}^{\rm qeq}+\sqrt{\eta}T^{(1)}(x,t)+O(\eta)
\label{T-exp-1}
\end{equation}
into   (\ref{t-evol-o}), we obtain
\begin{equation}
 c^{\rm o} \der{ T_X^{\rm qeq}}{X} \der{X}{t}
 = \sqrt{\eta}\kappa_X^{\rm o} \partial_x ^2 T^{(1)} +O(\eta),
\label{t-evol-o-2}
\end{equation}
where we have ignored $\sqrt{\eta}\partial_t T^{(1)}$ because this
term is estimated as $O(\eta)$. Hereafter, $c^{\rm o}$ is evaluated
at $T_{X(t)}^{\rm qeq}$. 
By solving this equation with the boundary condition
$\partial_x T=0$ at $x=0$, we derive $T^{(1)}$ as a quadratic
function in $x$. We thus obtain 
\begin{equation}
  T(x,t)= {T^{\rm int}_-}(t)+ \der{T_{X(t)}^{\rm qeq}}{t}
  \frac{c^{\rm o} }{2 \kappa_{X(t)}^{\rm o}}  (x^2-X^2) +O(\eta),
\label{t-sol-o}
\end{equation}
where   $ T_{-}^{\rm int}(t) =1/\beta_{-}^{\rm int}(t)$. 
Note that  $T_{-}^{\rm int}(t)-T_{X(t)}^{\rm qeq}=O(\sqrt{\eta})$
should hold from (\ref{T-exp-1}).

\begin{figure}[tb]
\centering
\includegraphics[width=7.5cm]{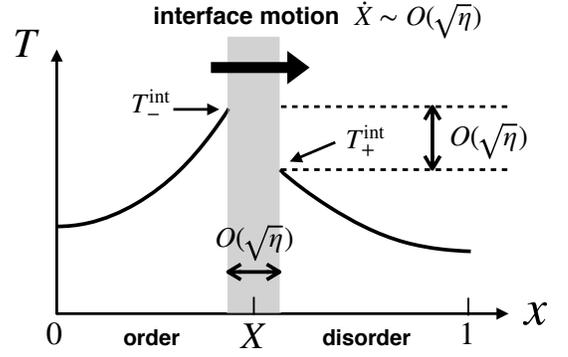}
\caption{Temperature configuration in the late stage of
  a relaxation process. Latent
heat is generated at the moving interface and it diffuses into the
bulk regions. See (\ref{t-sol-o}) and  (\ref{t-sol-d}) for the
expression of the profiles. 
A temperature gap appears in the interface region.
See (\ref{result-ons-a}) for the expression of the temperature gap.}
\label{fig:t-config}
\end{figure}

Similarly, in the disordered region $x \in [X_+(t),1]$, we obtain 
\begin{equation}
T(x,t)= {T_+^{\rm int}}(t)+ \der{T_{X(t)}^{\rm qeq}}{t}
  \frac{c^{\rm d} }{2 \kappa_{X(t)}^{\rm d}}
  [(1-x)^2-(1-X)^2]+O(\eta),
\label{t-sol-d}
\end{equation}
where  
$ T_{+}^{\rm int}(t) =1/\beta_{+}^{\rm int}(t)$ and   we have defined 
\begin{equation}
\kappa_X^{\rm d} \equiv \frac{\lambda^{\rm d}}{(T_X^{\rm qeq})^2}.
\end{equation}
In Fig. \ref{fig:t-config}, we show a schematic figure
of the temperature profiles in the two bulk regions.
An important observation is that the temperature of
the interface region is higher than that of the bulk regions
when $dX/dt >0$.  Physically, the slowly moving interface
in the relaxation process produces the latent heat which
acts as a heat source. This brings the distortion of the
temperature profiles in the bulk regions. Note that
$T^{\rm int}_{+}$ and $T^{\rm int}_{-}$ are not determined yet.

\subsection{Temperature gap}\label{t-gap-cal}

We define the average temperature in the ordered region as
\begin{equation}
   T_X^{\rm o}(t) \equiv \frac{1}{X_-} \int_0^{X_-} dx T(x,t).
\label{to-def}
\end{equation}  
By substituting
\begin{equation}
  u^{\rm o}(T(x,t))= u^{\rm o}(T_X^{\rm o}) +c^{\rm o}(T(x,t)-T_X^{\rm o})+O(\eta)
\end{equation}
into (\ref{psio-def}) and using (\ref{to-def}), we obtain
\begin{equation}
  \Psi^{\rm o} = \left( u^{\rm o}( T_X^{\rm o})-\frac{E}{A} \right) X_- +O(\eta).
\label{p-t}
\end{equation}
Similarly, by using
\begin{equation}
   T_X^{\rm d}(t) \equiv \frac{1}{1-X_+} \int_{X_+}^1 dx T(x,t),
\label{td-def}
\end{equation}  
we have 
\begin{equation}
  \Psi^{\rm d} = \left( u^{\rm d}( T_X^{\rm d})-\frac{E}{A} \right) (1-X_+)
+O(\eta).
\end{equation}  
We also obtain
\begin{eqnarray}
  S^{\rm o} &=&  X_- s^{\rm o}( T_X^{\rm o}) +O(\eta) , \label{So-def} \\
  S^{\rm d} &=&  (1-X_+) s^{\rm d}( T_X^{\rm d}) +O(\eta) . \label{Sd-def} 
\end{eqnarray}
We then define $T_0^{\rm int}$ as
\begin{equation}
  T_0^{\rm int} \equiv \der{U^{\rm int}}{S^{\rm int}},
\label{T0-def}  
\end{equation}
which represents the temperature in the interface region.

We here apply the Onsager theory to two macroscopic
quantities $\Psi^{\rm o}$ and $\Psi^{\rm d}$. We fix $\Psi^{\rm d}$
and consider the variation $\Psi^{\rm o} \to \Psi^{\rm o}+\delta \Psi^{\rm o}$.
From  energy conservation, we have
\begin{equation}
  \delta \Psi^{\rm o}  + \delta U^{\rm int}=0.
\label{246}
\end{equation}  
Since $\Psi^{\rm o}$ has the one-to-one correspondence with $T_X^{\rm o}$,
as shown in (\ref{p-t}), we have
\begin{equation}
  \delta \Psi^{\rm o}= X_- c^{\rm o} \delta  T_X^{\rm o}+O(\eta).
\label{247}
\end{equation}  
By using (\ref{246}) and (\ref{247}),  we derive
\begin{eqnarray}
  \delta S &=&  X_-  \frac{c^{\rm o} }{ T_X^{\rm o}}\delta  T_X^{\rm o}
             + \frac{1}{ T_0^{\rm int}} \delta U^{\rm int} \nonumber \\
&=& \left(\frac{1}{ T_X^{\rm o}}
             - \frac{1}{ T_0^{\rm int}} \right) \delta \Psi^{\rm o}. 
\end{eqnarray}  
Therefore, the equation of $\Psi^{\rm o}$ in (\ref{ons-psi-o}) is written as 
\begin{equation}
\der{\Psi^{\rm o}}{t}
                   = {L}^{\rm o}  \left(\frac{1}{T_X^{\rm o}}
      - \frac{1}{T_0^{\rm int}}  \right).
\label{psi-o-evo}   
\end{equation}
Similarly, (\ref{ons-psi-d}) becomes
\begin{equation}
\der{\Psi^{\rm d}}{t} = {L}^{\rm d}  \left(\frac{1}{T_X^{\rm d}}
      - \frac{1}{T_0^{\rm int}}   \right).
\label{psi-d-evo}   
\end{equation}
From (\ref{psi-o-evo}) and (\ref{psi-d-evo}), we obtain
\begin{equation}
\frac{1}{ T_X^{\rm d}}  -   \frac{1}{T_X^{\rm o}}
=
\frac{1}{{L}^{\rm d}} \der{\Psi^{\rm d}}{t}
  -
\frac{1}{{L}^{\rm o}} \der{\Psi^{\rm o}}{t} .
\label{diff-barT}    
\end{equation}

Let us express $T_X^{\rm o/d}$ in terms of $T_{+/-}^{\rm int} $.
By using (\ref{t-sol-o}), we calculate
\begin{equation}
   T_X^{\rm o}= T_-^{\rm int} 
  - \der{T_X^{\rm qeq}}{t}
  \frac{c^{\rm o}(T_X^{\rm qeq})^2}{3  \lambda^{\rm o}} X^2 +O(\eta).
  \label{to-cal}
\end{equation}  
Similarly, we have
\begin{equation}
  T_X^{\rm d}= T_+^{\rm int} 
    -\der{T_X^{\rm qeq}}{t}
  \frac{c^{\rm d}(T_X^{\rm qeq})^2}{3  \lambda^{\rm d}}
  (1-X)^2 +O(\eta).
\label{td-cal}
\end{equation}
Hereafter, we do not explicitly write $O(\eta)$. 
From (\ref{to-cal}) and (\ref{td-cal}), we obtain
\begin{align}
\frac{1}{T_X^{\rm d}}  -   \frac{1}{T_X^{\rm o}} 
=& \beta_+^{\rm int}-\beta_-^{\rm int} \nonumber \\
&  
+ \der{T_X^{\rm qeq}}{t}
  \left[ \frac{c^{\rm d}}{3 \lambda^{\rm d}} (1-X)^2
 -      \frac{c^{\rm o}}{3  \lambda^{\rm o}} X^2 \right] .
\label{tod-cal}
\end{align}
Substituting (\ref{tod-cal}) into (\ref{diff-barT}),
we have 
\begin{align}
  \beta_+^{\rm int}-\beta_-^{\rm int}
=&
 \der{T_X^{\rm qeq}}{t}
  \left[\frac{c^{\rm o}}{3  \lambda^{\rm o}} X^2 
    -   \frac{c^{\rm d}}{3 \lambda^{\rm d}} (1-X)^2  \right] 
\nonumber \\
&  +\frac{1}{{L}^{\rm d}} \der{\Psi^{\rm d}}{t}
  - 
  \frac{1}{{L}^{\rm o}} \der{\Psi^{\rm o}}{t}.
\label{tochu}
\end{align}

Next, we consider $d\Psi^{\rm o}/dt$.
From (\ref{psio-def}), we calculate
\begin{equation}
\der{\Psi^{\rm o}}{t}
= 
\left( u^{\rm o}(T_X^{\rm qeq})  -\frac{E}{A}  \right) \der{X}{t}
+c^{\rm o} X \der{T_X^{\rm qeq}}{t}.
\label{iv-93}
\end{equation}
  Here, by using  $u^{\rm o}(T)$  defined by (\ref{uo-def}),
we have the following identity:
\begin{align}
&\der{}{t}\left[ X^2 \left( u^{\rm o}(T_X^{\rm qeq})  -\frac{E}{A}  \right)\right]
  \nonumber \\
&= 2 X \der{X}{t}\left( u^{\rm o}(T_X^{\rm qeq})  -\frac{E}{A}  \right)
+X^2  c^{\rm o}
    \der{T_X^{\rm qeq}}{t}. 
\label{q-ident}
\end{align}
We also obtain
\begin{align}
&  \psi_X^{\rm qeq}(X) =  X \left( u^{\rm o}(T_X^{\rm qeq})-  \frac{E}{A} \right),
   \label{qeq-psi-X} 
\end{align}
from (\ref{psi-o-def}). 
By using (\ref{ons-cof-o}) and  (\ref{q-ident}) with (\ref{qeq-psi-X}),  we rewrite (\ref{iv-93}) as
\begin{align}
\frac{1}{{L}^{\rm o}} \der{\Psi^{\rm o}}{t} 
=&
\frac{g}{\lambda^{\rm o}}
\left\{ - \der{X}{t} \psi_X^{\rm qeq}(X) 
 +    \der{}{t}
     \left[ X \psi_X^{\rm qeq}(X)  \right]
     \right\}     \nonumber\\
=& \frac{g}{\lambda^{\rm o}}
\left(  X   \der{}{t} \psi_X^{\rm qeq}(X)  \right),
     \label{77}     
\end{align}
and we also have
\begin{align} 
\der{T_X^{\rm qeq}}{t}
 \frac{c^{\rm o}}{3 \lambda^{\rm o}} X^2 
 =
-\frac{1}{3\lambda^{\rm o}}
 \der{X}{t} \psi_X^{\rm qeq}(X)
+\frac{1}{3}
\left[ \frac{X}{\lambda^{\rm o}}  \der{}{t} \psi_X^{\rm qeq}(X)  \right],
\label{78}
\end{align}
where we have replaced $X_{+/-}$ in (\ref{ons-cof-o}) and 
(\ref{ons-cof-d}) by $X$ with ignoring $O(\eta)$ terms.

Here,   from (\ref{psi-d-def}),   we have
\begin{equation}
\left( u^{\rm d}(T_X^{\rm qeq})  -\frac{E}{A}  \right)  (1-X)
= -\psi_X^{\rm qeq}(X) .
\label{ud-psi}
\end{equation}
By using an identity similar to (\ref{q-ident}) and (\ref{ud-psi}),
we also have
\begin{equation}
\frac{1}{{L}^{\rm d}} \der{\Psi^{\rm d}}{t} 
= -\frac{g}{\lambda^{\rm d}}
\left(  (1-X)  \der{}{t} \psi_X^{\rm qeq}(X)  \right),
     \label{79}     
\end{equation}
and
\begin{align} 
\der{T_X^{\rm qeq}}{t}
 \frac{c^{\rm d}}{3 \lambda^{\rm d}} (1-X)^2 
 =&
-\frac{1}{3\lambda^{\rm d}}
 \der{X}{t}  \psi_X^{\rm qeq}(X)
\nonumber \\
& -\frac{1}{3}
\left[ \frac{1-X}{\lambda^{\rm d}} \der{}{t} \psi_X^{\rm qeq}(X) \right].
\label{80}
\end{align}
By substituting (\ref{77}), (\ref{78}), (\ref{79}) and (\ref{80})
into (\ref{tochu}),  we obtain
\begin{align}
&   \beta_+^{\rm int}-\beta_-^{\rm int} \nonumber \\
&=
  -\frac{1}{3} \left(\frac{1}{\lambda^{\rm o}}-
  \frac{1}{\lambda^{\rm d}} \right)\der{X}{t}
     \psi_X^{\rm qeq}(X) \nonumber \\
&\quad 
-  \left(g-\frac{1}{3} \right)
 \left( \frac{X}{\lambda^{\rm o}}+\frac{1-X}{\lambda^{\rm d}} \right)
  \der{}{t}\left[
 \psi_X^{\rm qeq}(X) \right] ,
\label{result-ons}
\end{align}
The formula (\ref{result-ons}) gives the inverse temperature gap of
$O(\sqrt{\eta})$.

Let us recall that $g$ is a phenomenological parameter and its
value is not specified yet. Here, we impose the condition
that the temperature gap vanishes when $X \to 0$  and
$X \to 1$.
Noting that $dX/dt \not = 0$ in the limit 
$X \to 0$ or $X \to 1$, this condition determines
the unique value of $g$ as $g=1/3$. We then have 
arrived at the formula of the inverse temperature gap:
\begin{equation}
 \beta_+^{\rm int}-\beta_-^{\rm int} 
= 
  -\frac{1}{3} \left(
  \frac{1}{\lambda^{\rm o}}-  \frac{1}{\lambda^{\rm d}} \right)
  \der{X}{t} \psi_X^{\rm qeq}(X)
\label{result-ons-2}
\end{equation}
up to the error of $O(\eta)$. By using (\ref{q-X}), we
can express (\ref{result-ons-2}) as 
\begin{equation}
\beta_+^{\rm int}-\beta_-^{\rm int} 
= 
  \frac{1}{3} \left(
  \frac{1}{\lambda^{\rm o}}-  \frac{1}{\lambda^{\rm d}} \right)
  \der{X}{t}
  X(1-X) q_X.
\label{result-ons-a}
\end{equation}
This formula clearly indicates that the temperature gap
is associated with the latent heat generated at the moving
interface. See Fig. \ref{fig:t-config} for the summary
of the result. 

\subsection{Final result}\label{res-correction}

We substitute (\ref{result-ons-2}) into (\ref{Ical-int-22}).
We then obtain 
\begin{equation}
\lim_{K \to \infty} \lim_{\eta \to 0} I_2^{\rm int} (\alpha_X)
=  - \frac{1}{3}\int_{X}^{X_{\rm eq}} \!\!\!\! dY \psi_Y^{\rm qeq}(Y)
 \left(  \frac{1}{\lambda^{\rm o}}-  \frac{1}{\lambda^{\rm d}} \right).
\label{Ical-int-result}
\end{equation}
By combining (\ref{I-bulk-result}) and (\ref{Ical-int-result})
in the formula (\ref{caI-start}), we complete the calculation
of the correction term as
\begin{eqnarray}
{\cal I}(\alpha_X)
&=&  \frac{A}{\lambda^{\rm o}}\int_0^X dx  \psi_X(x)
   + \frac{A}{\lambda^{\rm d}} \int_X^1 dx  \psi_X(x) \nonumber \\
& & - \frac{A}{\lambda^{\rm o}}\int_0^{X_{\rm eq}} dx  \psi_{X_{\rm eq}}^{\rm qeq}(x)
   - \frac{A}{\lambda^{\rm d}} \int_{X_{\rm eq}}^1 dx  \psi_{X_{\rm eq}}^{\rm qeq}(x)
   \nonumber \\
   & &  + \frac{2A}{3}\int_X^{X_{\rm eq}}dY \psi_{Y}^{\rm qeq}(Y)
   \left(
   \frac{1}{\lambda^{\rm o}} - \frac{1}{\lambda^{\rm d}}
   \right) .
\label{I-result}
\end{eqnarray}
 By substituting (\ref{I-result}) into (\ref{tildeS-0}),
we obtain 
\begin{align}
  \tilde {\cal S}(\alpha_X;& E,J) 
= A \int_0^1 dx  s(  u_X(x) ,m_X(x))   \nonumber \\
 &+  \frac{AJ}{\lambda^{\rm o}}\int_0^X dx  \psi_X(x)
    + \frac{AJ}{\lambda^{\rm d}}\int_X^1 dx  \psi_X(x)  \nonumber \\
&    - \frac{2AJ}{3}\int^X_0   dY \psi_{Y}^{\rm qeq}(Y)
   \left(
   \frac{1}{\lambda^{\rm o}} - \frac{1}{\lambda^{\rm d}}
   \right) 
\label{tildeS2}
\end{align}
  up to an additive constant independent of $X$.  
Combining it with (\ref{Zubarev-flux-0}), we finally obtain
the stationary distribution of interface configurations.

\section{Variational principle}
\label{Sec:var}

We consider the case $ E_1 \le E \le E_2$ with (\ref{53}) and (\ref{54}).
When $J=0$, the most probable configuration contains a single interface,
whose position is determined by the microcanonical ensemble. Explicitly,
the position $X_*$ maximizes the total entropy. Even when $J \not =0$,
the most probable configuration may contain a single interface. We then
expect that its position $X_*$ is determined by a  variational principle
that is obtained as an extension of the maximum entropy principle
when  $\epsilon=|J|$   is small. In this section, we study this variational
principle.  In Sec. \ref{Sec:var-1}, we present a formulation of the
problem. 
In Sec. \ref{4-formula}, we explicitly derive the  variational function.
After some preliminaries in Sec. \ref{5-prelimi}, 
we re-express the variational equation as the form of the free energy
difference at the interface   in Sec. \ref{5-val}.  
In Sec. \ref{5-result}, 
from this expression, we derive the temperature of the interface.
Throughout this section, we evaluate quantities neglecting $O(\epsilon^2)$
terms even without explicit remarks. 

\subsection{Formulation of the problem}
\label{Sec:var-1}

We assume that the most probable profile   in the steady
state    is independent of $(y,z)$
and possesses an interface
at $x=X_*$.  
Then, we observe the
ordered state in the region $0\le x < X_*$ and the disordered state in the
region $ X_* < x \le 1$. When $X_*$ is given, the most probable profile of
$(m(\bv{r}), v(\bv{r}), \phi(\bv{r}))$ in the limit $\eta \to 0$
is determined from 
the conditions $v(\bv{r})=0$, $\sigma(T,m)=0$, and $\lambda \partial_x\beta=J$
in each region.  It should be noted that $X_*$ is not obtained
by the stationary solution of (\ref{model-s1}), (\ref{model-s2}),
and  (\ref{model-s3}) with $\eta=0$. Thus, we determine  $X_*$
by considering the probability density $P(X;E,J)$ of the
interface position $X$ for small $\eta$. We expect that
$P(X;E,J)$ takes the form 
\begin{equation}
  P(X;E,J) = e^{\frac{1}{\eta^3} \left[{\cal V}(X;E,J)+O( \sqrt{\eta} )\right] }
\label{pot-def}
\end{equation} 
in the limit $\eta \to 0$. 
Here, the potential function
${\cal V}(X)$ is independent of $\eta$. Then, the most
probable position of the interface $X_*$ is given as
the maximizer of ${\cal V}(X;E,J)$, which is the
variational principle we expect.  

We consider the potential function ${\cal V}(X)$. 
For equilibrium cases $J=0$,  ${\cal V}(X)$ is given as
the total entropy for the quasi-equilibrium profile
with the interface position $X$ in the limit $\eta \to 0$.
We generalize this result to the case $J <0$. 

Let ${\cal C}_X$ be the set of  configurations with a
single interface with the interface position $X$.
Suppose that a configuration with a single interface
is observed. The probability density of the interface
position $X$ on this condition  is expressed as
\begin{equation}
  P(X;E,J)=\frac{ \int_{{\cal C}_X}  d \alpha_X
    {\cal P}_{\rm ss}(\alpha_X;E,J)}{
    \int_0^1  dY \int_{{\cal C}_Y}  d \alpha_Y
                 {\cal P}_{\rm ss}(\alpha_Y;E,J)},
\end{equation}
where ${\cal P}_{\rm ss}$ is given by (\ref{Zubarev-flux-0}).
Since we consider the limit $\eta \to 0$,  we reasonably
conjecture from (\ref{pot-def}) that 
\begin{equation}
 {\cal V}(X;E,J)
 =
 \max_{\alpha_X \in {\cal C}_X}\tilde {\cal S}(\alpha_X;E,J), 
 \label{pot}
\end{equation}
where fluctuations of $\alpha_X$ are assumed to be sub-leading
in the evaluation of  ${\cal V}(X;E,J)$.

\subsection{Formula of the potential}\label{4-formula}

\begin{figure}[tb]
\centering
\includegraphics[width=7cm]{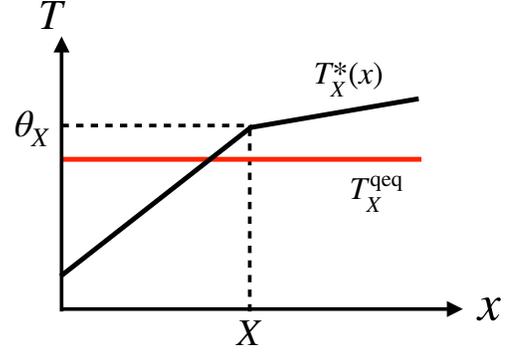}
\caption{Temperature profile $T_X^*(x)$
that maximizes
the modified entropy $\tilde {\cal S}(\alpha_X;E,J)$ for a given $X$.
$\kappa^{\rm d}>\kappa^{\rm o}$.}
\label{fig:T-NEQ}
\end{figure}

We calculate the right-hand side of (\ref{pot}).
Note that  the last line of (\ref{tildeS2}) 
is independent of $\alpha_X$, while it depends on $X$.
Thus, the last line is not relevant in the maximization
of $\tilde {\cal S}(\alpha_X;E,J)$, but necessary in the maximization
of ${\cal V}(X;E,J)$ in $X$.
Let $\alpha_X^*$ be the maximizer of $\tilde {\cal S}(\alpha_X;E,J)$
with  $X$ fixed. We then rewrite (\ref{pot}) as
\begin{equation}
{\cal V}(X; E,J) = \tilde {\cal S}( \alpha_X^*;E,J). 
\label{V-start}
\end{equation}
Now, we derive $\alpha_X^*$ by taking the variation of
$ \tilde {\cal S}(\alpha_X;E,J)$
in $m_X$, $v_X$ and $\psi_X$. The result of the variation
\begin{eqnarray}
&&  \int_0^1 dx \left[(\delta\psi_X)
  \left( -\partial_x \beta_X+ \frac{J}{\lambda}\right) \right. \nonumber \\
&&   +  \left. (\delta m_X) \beta \sigma(u_X,m_X) - (\delta  v_X) \beta v_X
 \right]
=0
\end{eqnarray}
leads to 
\begin{align}
& J= \lambda^{\rm o} \partial_x \beta^*_X
        \qquad {\rm for} \quad  x < X  \label{eq-t-o}, \\
& J = \lambda^{\rm d} \partial_x \beta^*_X 
  \qquad {\rm for} \quad  x > X,  \label{eq-t-d} \\
&  \sigma(T^*_X(x), m^*_X(x))= 0 , \label{eq-m} \\
&     v_X^*(x)= 0 ,
\end{align}
  where note that $\psi_X^{\rm qeq}(x)$ is independent of $\alpha_X$.
  Here, let $\theta$ be an interface temperature.
For given $X$ and $\theta$,
we define a new quantity $\tilde T^*_X(x;\theta)$ as   
the solution of (\ref{eq-t-o}) and (\ref{eq-t-d}) with
$\tilde T_X^*(X;\theta)=\theta$.  Obviously, $\tilde T^*_X(x;\theta)$
is equivalent to the stationary solution of the transportation
equation in the heat conduction.
Then, energy conservation 
\begin{equation}
 A \int_0^1 dx u(\tilde T^*_X(x;\theta), m^*_X(x))=E
\label{theta-T}
\end{equation}
provides the special value of $\theta$, which is denoted by $\theta_X$.
$T_X^*(x)$ is determined by $T_X^*(x)=\tilde T^*_X(x;\theta_X)$,
and then  $m^*_X(x)$ is determined from (\ref{eq-m}). 
In Fig.~\ref{fig:T-NEQ}, we display an example of the temperature
profile $T_X^*(x)$. Since $T_X^*(x)=T_X^{\rm qeq}+O(\epsilon)$, 
we also have 
\begin{eqnarray}
  \psi_X^*(x) &=&
  \int_0^x dy \left[ u^{\rm o}(T_X^*(y))- \frac{E}{A} \right]
  \nonumber    \\
   &=&  \psi_X^{\rm qeq}(x)+O(\epsilon) 
  \label{psi-o}
\end{eqnarray}
for $x <X$. Similarly, 
\begin{equation}
   \psi_X^*(x)=  \psi_X^{\rm qeq}(x)+O(\epsilon) 
\label{psi-d}
\end{equation}
for $x >X$. By substituting these results into (\ref{V-start})
  with(\ref{tildeS2})  ,
we obtain
\begin{align}
&{\cal V}(X; E,J)  \nonumber  \\
           &=  A \int_0^X dx   s^{\rm o}(T^*_X(x))
               + A \int_X^1 dx   s^{\rm d}(T^*_X(x))  \nonumber \\
&\quad +\frac{AJ}{\lambda^{\rm o}}\int_0^X dx \psi_X^{\rm qeq}(x)
   + \frac{AJ}{\lambda^{\rm d}}\int_X^1 dx  \psi_X^{\rm qeq}(x)
                      \nonumber \\
 &\quad    - \frac{2AJ}{3}\int^X_0  
                 dY \psi_{Y}^{\rm qeq}(Y)
   \left(  \frac{1}{\lambda^{\rm o}} - \frac{1}{\lambda^{\rm d}} \right).
\label{V-final}
\end{align}
Then, (\ref{theta-T}) is written as 
\begin{equation}
A\int_0^X dx u^{\rm o}(T^*_X(x))+ A \int_X^1 dx u^{\rm d}(T^*_X(x))=E.
\label{theta-det}
\end{equation}

\subsection{Preliminaries for maximization of the potential }
\label{5-prelimi}

In order to calculate $X_*$  that maximizes ${\cal V}(X)$
under the condition (\ref{theta-det}), we present
some preliminaries.  First,  noting
\begin{equation}
  \partial_x s_X^*(x)
= -T^*_X c^{\rm o}(T^*_X) \frac{J}{\lambda^{\rm o}}
\end{equation}
for $x < X$, we obtain
\begin{eqnarray}
 s_X^*(x) &=& s^{\rm o}(\theta_X)
  -\int_X^x dx T^*_X(x) c^{\rm o}(T^*_X) \frac{J}{\lambda^{\rm o}} \nonumber \\
  &=& s^{\rm o}(\theta_X) - (x-X)\theta_X c^{\rm o}(\theta_X)
  \frac{J}{\lambda^{\rm o}},
\end{eqnarray}
which leads to 
\begin{eqnarray}
  \int_0^X dx  s_X^*(x)
&=& Xs^{\rm o}(\theta_X) +\frac{X^2}{2}
\theta_X c^{\rm o}(\theta_X) \frac{J}{\lambda^{\rm o}}  \nonumber \\
&=& Xs^{\rm o}\left(\theta_X+ \theta_X^2\frac{X J}{2\lambda^{\rm o}}
\right).
\end{eqnarray}
Similarly, we have
\begin{equation}
  \int_X^1 dx  s_X^*(x)
  = (1-X)s^{\rm d}
  \left(\theta_X- \theta_X^2\frac{(1-X) J}{2\lambda^{\rm d}} \right).
\end{equation}
Here, it is convenient to introduce
\begin{eqnarray}
T_X^{\rm o}  & =&  \theta_X+ \theta_X^2\frac{X J}{2\lambda^{\rm o}}, \\
T_X^{\rm d}  & =&  \theta_X- \theta_X^2\frac{(1-X) J}{2\lambda^{\rm d}}.
\end{eqnarray}
It should be noted that
\begin{eqnarray}
  T_X^{\rm o}  &=&  \frac{1}{X} \int_0^X dx T_X^*(x)+O(\epsilon^2),
  \label{v-22}\\
  T_X^{\rm d}  &=&  \frac{1}{1-X} \int_X^1 dx T_X^*(x)+O(\epsilon^2).
  \label{v-23}
\end{eqnarray}
That is, $T_X^{\rm o}$ and $T_X^{\rm d}$ are the spatially averaged
temperatures in the ordered phase and in the disordered phase, respectively,
which are basically the  same as those in (\ref{to-def}) and (\ref{td-def}).

\subsection{Variational equation}
\label{5-val}

In this subsection, we simplify the variational equation.
Substituting   (\ref{psi-o-def}) and (\ref{psi-d-def}) 
into (\ref{V-final}), we have 
\begin{align}
 &\frac{{\cal V}(X;E,J)}{A}
=   
Xs^{\rm o}(T_X^{\rm o})+(1-X)s^{\rm d}(T_X^{\rm d}) \nonumber \\
& \quad +
\frac{X^2 J}{2\lambda^{\rm o}}\left( u^{\rm o}(\theta_X)-\frac{E}{A} \right)
-
\frac{(1-X)^2 J}{2\lambda^{\rm d}}
\left( u^{\rm d}(\theta_X)-\frac{E}{A} \right) \nonumber  \\
&\quad  
-\frac{2J}{3}\left(  \frac{1}{\lambda^{\rm o}} - \frac{1}{\lambda^{\rm d}} \right)
\int^X_0  dY \left( u^{\rm o}(\theta_Y)-\frac{E}{A} \right)Y ,
\label{Phi-def}
\end{align}
where $\theta_X$ in the right hand side is a function of $X$ whose
dependence is determined by 
\begin{equation}
X u^{\rm o}(T_X^{\rm o})+ (1-X) u^{\rm d}(T_X^{\rm d}) =\frac{E}{A},
\label{theta-det-2}
\end{equation}
where $T_X^{\rm o}$ and $T_X^{\rm d}$ are given by
(\ref{v-22}) and (\ref{v-23}). 

Then, the variational equation
\begin{equation}
  \der{\cal V}{X}=0
\end{equation}  
becomes
\begin{eqnarray}
& & s^{\rm o}(T_X^{\rm o})-s^{\rm d}(T_X^{\rm d}) \nonumber \\
& &
+X  \frac{c^{\rm o}(T_X^{\rm o})}{T_X^{\rm o}} \der{T_X^{\rm o}}{X}
+(1-X) 
\frac{c^{\rm d}(T_X^{\rm d})}{T_X^{\rm d}} \der{T_X^{\rm d}}{X}
\nonumber \\
&&
+\frac{X J}{\lambda^{\rm o}}\left( u^{\rm o}(\theta_X)-\frac{E}{A} \right)
+
\frac{(1-X) J}{\lambda^{\rm d}}
\left( u^{\rm d}(\theta_X)-\frac{E}{A} \right) \nonumber  \\
& &
+
\left[\frac{X^2 J}{2\lambda^{\rm o}} c^{\rm o}(\theta_X)
-
\frac{(1-X)^2 J}{2\lambda^{\rm d}}c^{\rm d}(\theta_X)\right]\der{\theta_X}{X}
\nonumber \\
&& -\frac{2J}{3}\left(\frac{1}{\lambda^{\rm o}} - \frac{1}{\lambda^{\rm d}} \right)
\left( u^{\rm o}(\theta_X)-\frac{E}{A} \right) X 
\nonumber \\
&&= 0 .
\label{Phi-1}
\end{eqnarray}
From (\ref{theta-det-2}), we also obtain
\begin{equation}
 u^{\rm o}(T_X^{\rm o})- u^{\rm d}(T_X^{\rm d})
 + X c^{\rm o}(T_X^{\rm o})     \der{T_X^{\rm o}}{X}
 + (1-X) c^{\rm d}(T_X^{\rm d}) \der{T_X^{\rm d}}{X}
 =0 .
\label{theta-det-3}
\end{equation}
The second line of (\ref{Phi-1}) is expressed as
\begin{align}
&X  \frac{c^{\rm o}(T_X^{\rm o})}{\theta_X} \der{T_X^{\rm o}}{X}
+(1-X)
  \frac{c^{\rm d}(T_X^{\rm d})}{\theta_X} \der{T_X^{\rm d}}{X}
\nonumber \\
&+
X (\theta_X-T_X^{\rm o})
\frac{c^{\rm o}(T_X^{\rm o})}{\theta_X^2} \der{T_X^{\rm o}}{X}
\nonumber \\
& +(1-X)(\theta_X-T_X^{\rm d})
\frac{c^{\rm d}(T_X^{\rm d})}{\theta_X^2} \der{T_X^{\rm d}}{X}.
\label{Phi-2nd}
\end{align}
By using (\ref{theta-det-3}), we find that the first line in (\ref{Phi-2nd})
is 
\begin{equation}
  - \frac{ u^{\rm o}(T_X^{\rm o}) -u^{\rm d}(T_X^{\rm d })}{\theta_X}.
\label{Phi-2nd-1}
\end{equation}
The combination with the first line in (\ref{Phi-1}) yields
\begin{align}
  &  s^{\rm o}(T_X^{\rm o})
        -\frac{u^{\rm o}(T_X^{\rm o})}{\theta_X}
          -\left[ s^{\rm d}(T_X^{\rm d})
            -\frac{u^{\rm d}(T_X^{\rm d})}{\theta_X}
           \right]
\nonumber \\      
&= s^{\rm o}(\theta_X)-\frac{u^{\rm o}(\theta_X)}{\theta_X}
-\left[ s^{\rm d}(\theta_X)-\frac{u^{\rm o}(\theta_X)}{\theta_X}\right]
\nonumber \\
&=
 - \frac{f^{\rm o}(\theta_X) -f^{\rm d}(\theta_X)}{\theta_X},
\label{Phi-1-1-2}
\end{align}
where we have defined
\begin{eqnarray}
f^{\rm o}(\theta_X)&=& u^{\rm o}(\theta_X)-\theta_X s^{\rm o}(\theta_X), \\
f^{\rm d}(\theta_X)&=& u^{\rm o}(\theta_X)-\theta_X s^{\rm d}(\theta_X) .
\end{eqnarray}
The second and third lines in (\ref{Phi-2nd}) become
\begin{equation}
-   \frac{X^2 J}{2\lambda^{\rm o}}
c^{\rm o}(T_X^{\rm o}) \der{T_X^{\rm o}}{X}
+\frac{(1-X)^2 J}{2\lambda^{\rm d}}
{c^{\rm d}(T_X^{\rm d})} \der{T_X^{\rm d}}{X},
\end{equation}
which cancels with the forth line in (\ref{Phi-1}).
The third line and the fifth line in (\ref{Phi-1}) are summarized as
\begin{equation}
\frac{J}{3}\left(  \frac{1}{\lambda^{\rm o}} - \frac{1}{\lambda^{\rm d}} \right)
\left( u^{\rm o}(\theta_X)-\frac{E}{A} \right) X ,
\label{Phi-3-5}
\end{equation}
where we have used
\begin{equation}
  X \left(u^{\rm o}(\theta_X) -\frac{E}{A} \right)
+ (1-X) \left(u^{\rm d}(\theta_X)  -\frac{E}{A} \right)=O(\epsilon),
\label{theta-det-22}
\end{equation}
which comes from (\ref{theta-det-2}). 
Furthermore, noting (\ref{q-X}), we re-express
(\ref{Phi-3-5}) as 
\begin{equation}
-\frac{J}{3}\left(  \frac{1}{\lambda^{\rm o}} - \frac{1}{\lambda^{\rm d}} \right)
{X(1-X)} q_X . 
\label{Phi-3-5-2}
\end{equation}
In this manner, (\ref{Phi-1-1-2}) and (\ref{Phi-3-5-2})
remain in the left-hand side of   (\ref{Phi-1}). Thus, 
the variational equation (\ref{Phi-1}) is simplified as
\begin{align}
f^{\rm o}(\theta_X)-f^{\rm d}(\theta_X) 
  =-\frac{\theta_X J}{3}\left( \frac{1}{\lambda^{\rm o}}
  - \frac{1}{\lambda^{\rm d}} \right)
{X(1-X)} q_X . 
\label{Phi-final}
\end{align}
This equation with (\ref{theta-det-2}) gives the most probable
value $(\theta_*, X_*)$ of the interface temperature $\theta$
and the interface position $X$.

\subsection{Result} \label{5-result}

\begin{figure}[tb]
\centering
\includegraphics[width=7cm]{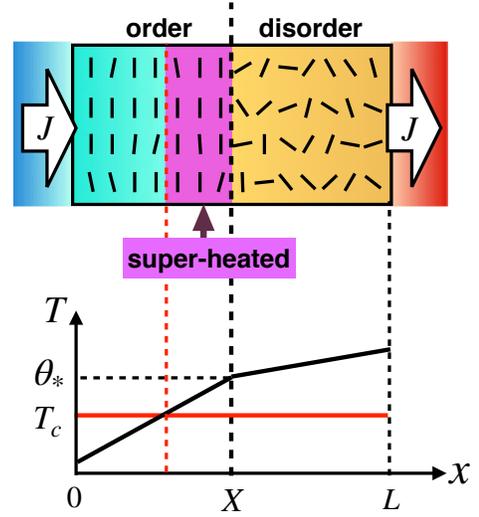}
\caption{Schematic of the main result.}
\label{fig:final}
\end{figure}

When we set $J=0$ in (\ref{theta-det-2}) and (\ref{Phi-final}),
we find that  $\theta_*=\Tc(=1)$ and $X_*=X_{\rm eq}$ given by (\ref{X-eq}).
When $J \not =0$, we derive the equation for
$\theta_*-\Tc$ from (\ref{Phi-final}) as 
\begin{align}
&-(s^{\rm o}(\Tc)-s^{\rm d}(\Tc))(\theta_*-\Tc) \nonumber \\
  &\quad =-\frac{\Tc J}{3}\left( \frac{1}{\lambda^{\rm o}}
  - \frac{1}{\lambda^{\rm d}} \right)
{X_{\rm eq}(1-X_{\rm eq})} q_X , 
\label{Phi-final-2}
\end{align}
which yields
\begin{equation}
\theta_*-\Tc
=-\frac{\Tc^2 J}{3}
\left(  \frac{1}{\lambda^{\rm o}} - \frac{1}{\lambda^{\rm d}} \right)
{X_{\rm eq}(1-X_{\rm eq})}.
\label{temp}
\end{equation}
When we use the standard thermal conductivity $\kappa$ defined by
(\ref{kappa-def}), we rewrite (\ref{temp}) as
\begin{equation}
\theta_*-\Tc
=-\frac{J}{3}
\left(  \frac{1}{\kappa^{\rm o}} - \frac{1}{\kappa^{\rm d}} \right)
{X_{\rm eq}(1-X_{\rm eq})}.
\label{temp-1}
\end{equation}
Suppose that $\kappa^{\rm d} > \kappa^{\rm o}$ (or $\kappa^{\rm d} < \kappa^{\rm o}$).
Noting $J <0$, we find $\theta_* > T_c$ (or $\theta_* < T_c$). This means
that the super-heated ordered state (or super-cooled disordered state)
stably appears near the interface in the heat conduction state.
See Fig. \ref{fig:final}. 
This phenomenon was predicted by an extended framework of
thermodynamics \cite{NS}, which is called {\it global thermodynamics}
\cite{NS2}. If the factor $1/3$ were $1/2$, the result
(\ref{temp-1}) would be equivalent to the quantitative prediction
by global thermodynamics. We conjecture that the discrepancy comes
from the approximation we used in Sec. \ref{t-gap}.
By comparing (\ref{temp-1}) with (\ref{result-ons-a}),
we find that $\theta_*-T_c$ is quantitatively connected
to the temperature gap $T_+^{\rm int}-T_-^{\rm int}$ when $J$ is
identified with $q_X dX/dt$. 

Finally, from the left-right symmetry, we notice that
$\theta_*$ is invariant for $(J,X) \to (-J,1-X)$.
Thus, we express (\ref{temp-1}) as 
\begin{equation}
\theta_*-\Tc
=\frac{|J|}{3}
\left(  \frac{1}{\kappa^{\rm o}} - \frac{1}{\kappa^{\rm d}} \right)
{X_{\rm eq}(1-X_{\rm eq})}
\label{temp-2}
\end{equation}
for any $J$.    Note that the symmetry breaking field
$\sigma^{\rm ex}(x)$
is also replaced by $\sigma^{\rm ex}(1-x)$ for the case $J >0$.  

\section{Concluding remarks}\label{remark}


We have proposed the stochastic model (\ref{model-s1}), 
(\ref{model-s2}), and (\ref{model-s3}) for describing
 phase coexistence in heat conduction.
As a special boundary condition, we  imposed
the non-equilibrium adiabatic condition
  (\ref{noise-21}) and (\ref{noise-22})  ,
which is a natural
extension of the adiabatic condition with $J=0$. For this system,
we formulated the variational principle for determining
the interface position $X$. We have shown that the variational
function ${\cal V}(X)$ given in (\ref{pot}) is calculated
as  (\ref{Phi-def}).  
By solving the variational problem, we  found that the interface
temperature deviates from $T_c$, which implies that
quasi-equilibrium states stably appear near the interface.
Before ending this paper, we discuss possible directions for studies.


First, we consider a liquid-gas transition, which is the most popular
first-order transition. The generalized hydrodynamics with the
interface thermodynamics was proposed \cite{Anderson98,Bedeaux03,Onuki},
and the fluctuating hydrodynamics without interfaces is well-established
\cite{Schmitz}. Thus,  a stochastic model
could be constructed through a combination of the two models.
By imposing the non-equilibrium adiabatic boundary conditions,
we may derive a potential function for determining the
liquid-gas interface. It is reasonable to conjecture that the
potential function is calculated from the modified 
entropy for the stationary profile of the interface position
$X$, because the method developed in this paper can be used
for  liquid-gas coexistence in  heat conduction.
The main difference is that the density is conserved, which 
causes an additional contribution to the interface temperature,
as shown in Ref. \cite{NS2}. 
Explicit calculation of the interface temperature may
be an important exercise. 


Secondly, the variational formula we have derived in this paper
may be related to global thermodynamics for heat conduction \cite{NS2}.
Both formulas predict that  the interface temperature
deviates from the transition temperature at equilibrium.
To find the direct connection between the two
theories, one may construct
a thermodynamic framework by employing an extended
Clausius relation for the stochastic order parameter
dynamics. See Refs.
\cite{Hatano-Sasa,KNST,NN,Jona-thermo,Maes-thermo,
Chiba-Nakagawa} for studies related to an extended Clausius relation.
This is the next subject in developing the theory.


Here, we briefly   review  
the global thermodynamics. The theory 
describes spatially inhomogeneous systems by a few global quantities,
such as the global temperature,
which is defined such that the fundamental relation in thermodynamics
is satisfied. 
This idea is simple and natural but has never
been considered in previous studies seeking an extended framework
of thermodynamics \cite{Keizer,Eu,Jou,Oono-paniconi,Sasa-Tasaki,
Bertin,Seifert-contact,Dickman}.
More importantly, this framework
naturally leads to a quantitative prediction of the interface
temperature $\theta$ different from $\Tc$. Therefore,
experiments can judge the validity of the fundamental hypothesis
on which global thermodynamics is built. See Ref. \cite{NS2} for
an explanation of the theory, including a comparison with
other  extended frameworks of thermodynamics.


Thirdly, the result on the interface temperature is obtained only for
the special boundary condition. Naturally, one may want to derive
the interface temperature for more standard cases
where two heat baths of different temperatures contact with
the system. Even for this case, we can use the stochastic
dynamics (\ref{model-s1}), 
(\ref{model-s2}), and (\ref{model-s3}) with the boundary
conditions $T(0,t)=T_{\rm L}$ and $T(1,t)=T_{\rm R}$. We can derive
the Zubarev-Mclennan representation, which includes the
time integration of the entropy production rate. This
term can hardly be evaluated theoretically without knowing
the steady state profile. Although we physically conjecture
that the interface temperature is independent of boundary conditions when the
value of the heat flux is the same, we do not have a
proof of this conjecture.  It is challenging 
to calculate the interface temperature for the boundary
conditions $T(0,t)=T_{\rm L}$ and $T(1,t)=T_{\rm R}$.


Fourthly, to the best of our knowledge, the first-order transition in  heat
conduction has never been studied by  systematic numerical
experiments. One reason for this is that there are no paradigmatic models for
describing the phase coexistence in heat conduction. It
may be useful if such a numerical model was devised.
Furthermore, by performing numerical simulations of such
models, one may obtain  a phase diagram of the system.
In particular, the numerical determination of the interface
temperature may be stimulating. The results will be
compared with our theoretical results quantitatively.


Fifthly, related to the fourth problem, one may recall that the molecular
dynamics simulations were performed in order to study the phase 
coexistence in  heat conduction \cite{Bedeaux00,Ogushi}.
However, no deviation of the interface temperature from the transition
temperature was observed.
We conjecture that this is due to  insufficient  separation
of scales. For example,
when $\eta=10^{-2}$, the dimensionless interface width in our
description is $10^{-1}$. Such a system may be well described by
a deterministic equation, and thus $\theta=\Tc$ holds.
Even for such small systems, the precise measurement of fluctuating
quantities may reveal the true behavior in the limit $\eta \to 0$. 
Formulating such statistical properties is an important
theoretical problem. 


Finally, the most important future study is to stably observe
the super-heated ordered (or super-cooled disordered) state
in laboratory experiments. 
Even qualitative observation of the stabilization of such
states is quite interesting.  To observe this phenomenon,
a precise temperature profile should be measured. A novel concept
must be designed for such an experimental setup. 

After studying these subjects, we will aim to construct
a universal theory for phase coexistence out of equilibrium.
We hope that this paper is a starting point for studying various
dynamical behaviors associated with phase coexistence out of equilibrium.


\section*{Acknowledgment}
The authors thank  Christian Maes, Kazuya Saito, Satoshi Yukawa,
Michikazu Kobayashi, Yuki Uematsu, Masafumi Fukuma, Kyosuke Tachi,
Akira Yoshida and Hiroyoshi Nakano for their useful comments.
The authors also specially thank Michikazu Kobayashi for his
informal communication on numerical simulations of an order-disorder
transition under heat conduction. 
The present study was supported by KAKENHI (Nos. 17H01148, 19H05496,
19H05795, 19K03647, 17K14355, 19H01864, 20K20425, 20J00003). 



\appendix

\section{Example of entropy functional}
\label{example}

In this Appendix, we provide a specific example of
$s(u,m)$ that exhibits the first-order transition at
$T=T_c$. Although our theory is formulated regardless of
specific forms of $s(u,m)$, one may  consider the example
  in the argument of the main text.   

\subsection{Landau theory}

\begin{figure}[tb]
\centering
\includegraphics[width=6cm]{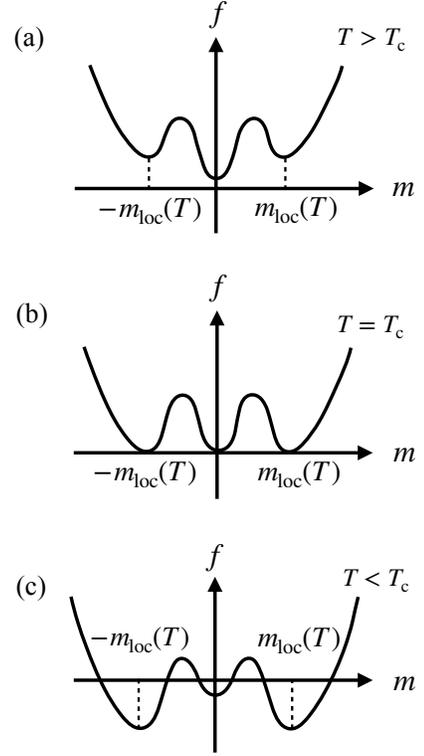}
\caption{Free energy as a function of $m$ for $T$ fixed.}
\label{fig:f-m}
\end{figure}

We start with a Landau free energy density
\begin{equation}
  f(T,m)=\frac{a_1}{2}(T-T_0)m^2-\frac{a_2}{4}m^4
      +\frac{a_3}{6}m^6 +\varphi(T),
\label{GL-ex}
\end{equation}
which describes the first order transition at some temperature 
$T_c$. 
Here, $a_1$, $a_2$, $a_3$, and $T_0$ are
positive constants. The functional form of
  $\varphi(T)$ will be determined later. See (\ref{phid}). 
For a given $T$, the equilibrium value $m_{\rm eq}(T) \ge 0$ is
determined as the minimizer of $f(T,m)$ with respect to $m$.
As shown in Fig. \ref{fig:f-m}, $m_{\rm eq}(T)$ is expressed
in terms of positive $m_{\rm loc}(T)$ in the locally stable state:
\begin{align}
 & m_{\rm eq}(T) = 0  &{\rm for} ~~  T > \Tc,  \\
&  m_{\rm eq}(T) = m_{\rm loc}(T) &  {\rm for} ~~  T < \Tc,
\end{align}
where $\Tc$ is determined as 
\begin{equation}
f(\Tc, m_{\rm loc}(\Tc))=f(\Tc, 0).
\label{21}
\end{equation}
Since $m_{\rm loc}(\Tc) >0$, $m_{\rm eq}(T)$ is discontinuous
at $T=\Tc$.

\begin{figure}[tb]
\centering
\includegraphics[width=6.5cm]{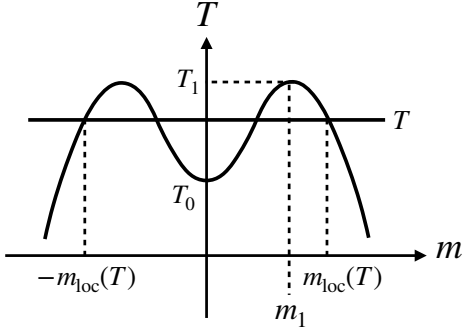}
\caption{$T(m)$ as  a function of $m$.}
\label{fig:T-m}
\end{figure}

We derive $m_{\rm loc}(T)$ explicitly. We define $\sigma(T,m)$ as
\begin{equation}
\sigma(T,m)\equiv -\pderf{f}{m}{T}.
\label{h-def}
\end{equation}
The locally stable states satisfy $\sigma(T,m)=0$:
\begin{equation}
a_1 (T-T_0)m-a_2 m^3+ a_3 m^5=0.
\label{ext}
\end{equation}  
Non-trivial solutions other than $m=0$ satisfy
\begin{equation}
T=T_0+\frac{a_2}{a_1} m^2-\frac{a_3}{a_1} m^4,
\label{T-m}
\end{equation}
where the right-hand side is written as $T(m)$.
See Fig. \ref{fig:T-m}. 
In order to seek the solutions, we consider
\begin{equation}
 a_1 T'(m)= 2a_2 m-4a_3 m^3=0,
\end{equation}
which gives $m=0$ and $m=\pm m_1$ with
\begin{equation}
m_1=\sqrt{\frac{a_2}{2a_3}}.
\end{equation}
By setting
\begin{equation}
T_1=T(m_1)= T_0+\frac{a_2^2}{4a_1a_3},
\end{equation}
we find three locally stable states
$m=0$ and $m=\pm m_{\rm loc}(T)$ when $T_0 \le T \le T_1$,
where $m_{\rm loc}(T) >0 $ is given by 
\begin{equation}
m_{\rm loc}(T)= \sqrt{\frac{a_2+ \sqrt{a_2^2-4a_1a_3 (T-T_0)} }{2a_3}}.
\label{m-loc}
\end{equation}

\subsection{Entropy density}

The entropy density $s(T,m)$ is given by
\begin{eqnarray}
 s&=& -\pderf{f}{T}{m}  \label{sfm} \\
  &=& -\frac{a_1}{2}m^2 -\varphi'(T).
\label{s-m-T}
\end{eqnarray}
The internal energy density $u(T,m)$ is determined as
\begin{equation}
u(T,m)=-\frac{a_1}{2}T_0m^2-\frac{a_2}{4}m^4
    +\frac{a_3}{6}m^6 +\varphi(T)-T\varphi'(T).  
\label{u-t-m}
\end{equation}
For simplicity, we assume that the heat capacity per unit volume,
which is defined as 
\begin{equation}
c_m=\pderf{u}{T}{m},
\end{equation}
is constant. Then, the last two terms of $u(T,m)$ should be $c_m T$
up to an additive constant. 
This leads to 
\begin{equation}
  \varphi'(T)=-c_m \log T+{\rm const}.
\label{phid}
\end{equation}
From (\ref{u-t-m}), we then derive
\begin{equation}
T(u,m)= \frac{1}{c_m}\left[u+\frac{a_1}{2}T_0m^2+\frac{a_2}{4}m^4
    -\frac{a_3}{6}m^6\right].
\end{equation}
By substituting this into (\ref{s-m-T}) with (\ref{phid}),
we obtain the entropy density as a function of $(u,m)$:
\begin{align}
  s(u,m)&= -\frac{a_1}{2}m^2 \nonumber  \\
 &+   c_m \log 
     \left[u+\frac{a_1}{2}T_0m^2+\frac{a_2}{4}m^4
    -\frac{a_3}{6}m^6 \right]
\label{entropy-ex}
\end{align}
up to an additive constant.

By rewriting (\ref{sfm}) as
\begin{equation}
  s(u,m)= -\left. \pder{f(T,m)}{T}\right|_{T=T(u,m)},
\end{equation}
we obtain
\begin{equation}
  \pderf{s}{u}{m} = \frac{1}{T}.
  \label{sum}
\end{equation}
By noting $f=u-Ts$, we also rewrite (\ref{h-def}) as
\begin{equation}
  \sigma=  T(u,m) \pderf{s}{m}{u}.
\label{h-sder}
\end{equation}
These relations, (\ref{sum}) and (\ref{h-sder}), are
summarized as  (\ref{f-relation}).

\section{Precise form of the stochastic model}
\label{s-model:detail}

A formal expression of the stochastic model
was immediately obtained in Sec. \ref{Sec:formal-model}.
However, due to the multiplicative nature of the noise,
the formal model exhibits a singular behavior.
Therefore, we must perform a  careful analysis of
the stochastic process by appropriately choosing the
short-length cut-off of the noise. 
It should be noted that the singularity is specific
to the dynamics of non-conserved quantities and that
it does not appear in the standard 
fluctuating hydrodynamics \cite{Zubarev-Morozov,Morozov}.
In this section, by a theoretical argument
using the separation of scales, we obtain a consistent
stochastic model. We do not find references that
mention this remark, but this is not  surprising even if
it was  well-recognized by specialists in the 1970's. 
In Appendix \ref{s-preliminary}, after some preliminaries,
we write a normal form of the Onsager theory.  
In Appendix \ref{s-model}, we  derive  the stochastic
model with precisely specifying the  noise property.

\subsection{Preliminaries for the derivation}\label{s-preliminary}

In order to derive the stochastic model, we rewrite the set of
deterministic equations,  (\ref{ons-m}), (\ref{ons-v}), and (\ref{ons-phi}),
as the simplest form. The key concept here is to introduce $\bv{q}$  by
\begin{equation}
  \phi= \frac{E}{LL_yL_z} + \bv{\nabla}\bv{q},
\label{q-def}
\end{equation}
where we impose $\bv{q}\bv{n}=0$ at the boundaries so as to satisfy
(\ref{e-con}).
We express (\ref{q-def})
as $\phi=\phi(\bv{q})$. We here note
\begin{align}
 & {\cal S}(m,v,\phi(\bv{q}+\delta \bv{q})) 
                        - {\cal S}(m,v,\phi(\bv{q})) \nonumber \\
                 &= \int d^3\bv{r} 
                       \left. \var{\cal S}{\phi(\bv{r})}
                       \right|_{\phi=\phi(\bv{q})}\bv{\nabla}
                         \delta \bv{q}(\bv{r}) \nonumber \\
                 &= \int d^3\bv{r} \bv{\nabla}\left[
                       \left. \var{\cal S}{\phi(\bv{r})}
                       \right|_{\phi=\phi(\bv{q})}
                       \delta \bv{q}(\bv{r}) \right] \nonumber \\
                 &\qquad\qquad - \int d^3\bv{r} \bv{\nabla}\left[
                       \left. \var{\cal S}{\phi(\bv{r})}
                       \right|_{\phi=\phi(\bv{q})}
                       \right]\delta \bv{q}(\bv{r}) \nonumber \\
                 &= - \int d^3\bv{r} \bv{\nabla}\left[
                       \left. \var{\cal S}{\phi(\bv{r})}
                       \right|_{\phi=\phi(\bv{q})}
                       \right]\delta \bv{q}(\bv{r}) ,
\label{deltaS-q}
\end{align}
where we have used the boundary condition $\bv{q}\bv{n}=0$. We simply
express the result (\ref{deltaS-q}) as 
\begin{equation}
\var{\cal S}{\bv{q}(\bv{r})}= - 
\bv{\nabla}\var{\cal S}{\phi(\bv{r})}.
\end{equation}
By using this expression and substituting  (\ref{q-def}) into
(\ref{ons-phi}), we  rewrite (\ref{ons-phi}) as 
\begin{equation}
\partial_t \bv{q} = 
   \lambda \var{\cal S}{\bv{q}(\bv{r})} +\bv{B},
   \label{ons-q-0}
\end{equation}
where $\bv{B}$ satisfies
\begin{equation}
  \bv{\nabla}\bv{B}=0.
\label{B-con-1}
\end{equation}  
For a given $\phi$, $\bv{\nabla} \times \bv{q}$ may take arbitrary values.
We fix this value at time $t$ by the solution of the equation
\begin{equation}
\partial_t (\bv{\nabla} \times \bv{q}) = 
\bv{\nabla} \times  \lambda \var{\cal S}{\bv{q}(\bv{r})}
\label{rot-q}
\end{equation}
with the initial value $\bv{\nabla} \times \bv{q} =\bv{0}$ at $t=0$.
Under this fixing condition, we have $\bv{\nabla} \times \bv{B}= \bv{0}$.
Together with (\ref{B-con-1}), we find that $\bv{B}$ is constant in $\bv{r}$.
Finally, noting the condition that $\bv{q}\bv{n}=0$ and 
$\bv{\nabla}\beta \bv{n}=0$ at the boundary, we have $\bv{B} \bv{n}=0$
from  (\ref{ons-q-0}). We thus derive
\begin{equation}
  \bv{B}=\bv{0}. 
\end{equation}
Substituting this result into  (\ref{ons-q-0}), we obtain
\begin{equation}
\partial_t \bv{q} = 
   \lambda \var{\cal S}{\bv{q}(\bv{r})}.
   \label{ons-q}
\end{equation}
As shown below, the variable $\bv{q}$ is convenient to analyze
the stochastic model. 
As far as we checked, there are no references that introduce
the variable $\bv{q}$ instead of a locally conserved quantity.

Here, we define the five components field
\begin{equation}
  \chi \equiv (m,v,q_x,q_y,q_z ),
\end{equation}
and $\chi^a$  $(a=1,2,\cdots, 5)$ denotes each component.
For any functional of $\alpha=(m,v,\phi)$, such as ${\cal S}(\alpha)$
and ${\cal P}_{\rm eq}(\alpha)$, we define the functional of $\chi$
through $\alpha=\alpha(\chi)$. For example,
${\cal P}_{\rm eq}(\chi)$ represents ${\cal P}_{\rm eq}(\alpha(\chi))$.
The set of equations
(\ref{ons-m}), (\ref{ons-v}), and (\ref{ons-q}) is expressed
as
\begin{equation}
  \partial_t \chi^a = \sum_{b=1}^5 L^{ab}(\chi(\bv{r}),\nabla \chi(\bv{r}))
  \var{\cal S}{\chi^b (\bv{r})},
\label{ons-chi}
\end{equation}  
where $L^{12}=-L^{21}=-T$, $L^{22}=\gamma  T$, 
$L^{33}= L^{44}=L^{55}=\lambda$, and $L^{ab}=0$
for the other components. It should be noted that 
$T$ and $\lambda$ are functions of $(u,m)$,
while $\gamma$ is a constant. Since 
\begin{equation}
 u=\frac{E}{LL_yL_z}+\bv{\nabla}\bv{q}- \frac{v^2}{2}
                  -d_e  \frac{|\bv{\nabla} m|^2}{2},
\end{equation}
$L^{ab}(\chi(\bv{r}),\nabla \chi(\bv{r}))$ is determined from
$\chi$ and $\nabla \chi$ for each $\bv{r}$. 


Now, the stochastic model is constructed so as to satisfy the
detailed balance condition with respect to the stationary
distribution ${\cal P}_{\rm eq}(\chi)$. If we ignore  $v$
dependence of $T$ with fixed $(m, \bv{q})$, the model would be
immediately obtained as
\begin{equation}
  \partial_t \chi^a = \sum_{b=1}^5 L^{ab}(\chi(\bv{r}),\nabla \chi(\bv{r}))
  \var{\cal S}{\chi^b (\bv{r})}+\sqrt{2L^{aa}}\xi^{a}.
\label{ons-chi-noise}
\end{equation}
See e.g. \cite{Graham}.
The model is identical to the formal model
introduced in Sec. \ref{Sec:formal-model}.
Unfortunately, however, we cannot ignore $v$ dependence of $T$ so
as to satisfy the detailed balance condition. To make the matter
worse, the contribution gives a spurious divergence, as will be
seen in the next subsection.  


In order to resolve this problem,  
we notice that the noises should have a finite correlation length
because the noises appear as the result of  coarse-graining
of microscopic mechanical degrees of freedom \cite{Zwanzig}.
We describe this property by introducing a cutoff $\Lambda_c$ for the
noise and replace  (\ref{ons-naive-noise}) by 
\begin{eqnarray}
  \bra \xi^{a}(\bv{r},t) \xi^b(\bv{r}',t') \ket&=& 
  \delta^{ ab}\delta_{\Lambda_c}(\bv{r}-\bv{r'})\delta(t-t')
    \label{nxi}
\end{eqnarray}
with
\begin{equation}
\delta_{\Lambda_c}(\bv{r})=\int_{|\bv{k}|\Lambda_c <1} \frac{d^3\bv{k}}{(2\pi)^3}
e^{i \bv{k}\bv{r} } .
\end{equation}
Here, the cut-off length $\Lambda_c$ is much
larger than the microscopic length scale $\ell$ and much
shorter than the coarse-grained size $\Lambda$. We thus 
impose 
\begin{equation}
\ell \ll \Lambda_c \ll \Lambda \ll L.
\label{separation2}
\end{equation}  
The condition $\ell \ll \Lambda_c$ is necessary
to remove a singular term associated with the multiplicative
nature of the noise, which will be discussed below.
This cut-off  induces the non-local coupling between the
Onsager coefficients and the thermodynamic forces.  Since
the length of the non-local coupling is $\Lambda_c$
and the spatial variation of the variables is larger
than $\Lambda$,  we can
approximate it by the local coupling ignoring the contribution
of $O(\Lambda_c/\Lambda)$. We will give a precise
argument for the derivation of the model in Appendix \ref{s-model}.

Summarizing these results, we write the stochastic model as
\begin{align}
&  \partial_t m = T \otimes {\beta v} ,
  \label{ons-m-n-0} \\
&  \partial_t v = -\gamma T \otimes  \beta v 
  +T\otimes \pderf{s}{m}{u} 
  +d_e   T\otimes (\bv{\nabla}\beta)(\bv{\nabla} m)\nonumber \\
&  \qquad\qquad +  T \otimes d_f\beta \Delta m  
 +\sqrt{2\gamma  T}\otimes \xi^v ,
  \label{ons-v-n-0} \\
&  \partial_t \phi =
  - \bv{\nabla}\left( \lambda \otimes \bv{\nabla} \beta 
 +  \sqrt{2\lambda  }\otimes \bv{\xi}^\phi \right) ,
   \label{ons-phi-n-0}
\end{align}
where  $f \otimes g$ is defined as
\begin{equation}
  f\otimes g = \int\! d^3 \bv{r}' \! \!\int \! d^3 \bv{r}''
  f(\bv{r''})
  \delta_{\Lambda_c}(\bv{r}-\bv{r}'')
  \delta_{\Lambda_c}(\bv{r}'-\bv{r}'')
  g(\bv{r}').
\end{equation}  
Since $\Lambda_c \ll \Lambda$, (\ref{ons-m-n-0}), (\ref{ons-v-n-0}),
(\ref{ons-phi-n-0}) may be interpreted as a physical model of the
the formal  model (\ref{ons-m-n}), (\ref{ons-v-n}), and
(\ref{ons-phi-n}). It should be noted that 
the  unsatisfactory properties of the formal model are not
observed in the physical model  
(\ref{ons-m-n-0}), (\ref{ons-v-n-0}), and  (\ref{ons-phi-n-0})
with (\ref{nxi}). Therefore, we should study the physical
model. Although the expression of the physical model is
rather complicated, the theoretical analysis can be done similarly
to that of the formal model. Keeping this in mind, 
we study the formal model in the main text.

\subsection{Derivation}\label{s-model}

\begin{figure}[tb]
\centering
\includegraphics[width=6cm]{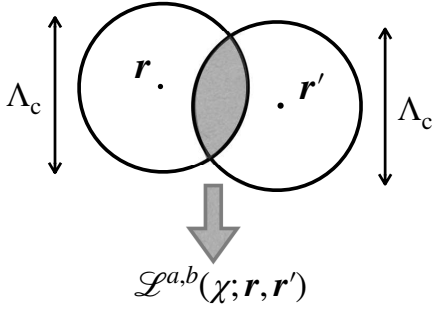}
\caption{Schematic figure of non-local Onsager coefficient ${\cal L}^{ab}$.}
\label{fig:coeff}
\end{figure}

Since we assume the cut-off length in the noise, 
(\ref{ons-chi-noise})    becomes a non-local form with
using a functional of $\chi$ as 
\begin{eqnarray}
  {\cal L}^{ab}(\chi;\bv{r},\bv{r}')
&\equiv& \int d^3 \bv{r}'' 
  L^{ab}(\chi(\bv{r}''), \nabla \chi(\bv{r}''))
  \nonumber \\
&& \times
 \delta_{\Lambda_c}(\bv{r}-\bv{r}'') \delta_{\Lambda_c}(\bv{r}'-\bv{r}''),
\end{eqnarray}  
which is illustrated in Fig. \ref{fig:coeff}.
Further, since the Onsager coefficients $L^{ab}$ in (\ref{ons-chi})
depend on $\chi$, we have to consider multiplicative
nature of the noise in the stochastic dynamics.
From these, the stochastic model
 (\ref{ons-chi-noise})  
is replaced by  
\begin{eqnarray}
  \partial_t \chi^a &=& \sum_b \int d^3\bv{r}'
  \left[ {\cal L}^{ab}(\chi;\bv{r},\bv{r}')
  \var{\cal S}{\chi^b (\bv{r}')} \right. \nonumber \\
& & \left.  + \var{{\cal L}^{ab}(\chi;\bv{r},\bv{r}')}
  {\chi^b (\bv{r}')}\delta^{ab} \right] \nonumber \\
 & &   + \int d^3\bv{r}' {\cal G}^a (\chi;\bv{r},\bv{r}')
  \cdot \xi^a(\bv{r}'),
\label{ons-q-s}
\end{eqnarray}  
where the functional ${\cal G}^a (\chi;\bv{r},\bv{r}')$ is determined
later and the symbol `$\cdot$' in front of $\xi^a$
represents the Ito multiplication.  The second term on the right-hand
side of (\ref{ons-q-s}) is necessary to yield the equilibrium
stationary distribution (\ref{micro-canonical})
\cite{Graham, Itami-Sasa}. Here, it should be noted that
the off-diagonal components of ${\cal L}^{ab}$ do not appear
in the second term, because the terms with off-diagonal
components of ${\cal L}^{ab}$ do not contribute to
the entropy production. See \cite{Itami-Sasa} for the detail.

The Fokker-Planck equation for the probability density
${\cal P}(\chi,t)$ corresponding to (\ref{ons-q-s}) is written as 
\begin{align}
&   \partial_t {\cal P}(\chi,t) 
+   \sum_{ab} \int d^3 \bv{r} d^3 \bv{r}'
  \var{ }{\chi^a(\bv{r})} \left[
     {\cal A}^{ab}  (\chi; \bv{r},\bv{r}')
      {\cal P}(\chi,t)    \right] \nonumber \\
&=   \frac{1}{2}
\sum_a \int d^3\bv{r} d^3 \bv{r}'
    \vart{}{\chi^a(\bv{r})}{\chi^a(\bv{r}')}
         [ {\cal B}^a (\chi; \bv{r},\bv{r}')
           {\cal P}(\chi,t)]
\end{align}
with
\begin{align}
&  {\cal A}^{ab}  (\chi; \bv{r},\bv{r}')
\equiv
 {\cal L}^{ab}(\chi;\bv{r},\bv{r}') \var{\cal S}{\chi^b(\bv{r}') }
+ 
\delta^{ab}\var{ {\cal L}^{ab}(\chi;\bv{r},\bv{r}')}{\chi^b(\bv{r}')},
 \\
& {\cal B}^a (\chi; \bv{r},\bv{r}') \nonumber \\
&\equiv
 \int d^3\bv{r''}d^3 \bv{r}'''
 {\cal G}^a(\chi;\bv{r},\bv{r}''){\cal G}^a(\chi;\bv{r}',\bv{r}''' )
\delta_{\Lambda_c}(\bv{r}''-\bv{r}''') .
\end{align}
Here, as shown in \cite{Itami-Sasa,Graham}, the detailed balance
condition is expressed as 
\begin{align}
 &   
\int d^3 \bv{r}  \var{ }{\chi^1(\bv{r})} \left[
      {\cal L}^{12}(\chi;\bv{r},\bv{r}') \var{\cal S}{\chi^2(\bv{r}') }
      {\cal P}_{\rm eq}(\chi)    \right] \nonumber \\
 &+
\int d^3 \bv{r}
  \var{ }{\chi^2(\bv{r})} \left[
      {\cal L}^{21}(\chi;\bv{r},\bv{r}') \var{\cal S}{\chi^1(\bv{r}') }
      {\cal P}_{\rm eq}(\chi)    \right]
=0 , 
 \label{itami} \\
& 2 {\cal L}^{aa}(\chi;\bv{r},\bv{r}')
  =  {\cal B}^a (\chi; \bv{r},\bv{r}')
  \label{con-L} , 
\end{align}
which leads to the stationary distribution (\ref{micro-canonical}).
We thus have to confirm (\ref{itami}) and  (\ref{con-L}).

First, we estimate the left-hand side of (\ref{itami}).
From the anti-symmetric property
\begin{equation}
      {\cal L}^{12}(\chi;\bv{r},\bv{r}')=-{\cal L}^{21}(\chi;\bv{r}',\bv{r}),
\end{equation}
the left-hand side of (\ref{itami}) is written as
\begin{align}
&   
  \int d^3 \bv{r}  \var{{\cal L}^{12}(\chi;\bv{r},\bv{r}')}
                   {\chi^1(\bv{r})} 
                  \var{\cal S}{\chi^2(\bv{r}') }
        {\cal P}_{\rm eq}(\chi)   \nonumber \\
&\quad
+ \int d^3 \bv{r}
  \var{{\cal L}^{21}(\chi;\bv{r},\bv{r}')}{\chi^2(\bv{r})} 
       \var{\cal S}{\chi^1(\bv{r}') }
      {\cal P}_{\rm eq}(\chi)     .
      \label{itami2}
\end{align}      
We here explicitly calculate 
\begin{eqnarray}
\int d^3 \bv{r}\var{ {\cal L}^{12}(\chi;\bv{r},\bv{r}')}{\chi^1(\bv{r})}
&=& -\pderf{T}{m}{u}\delta_{\Lambda_c}(0)   \nonumber \\
&=& \frac{1}{c_m} \pderf{u}{m}{T}\delta_{\Lambda_c}(0),
\end{eqnarray}
where we have used $\delta_{\Lambda_c}'(0)=0$.
Similarly, we have 
\begin{eqnarray}
\int d^3 \bv{r} \var{ {\cal L}^{21}(\chi;\bv{r},\bv{r}')}{\chi^2(\bv{r})}
&=& \pderf{T}{u}{m}\pderf{u}{v}{\bv{q},m}\delta_{\Lambda_c}(0) \nonumber \\
&=& - \frac{v}{c_m }\delta_{\Lambda_c}(0).
\end{eqnarray}
These expressions involve the dimensionless quantity
$\delta_{\Lambda_c}(0)/c_m$.
Since $\delta_{\Lambda_c}(0)=O(\Lambda_c^{-3})$ and $c_m =O(\ell^{-3})$, 
$\delta_{\Lambda_c}(0)/c_m$ is estimated as $O(\ell^3/\Lambda_c^3)$.
This leads to 
\begin{eqnarray}
  \int d^3 \bv{r}
  \var{ {\cal L}^{12}(\chi;\bv{r},\bv{r}')}{\chi^1(\bv{r})}
  &=& \pderf{u}{m}{T}O\left( \frac{\ell^3}{\Lambda_c^3} \right),
  \label{itami-con-1}\\
 \int d^3 \bv{r}
  \var{ {\cal L}^{21}(\chi;\bv{r},\bv{r}')}{\chi^2(\bv{r})}
  &=&   {v}\left( \frac{\ell^3}{\Lambda_c^3} \right)
  \label{itami-con-2} 
\end{eqnarray}
in the asymptotic limit $\ell/\Lambda_c \to 0$.  By substituting
(\ref{itami-con-1}) and (\ref{itami-con-2}) into (\ref{itami2}),
we find that (\ref{itami2}) is proportional to  $O(\ell^3/\Lambda_c^3)$,
which is zero in the limit (\ref{separation2}). Then, we have
confirmed (\ref{itami}).  Note that (\ref{itami-con-1}) and
(\ref{itami-con-2}) exhibit the divergence without the cutoff $\Lambda_c$.
This apparent divergence becomes zero in the appropriate limit
after introducing the cut-off $\Lambda_c$. Such an asymptotic estimate
using a similar cut-off was used in Ref. \cite{Nakano}.

Next, we determine ${\cal G}^a$ from the condition (\ref{con-L}).
We note that (\ref{con-L}) is satisfied when 
\begin{align}
&  2 L^{aa}(\chi(\bv{r}''), \nabla \chi(\bv{r}''))
  \delta_{\Lambda_c}(\bv{r}-\bv{r}'')\delta_{\Lambda_c}(\bv{r}'-\bv{r}'')
\nonumber \\
& =
\int d^3 \bv{r}'''
{\cal G}^a(\chi;\bv{r},\bv{r}''){\cal G}^a(\chi;\bv{r}',\bv{r}''' )
\delta_{\Lambda_c}(\bv{r''}-\bv{r'''}).
\label{con-L2}
\end{align}
By substituting 
\begin{align}
&{\cal G}^a(\chi;\bv{r},\bv{r}') 
= \int d^3\bv{r}''
  \sqrt{2L^{aa}(\chi(\bv{r}''),\nabla \chi(\bv{r}''))} \nonumber \\
 &\qquad
  \times  \delta_{\Lambda_c}(\bv{r}-\bv{r}'')\delta_{\Lambda_c}(\bv{r}'-\bv{r}'')
                                                     \nonumber \\
 &\qquad                                       
   \times \left[1+O\left( \frac{\Lambda_c^3}{\Lambda^3} \right) \right]
\label{G-sol}
\end{align}
into the right-hand side of (\ref{con-L2}), we confirm that
the right-hand side is equal to the left-hand side of (\ref{con-L2})
with an error of $O((\Lambda_c/\Lambda)^3 )$. Therefore, 
we claim that the condition (\ref{con-L}) holds.

Finally, we investigate the second term in the right-hand
side of (\ref{ons-q-s}). We concretely calculate each term
as follows.  
\begin{eqnarray}
\int d^3 \bv{r}'
\var{ {\cal L}^{22}(\chi;\bv{r},\bv{r}')}{\chi^2(\bv{r}')}
&=& \gamma  \pderf{T}{u}{m} \pderf{u}{v}{\bv{q},m}\delta_{\Lambda_c}(0)
\nonumber \\
&=& -\gamma  v  O\left( \frac{\ell^3}{\Lambda_c^3} \right),
\label{tui-1}
\end{eqnarray}
and
\begin{eqnarray}
\int d^3 \bv{r}'
\var{ {\cal L}^{33}(\chi;\bv{r},\bv{r}')}{\chi^3(\bv{r}')}
&=& 0,
\label{333}
\end{eqnarray}
where we have used $\delta_{\Lambda_c}'(0)=0$. 
(\ref{tui-1}) provides a correction of the momentum dissipation term
$-\gamma v$. This correction can be negligible from the condition
(\ref{separation2}).
Therefore, the second term in the right-hand side of (\ref{ons-q-s})
can be ignored. We here remark that the equality (\ref{333}) leads to the
statement that the multiplication rule of the noise, Ito or Stratonovich,
is irrelevant for the standard fluctuating hydrodynamics
\cite{Zubarev-Morozov,Morozov}.

More explicitly, by considering a physical situation, we may estimate
$\eta=10^{-8}$. Recalling $\Lambda /L= O(\sqrt{\eta})$, we express 
(\ref{separation2}) by
\begin{equation}
\sqrt{\eta} \ll  \frac{\Lambda_c}{\Lambda} \ll 1.
\label{separation22}
\end{equation}
As one example, we  choose $\Lambda_c/\Lambda =10^{-2}$,  
which makes the theory consistent. It should be noted that
we consider the case that the interface width  also vanishes
in the limit $\eta \to 0$, which is in contrast to the standard
weak  noise limit \cite{Bertini-rev}.  This aspect brings non-trivial
noise effects even in the limit $\eta \to 0$.

\section{Derivation of (\ref{zubarev})}
\label{der-zubarev}

In this Appendix, we derive (\ref{zubarev}). In order to
simplify the notation, we omit $E$ dependence 
such that ${\cal P}(\alpha,t_f;E,J)$ is expressed as
${\cal P}(\alpha,t_f;J)$. 
Following the notation in the main text,
we define $\alpha^\dagger$ for $\alpha=(m,v,\phi)$ as
$\alpha^\dagger=(m,-v,\phi)$ and $\hat \alpha^\dagger$ denotes
the time-reversal of $\hat \alpha$. That is,
$$\hat \alpha^\dagger(t)=(m(t_f-t),-v(t_f-t),\phi(t_f-t)).$$

We first substitute 
the dimensionless version of (\ref{37}) and (\ref{36}),
\begin{align}
&  \var{\cal S}{m(\bv{r})} =  \beta \sigma
  +\eta [d_e (\bv{\nabla}\beta)(\bv{\nabla} m) 
  + \beta d_f\Delta m] , \label{37-2} \\
&  \var{\cal S}{v(\bv{r})} =  - {\beta v}, \label{36-2}
\end{align}
into (\ref{III-65}) and a similar expression of 
$\hat{\cal I}(\hat \alpha^\dagger|(\alpha(t_f))^\dagger;-J)$.
By noting
\begin{eqnarray}
  \bv{j}\bv{\nabla}\beta &=&  \bv{\nabla}( \bv{j}\beta)-
                              \beta \bv{\nabla}\bv{j} \nonumber \\
             &=& \bv{\nabla}( \bv{j}\beta)+
                  \var{\cal S}{\phi(\bv{r})} \partial_t \phi,
\end{eqnarray}
we obtain
\begin{align}
 & \hat{\cal I}(\hat \alpha|\alpha(0);J)-
  \hat{\cal I}(\hat \alpha^\dagger|(\alpha(t_f))^\dagger;-J) \nonumber \\
&=
  -\int_0^{t_f}dt\int d^3\bv{r}
 \left[(\partial_t m)\var{\cal S}{m(\bv{r})}
     + (\partial_t v)\var{\cal S}{v(\bv{r})} \right.
  \nonumber \\
 &    
 \hskip1cm +\left.
  \bv{\nabla} \left(\bv{j} \beta \right)
   + (\partial_t \phi)
   \var{\cal S}{\phi(\bv{r})}\right],
\end{align}
  which leads to
\begin{align}
&  \hat{\cal I}(\hat \alpha|\alpha(0);J)
  -  \hat{\cal I}(\hat \alpha^\dagger|(\alpha(t_f))^\dagger;-J) \nonumber \\
&=
  -{\cal S}(\alpha(t_f))+{\cal S}(\alpha(0))
\nonumber \\  
&  - J \int d^2\bv{r}_\perp\int_0^{t_f}dt  (\beta(1,\bv{r}_\perp,t)
  -\beta(0,\bv{r}_\perp,t)),
\label{LDP}
\end{align}
where $\bv{r}_\perp=(y,z)$. 

Now, for an initial distribution ${\cal P}_0$, 
the distribution at $t=t_f$ is  expressed as
\begin{equation}
{\cal P}(\alpha,t_f;J)=
\int {\cal D}\hat \alpha {\cal P}_{0}(\alpha(0))
\hat {\cal P}(\hat \alpha|\alpha(0);J) \delta(\alpha(t_f)-\alpha).
\label{td-dis}  
\end{equation}
Here, as a special choice, we take
\begin{equation}
 {\cal P}_0(\alpha)={\cal N}
  \exp\left( \frac{1}{\eta^3} {\cal S}(\alpha) \right)
  \delta\left( \int d^3\bv{r} \phi(\bv{r})-E \right).
\label{sp-choi}  
\end{equation}
From (\ref{LDP}), we find
\begin{align}
&  \frac{  \hat {\cal P}(\hat \alpha|\alpha(0);J)
    {\cal P}_0(\alpha(0))}
       {\hat {\cal P}(\hat \alpha^\dagger|(\alpha(t_f))^\dagger;-J)
     {\cal P}_0((\alpha(t_f))^\dagger)} \nonumber \\
& =
     \exp \left(
     \frac{J}{\eta^3} \int d^2\bv{r}_\perp
     \int_0^{t_f} dt  (\beta(1,\bv{r}_\perp, t)-\beta(0,\bv{r}_\perp,t)) \right).
\label{LDP2}
\end{align}
We then rewrite (\ref{td-dis}) as
\begin{align}
& {\cal P}(\alpha,t_f;J) =
\int {\cal D}\hat \alpha
     {\cal P}_{0}((\alpha(t_f))^\dagger)
     \hat {\cal P}(\hat \alpha^\dagger|(\alpha(t_f)^\dagger;-J)  \nonumber \\
& \times \frac{{\cal P}_{0}(\alpha(0))\hat {\cal P}(\hat \alpha|\alpha(0);J) }
     {{\cal P}_{0}((\alpha(t_f))^\dagger)\hat {\cal P}(\hat \alpha^\dagger|
         (\alpha(t_f))^\dagger;-J) }  \delta(\alpha(t_f)-\alpha)  .
\end{align}
The substitution of (\ref{LDP2}) into the right-hand side yields
\begin{align}
&\int {\cal D}\hat \alpha^\dagger
     {\cal P}_{0}((\alpha(t_f))^\dagger)
     \hat {\cal P}(\hat \alpha^\dagger|(\alpha(t_f))^\dagger;-J)  \nonumber \\
& \quad
    \times  \e^{J/ \eta^3
       \int d^2\bv{r}_\perp
       \int_0^{t_f} dt  (\beta(1,\bv{r}_\perp,t)-\beta(0,\bv{r}_\perp,t)) }
\nonumber \\
&   \qquad \times   
\delta((\alpha(t_f))^\dagger-\alpha^\dagger) .
\label{e-10}
\end{align}
By using the transformation $\hat \alpha \to \hat \alpha^\dagger$ in the
path integral variable,  (\ref{e-10}) is written as   
\begin{align}
& \int {\cal D}\hat \alpha
     {\cal P}_{0}(\alpha(0))
     \hat {\cal P}(\hat \alpha|\alpha(0);-J) \nonumber \\
&\quad \times  \e^{J/ \eta^3  \int d \bv{r}_\perp
       \int_0^{t_f} dt  (\beta(1,\bv{r}_\perp,t)-\beta(0,\bv{r}_\perp,t)) }
\nonumber \\     
 &\qquad \times    
\delta(\alpha(0)-\alpha^\dagger),
\label{e-11}
\end{align}
where we have used
\begin{align}
\int_0^{t_f} dt \beta(1, \bv{r_\perp},t_f-t)
 &=  -\int_{t_f}^0  dt' \beta(1, \bv{r_\perp},t') \nonumber \\
&=  \int_0^{t_f}  dt' \beta(1, \bv{r_\perp},t').
\end{align}

 By substituting (\ref{sp-choi}) into (\ref{e-11}),    
we finally
obtain
\begin{align}
&{\cal P}(\alpha,t_f;J)=  {\cal N} \e^{
     {\cal S}(\alpha)/\eta^3
   } \nonumber \\
& \times    
\bra  \e^{
  J/\eta^3\int d^2\bv{r}_\perp
  \int_0^{t_f} dt  (\beta(1,\bv{r}_\perp,t)-\beta(0,\bv{r}_\perp,t))
          }
\ket_{\alpha^\dagger \to *}^{-J} 
\nonumber \\
& \times 
\delta\left( \int d^3\bv{r} \phi(\bv{r})-E \right) ,
\label{a:zubarev}
\end{align}
where $ \bra \ \ket_{\alpha \to *}^{-J}$ represents
the expectation value over trajectories $\alpha(t)$ starting from
$\alpha(0)=\alpha$ with respect to the path probability density
in the system with $-J$.

We here remark that (\ref{LDP2}) is referred to
as the {\it local detailed balance condition} which connects
the ratio of path probabilities 
of forward and backward trajectories with the entropy production
along the trajectory. This is the key relation for deriving
many universal relations.

\section{ Estimation of $\tau_{\rm int}$   }
\label{i-motion}

  In this Appendix, we estimate the typical time scale of the
interface motion by analyzing the deterministic model
(\ref{ons-m-d}), (\ref{ons-v-d}), and (\ref{ons-phi-d}) 
with the condition that the heat conduction is sufficiently fast.
Note that the model is not the dimensionless version in order to
clarify the physical argument for the estimation.

\begin{figure}[tb]
\centering
\includegraphics[width=8cm]{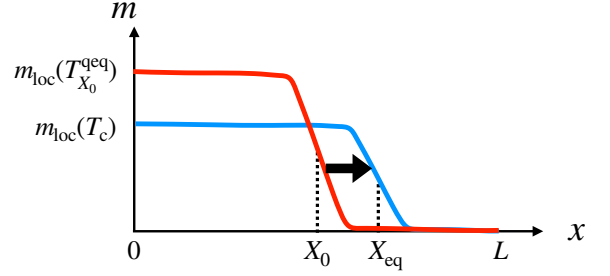}
\caption{Schematic figure of interface motion from the
quasi-equilibrium state to the equilibrium state.}
\label{fig:interface}
\end{figure}

We assume an initial state with an interface position $X_0$ 
and a uniform temperature $T^{\rm qeq}_{X_0} \not = \Tc$, which satisfies
\begin{equation}
  X_0 u^{\rm o}(T^{\rm qeq}_{X_0})
  +  (L-X_0) u^{\rm d}(T^{\rm qeq}_{X_0})=\frac{E}{L_yL_z}, 
\end{equation}
and
\begin{eqnarray}
m(x,0) &=&  m^{\rm qeq}(x-X_0;X_0), \\
v(x,0) &=&  0
\end{eqnarray}
with
\begin{equation}
  m^{\rm qeq}(x-X;X)\equiv \bar{\bar m} \left(  \frac{x-X}{\sqrt{\eta}} \right)
   m_{\rm loc}(T^{\rm qeq}_{X}).
\label{m-qe}
\end{equation}
When $0 < \eta \ll 1$, the interface slowly moves to the equilibrium
position $X_{\rm eq}$, as shown in Fig.~\ref{fig:interface}.
The initial state corresponds to the quasi-equilibrium
state in thermodynamics, because $T^{\rm qeq}_{X_0} \not = \Tc$.
The time evolution describes the transition from the quasi-equilibrium state
$T^{\rm qeq}_{X_0}$ to the true equilibrium state $\Tc$.
We describe this interface motion quantitatively.

Let $X(t)$ be the position of the interface at time $t$.
We assume that the interface motion is slowest which will be
confirmed by (\ref{tau-int-scale}) in a self-consistent manner.
Then,  the   other dynamical variables are slaved to the slow
variable $X(t)$. Based on this picture, we set
\begin{align}
&m(x,t) = m^{\rm qeq}(x-X(t);X(t))+m'(x,t) , \label{m-1} \\
&T(x,t) = T^{\rm qeq}_{X(t)}+T'(x,t) , \label{T-1} \\
&v(x,t) =  \partial_t m(x,t) , \label{v-1}
\end{align}
where $m'$ and $T'$ are small corrections,
which are neglected in the lowest order calculation.

The temperature $T^{\rm qeq}_{X(t)}$ satisfies 
\begin{equation}
  X(t) u^{\rm o}(T^{\rm qeq}_{X(t)}) +  (L-X(t)) u^{\rm d}(T^{\rm qeq}_{X(t)})
  =\frac{E}{L_yL_z}
\label{xt-theta}
\end{equation}
for the interface position $X(t)$. We  now attempt to
determine   $m^{\rm qeq}(x-X(t);X(t))$ for small $\eta$.
Here, since $m^{\rm qeq}(x-X(t);X(t))$ is slowly evolving,
\begin{equation}
  \partial_t^2 m^{\rm qeq} \ll \gamma \partial_t m^{\rm qeq} ,
\label{o-damp}
\end{equation}
which will be checked by (\ref{tau-int-scale}).
By substituting (\ref{m-1}), (\ref{T-1}), and (\ref{v-1}) into
(\ref{ons-v-d}), we obtain
\begin{align}
  -\gamma  \der{X}{t} \partial_x m^{\rm qeq}=
  -\left.
  \pder{f(T^{\rm qeq}_{X(t)},m)}{m}\right|_{m=m^{\rm qeq}} 
  +d_f \partial_x^2 m^{\rm qeq},
\label{x-m-q}
\end{align}
where we have used (\ref{o-damp}). More precisely, although the left-hand
side should be
\begin{align}
  -\gamma  \der{X}{t}
  \left[ \partial_x m^{\rm qeq} -m^{\rm qeq}\pder{}{X}\log m_{\rm loc}(T_X^{\rm qeq})
    \right],
\end{align}
the second term can be ignored for small $\eta$, because the first term
in the square bracket is $O(m_{\rm loc}/\Lambda)$ and the second term is 
$O(m_{\rm loc}/L)$.  In the limit $\eta \to 0$, we consider
(\ref{x-m-q}) as the differential equation defined in
$x-X(t) \in [-\infty, \infty]$ with the boundary condition
\begin{equation}
  m^{\rm qeq}(x-X(t);X(t)) \to m_{\rm loc}(T^{\rm qeq}_{X(t)})
\end{equation}
for $x-X(t) \to -\infty$, and 
\begin{equation}
  m^{\rm qeq}(x-X(t);X(t)) \to 0
\end{equation}
for $x-X(t) \to +\infty$. We here note that a solution of 
the differential equation (\ref{x-m-q}), $m^{\rm qeq}(x-X(t);X(t))$,
exists only for a special value of $dX/dt$. In other words, 
by solving the non-linear eigenvalue equation (\ref{x-m-q})
with  $T^{\rm qeq}_{X(t)}$  
given by (\ref{xt-theta}),  we determine 
the eigenvalue $dX/dt$ and the solution $m^{\rm qeq}(x-X(t);X(t))$,
simultaneously.

The solution of the equation (\ref{x-m-q}) is understood by
identifying (\ref{x-m-q}) with a Newton equation for the
coordinate $m^{\rm qeq}$ with a fictitious time $x'=x-X(t)$,
where the fictitious mass is $d_f$, the fictitious friction $\gamma dX/dt$,
and the potential $-f(T^{\rm qeq}_{X(t)}, m^{\rm qeq})$. The precise form
of $m^{\rm qeq}(x-X(t);X(t))$ and the eigenvalue $dX/dt$ can be
numerically determined by solving (\ref{x-m-q}). Here,
assuming the form $m^{\rm qeq}(x-X(t);X(t))$, we express $dX/dt$
in terms of $m^{\rm qeq}(x-X(t);X(t))$. Indeed, multiplying
$\partial_x m^{\rm qeq}(x-X(t);X(t))$ to  both sides of
(\ref{x-m-q}) and integrating them over the whole region, we
obtain
\begin{align}
& -\gamma \der{X}{t} \int_{-\infty}^\infty dx (\partial_x m^{\rm qeq})^2
  \nonumber \\
&\qquad = 
  f(T^{\rm qeq}_{X(t)},m_{\rm loc}(T^{\rm qeq}_{X(t)})) - f(T^{\rm qeq}_{X(t)},0).
\label{x-m-q-int}
\end{align}
This is a rather standard analysis. See for example Ref. \cite{Pomeau}.
The equation (\ref{x-m-q-int}) represents the equation of motion for $X$.
It means that the interface moves so as to decrease the total free energy.
The driving force is the free energy difference given in the right-hand
side, and the left-hand side describes the friction force for the interface
motion.
Now, when $X(t)$ is close to $X_{\rm eq}$, we have a linear equation 
\begin{equation}
  \tau_{\rm int} \frac{dX}{dt} = - (X-X_{\rm eq}).
\label{eq-m-int-2}
\end{equation}
Then, $\tau_{\rm int}$ provides the time scale of the interface motion.
Below, by analyzing (\ref{x-m-q-int}), we derive $\tau_{\rm int}$.

We specifically study the case that $(X(t)-X_{\rm eq})/L$ is
small. In this case, $(T^{\rm qeq}_X - T_c)/T_c$ is also small.
By recalling (\ref{21}), we notice
\begin{align} 
&  f(T^{\rm qeq}_{X(t)},m_{\rm loc}(T^{\rm qeq}_{X(t)})) - f(T^{\rm qeq}_{X(t)},0) \nonumber \\
  &~~=  f(T^{\rm qeq}_{X(t)},m_{\rm loc}(T^{\rm qeq}_{X(t)}))
               - f(T_c,m_{\rm loc}(T_c))  \nonumber \\
& \qquad-[f(T^{\rm qeq}_{X(t)},0) -f(T_c,0)] . 
\end{align}
We thus estimate 
\begin{align} 
&  f(T^{\rm qeq}_{X(t)},m_{\rm loc}(T^{\rm qeq}_{X(t)})) - f(T^{\rm qeq}_{X(t)},0) \nonumber \\
  &= -[s(T_c,m_{\rm loc}(T_c))-s(T_c,0)]  (T^{\rm qeq}_{X(t)}-T_c) \nonumber \\
&=
  -\frac{u^{\rm o}(T_c)-u^{\rm d}(T_c)}{T_c}
  \left. \der{T^{\rm qeq}_{X}}{X}\right|_{X=X_{\rm eq}} \!\!\!\!(X(t)-X_{\rm eq}),
\label{f-diff}
\end{align}
where we have ignored higher-order terms of 
$(X(t) - X_{\rm eq})/L$. 
We first  notice that
$u^{\rm o}(T_c)$ and $u^{\rm d}(T_c)$ are proportional to $T_c \ell^{-3}$
up  to a multiplicative numerical constant, because of the
equipartition law. Furthermore, the derivative of (\ref{xt-theta})
in $X(t)$ provides an expression of $dT^{\rm qeq}_X/d X$, from
which we find 
\begin{equation}
 \der{T^{\rm qeq}_X}{X}\simeq  \frac{T_c}{L}.
\end{equation}
We  thus estimate the right-hand side
of (\ref{f-diff}) as 
\begin{eqnarray} 
   \frac{\ell^{-3}T_c}{L} (X-X_{\rm eq}),
\label{f-diff-2}
\end{eqnarray}
up to a multiplicative numerical constant. Furthermore,
$m^{\rm qeq}(x-X(t);X(t))$ may be replaced by $\bar{\bar m} (\xi) m_{\rm loc}(T_c)$
in (\ref{m-st}) in this description.
We then re-write (\ref{x-m-q-int}) as
\begin{equation}
  \gamma_{\rm int} \frac{dX}{dt} \simeq -  \frac{\ell^{-3}T_c}{L}(X-X_{\rm eq}) ,
\end{equation}
with 
\begin{equation}
\gamma_{\rm int}
\equiv \frac{\gamma  m_{\rm loc}^2(T_c)}{\Lambda}\int_{-\infty}^\infty d\xi
(\partial_\xi \bar{\bar m})^2,
\end{equation}
Thus, the time scale of the interface motion is estimated as
\begin{equation}
  \tau_{\rm int} = \frac{\gamma  m_{\rm loc}^2(T_c) L}{\ell^{-3} T_c \Lambda}.
\label{74}  
\end{equation}

Let $\tau$ be a macroscopic time scale characterizing the change of the
order parameter density field $m$,  as defined in the main
text.  
From (\ref{vdef})
  and (\ref{phi-wt-2}), we have
\begin{equation}
\phi \simeq  v^2  \simeq  \left(\frac{m}{\tau} \right)^2 , 
\end{equation}
which yields
\begin{equation}
  m^2 \simeq \phi \tau^2 .
\end{equation}  
This estimate allows us to further rewrite (\ref{74})  as
\begin{equation}
  \tau_{\rm int} \simeq  {\gamma \tau} \frac{L}{\Lambda} \tau.
\label{tau-int-scale}
\end{equation}
The time scale of momentum dissipation $\gamma^{-1}$
is  shorter than the macroscopic time scale $\tau$, because
the momentum of the order parameter is not a conserved quantity. This means
that $\gamma \tau > 1$. Therefore, 
it generally holds that 
\begin{equation}
  \frac{\tau_{\rm int}}{\tau} =\gamma \tau O(\eta^{-\frac{1}{2}}) \to \infty
\label{tau-int-0}  
\end{equation}  
in the limit $\eta \to 0$. 
That is, the interface motion is singularly slow.
Below, we assume that $\gamma \tau=O(\eta^{0})$, which leads 
to 
\begin{equation}
  \frac{\tau_{\rm int}}{\tau} = O(\eta^{-\frac{1}{2}}).
\label{tau-int}  
\end{equation}



\begin{thebibliography}{99}

  
\bibitem{Callen} H. B. Callen,
{\it Thermodynamics and an Introduction to Thermostatistics}, 2nd ed.
(Wiley, New York, 1985). 


\bibitem{boiling}
 J. R. Thome, Boiling in microchannels: a review of experiment and theory,
Int. J. Heat and Fluid flow {\bf 25}, 128-139 (2004).
  
\bibitem{crystal}
  E. Ben-Jacob and P. Garik, The formation of patterns in nonequilibrium
  growth, Nature {\bf 343}, 523-530 (1990).
  
\bibitem{Cannell} 
G. Ahlers, L. I.  Berge, and D. S. Cannell,
Thermal convection in the presence of a first-order phase change,
Phys. Rev. Lett. {\bf 70}, 2399 (1993).  

\bibitem{Zhong}
J.-Q. Zhong, D. Funfschilling, and G. Ahlers,
Enhanced heat transport by turbulent two-phase Rayleigh-Benard convection,
Phys. Rev. Lett. {\bf 102}, 124501 (2009).  


\bibitem{Ahlers}
S. Weiss and G. Ahlers,
Nematic-isotropic phase transition in turbulent thermal convection,
J. Fluid Mech. {\bf 737}, 308-328 (2013).

\bibitem{mips}
M. E. Cates and J. Tailleur, Motility-Induced Phase Separation,
Annual Review of Condensed Matter Physics {\bf 6}, 219-244 (2015).

\bibitem{Urban}
P. Urbana, D. Schmoranzerb,
P. Hanzelkaa, K. R. Sreenivasanc, and L. Skrbekb,
Anomalous heat transport and condensation in convection of cryogenic helium,
Proc. Nat. Acad. Sci. {\bf 110},  8036-8039 (2013).



\bibitem{Anderson98}
D. M. Anderson, G. B. McFadden, and A. A. Wheeler,
Diffuse-interface methods in fluid mechanics, 
Annual Review of Fluid Mechanics {\bf 30}, 139-165 (1998).


\bibitem{Bedeaux03}
  D. Bedeaux, E. Johannessen, and A. R{\o}sjorde,
  The nonequilibrium van der Waals square
  gradient model.(I). The model and its numerical solution,
  Physica A {\bf 330}, 329-353 (2003).
  
\bibitem{Onuki}
  A. Onuki, Dynamic van der Waals theory,
  Phys. Rev. E {\bf 75}, 036304 (2007).


  
\bibitem{Schmitz}
R. Schmitz, Fluctuations in nonequilibrium fluids,
Physics Reports {\bf 171}, 1-58 (1988).



\bibitem{Bertini-rev}
L. Bertini, A. De Sole, D. Gabrielli, G. Jona-Lasinio, and C. Landim,
Macroscopic fluctuation theory, Rev. Mod. Phys. {\bf 87}, 593 (2015).





\bibitem{FNS}
D. Forster, D. R. Nelson, and M. J. Stephen,
Large-distance and long-time properties of a randomly stirred fluid,
Phys. Rev. A {\bf 16}, 732 (1977).


  
\bibitem{HH}  
P.C. Hohenberg and B. I. Halperin,
 Theory of dynamic critical phenomena, Rev. Mod. Phys.
 {\bf 49} 435-479 (1977). 


 \bibitem{NS}
  N. Nakagawa and  S.-i. Sasa,
  Liquid-gas transitions in steady heat conduction,
  Phys. Rev. Lett. {\bf 119}, 260602 (2017).

\bibitem{NS2}
N. Nakagawa and S.-i. Sasa, Global thermodynamics for heat
conduction states, 
J. Stat. Phys. {\bf 177}, 825-888 (2019).



\bibitem{nematic-isotropic}
E. F. Gramsbergen, L. Longa, and W.H. de Jeu,
Landau theory of the nematic-isotropic phase transition,
Physics Report {\bf 135}, 195-257 (1986).


  

\bibitem{Sekimoto-book}
K. Sekimoto,  {\it Stochastic Energetics}, Lect. Notes Phys. 799
(Springer-Verlag, Berlin, 2010).
  
\bibitem{SeifertRPP} U. Seifert,
  Stochastic thermodynamics, fluctuation theorems
  and molecular machines, Rep. Prog. Phys. {\bf 75}, 126001 (2012).

  

\bibitem{Evans-Cohen-Morriss}
  D. J. Evans, E. G. D. Cohen,  G. P. Morriss,
  Probability of second law violations in shearing
  steady states,  Phys. Rev. Lett. {\bf 71}, 2401--2404 (1993).


\bibitem{Gallavotti}
G. Gallavotti and E. G. D. Cohen,
Dynamical Ensembles in Nonequilibrium Statistical Mechanics,
{ Phys. Rev. Lett.}
\textbf{74},  2694 (1995).

  
\bibitem{Kurchan}
J. Kurchan,  Fluctuation theorem for stochastic dynamics,
{J. Phys. A: Math. Gen.}
\textbf{31},  3719 (1998).

\bibitem{LS}
J. L. Lebowitz and  H. Spohn, 
A Gallavotti--Cohen-type symmetry in the large
deviation functional for stochastic dynamics,
J. Stat. Phys.\textbf{95}, 333 (1999). 

\bibitem{Maes}
C. Maes,
The fluctuation theorem as a Gibbs property,
J. Stat. Phys. \textbf{95},  367-392 (1999).

\bibitem{Crooks}
G. E. Crooks, 
Entropy production fluctuation theorem and the nonequilibrium
work relation for free energy differences,
{Phys. Rev. E} \textbf{60}, 2721 (1999).
  
  

\bibitem{JarzynskiPRL}
  C. Jarzynski,
  Nonequilibrium equality for free energy differences,
  Phys. Rev. Lett. {\bf 78}, 2690--2693 (1997).


\bibitem{Sasa-fluid}
  S.-i. Sasa,  Derivation of hydrodynamics from the Hamiltonian
  description of particle systems, Phys. Rev. Lett. {\bf 112},
  100602 (2014).
  
\bibitem{Sasa-oscillator}  
  S.-i. Sasa,
  Collective dynamics from stochastic thermodynamics,
  New Journal of Physics {\bf 17}, 045024 (2015).
  

\bibitem{Zubarev} D. N. Zubarev,  {\it
 Nonequilibrium Statistical Thermodynamics}, 
(Consultants Bureau, New York, 1974).

\bibitem{Mclennan}
J. A. Mclennan, Phys. Fluids {\bf 3}, 493 (1960);
{\it Introduction to Non-equilibrium Statistical Mechanics}
(Prentice-Hall, 1988).


\bibitem{KN}
T. S. Komatsu and N. Nakagawa,
Expression for the stationary distribution in nonequilibrium steady states,
Phys. Rev. Lett. {\bf 100}, 030601 (2008).

\bibitem{KNST-rep}
T. S. Komatsu, N. Nakagawa, S.-i. Sasa, and H. Tasaki
Representation of nonequilibrium steady states in large mechanical systems,
J. Stat. Phys. {\bf 134}, 401-423 (2009).


\bibitem{Maes-rep}
C.  Maes and  K. Neto\v{c}n\'{y},
Rigorous meaning of McLennan ensembles,
J. Math. Phys. {\bf 51}, 015219 (2010).



\bibitem{gap}
  G. Fang and C. A. Ward, Temperature measured close to
  the interface of an evaporating liquid, Phys. Rev. E {\bf 59},
  417-428 (1999).


\bibitem{min-ent}
  E. T. Jaynes, 
  The minimum entropy production principle,
  Ann. Rev. Phys. Chem. {\bf 31}, 579- 601 (1980).

\bibitem{Klein}
  M. J. Klein and P. H. E. Meijer, Principle of minimum entropy production,
  Phys. Rev. {\bf 96}, 250-255 (1954).

\bibitem{Maes-LD}
C.  Maes and  K. Neto\v{c}n\'{y},
Minimum entropy production principle from a dynamical fluctuation law,
J.  Math.  Phys. {\bf 48}, 053306 (2007).


\bibitem{Derrida}
B. Derrida, 
Non-equilibrium steady states: fluctuations and
large deviations of the density and of the current,
J. Stat. Mech. P07023 (2007).





\bibitem{Nemoto}
T. Nemoto and S.-i. Sasa,
Thermodynamic formula for the cumulant generating function of
time-averaged current, 
Phys. Rev. E {\bf 84}, 061113 (2011).



\bibitem{lps} J. L. Lebowitz,  E. Presutti, H.  Spohn,
  Microscopic Models of Hydrodynamic Behavior,
  J. Stat. Phys. \textbf{51}, 841 (1988). 


\bibitem{Landau-Lifshitz-Fluid}
L. D. Landau and E. M. Lifshitz, 
{\it Fluid Mechanics}, 
(Pergamon Press, Oxford 1959).


\bibitem{HHM}  
 B. I. Halperin, P.C. Hohenberg, and S. K. Ma,
 Renormalization group methods for critical dynamics I.
 Recursion relations and effects of energy conservation,
 Phys. Rev. {\bf B 10} 139-153, (1974).

\bibitem{Penrose}
  O. Penrose and P. C. Fife, 
  Thermodynamically consistent models of phase-field type
  for the kinetic of phase transitions, Physica D {\bf 43}, 44-62 (1990). 

  
\bibitem{Fujitani}
  Y. Fujitani, Perturbation calculation for the density
  profile across the flat liquid-vapor interface in the
  steady heat-flow state, J. Phys. Soc. Jpn. {\bf 79}, 074002 (2010). 

\bibitem{Fukuma-Sakatani}
  M. Fukuma and Y. Sakatani, Entropic formulation
  of relativistic continuum mechanics, Phys. Rev. E{\bf 84},
  026315 (2011).



\bibitem{width}
R. M. Townsend and S. A. Rice,
Molecular dynamics studies of the liquid-vapor interface of water,
J. Chem. Phys. {\bf 94}, 2207 (1991).
  
\bibitem{Triez-Zwanzig}
D. G. Triezenberg and R. Zwanzig,
Fluctuation theory of surface tension,
Phys. Rev. Lett. {\bf 28} 1183-1185 (1972). 

\bibitem{Weeks}
J.D. Weeks, Structure and thermodynamics of the liquid-vapor interface,
J. Chem. Phys. {\bf 67} 3106 (1977). 
  
 

\bibitem{Feynman}
R. P. Feynman, R. B. Leighton, and M. Sands,
{\it The FeynmanLectures on Physics},
Vol. I (Addison-Wesley, Reading, Mas-sachusetts, 1963) Chap. 39-4.

\bibitem{Lieb}
E. H. Lieb,
Some problems in statistical mechanics
that I would like to see solved,
Physica A {\bf 263}, 491 (1999).

\bibitem{Gruber}
C. Gruber and J. Piasecki,
Stationary motion of the adiabaticpiston,
Physica A {\bf 268}, 412 (1999).

\bibitem{Gruber2}
 C. Gruber and L. Frachebourg,
 On the adiabatic properties of a stochastic adiabatic wall:
 Evolution, stationary non-equilibrium, and equilibrium states,
 Physica A {\bf 272}, 392 (1999).
 


  
  
\bibitem{Hatano-Sasa} T. Hatano and S.-i. Sasa,
  Steady-state thermodynamics of Langevin systems,
  Phys. Rev. Lett. {\bf 86}, 3463--3466 (2001).

\bibitem{KNST}
T. S. Komatsu, N. Nakagawa, S.-i. Sasa, and H. Tasaki,
Steady-state thermodynamics for heat conduction: Microscopic derivation,
Phys. Rev. Lett. {\bf 100}, 230602 (2008).



\bibitem{NN}
N. Nakagawa,
Work relation and the second law of
thermodynamics in nonequilibrium steady states,
Phys. Rev. E {\bf 85}, 051115 (2012).
  
  
\bibitem{Jona-thermo}
L. Bertini, D. Gabrielli, G. Jona-Lasinio, and C. Landim,
Clausius inequality and optimality of
quasistatic transformations for nonequilibrium stationary states,
Phys. Rev. Lett. {\bf 110}, 020601  (2013).

  
\bibitem{Maes-thermo}
C. Maes and K. Neto\v{c}n\'{y},
A nonequilibrium extension of the Clausius heat theorem,
J. Stat. Phys. {\bf 154}, 188-203 (2014).


\bibitem{Chiba-Nakagawa}
Y. Chiba and N. Nakagawa,
Numerical determination of entropy associated with excess heat in steady-state thermodynamics,
Phys. Rev. E {\bf 94}, 022115 (2016).



\bibitem{Keizer}  
J.  Keizer,
Thermodynamics at nonequilibrium steady states,
J. Chem. Phys.  {\bf 69}, 2609 (1978).

\bibitem{Eu}
B. C. Eu,
Irreversible thermodynamics of fluids,
Annals of Physics {\bf 140}, 341-371 (1982).

\bibitem{Jou}
D. Jou, J. Casas-V\'{a}zquez, and G. Lebon,
Extended irreversible thermodynamics,
Rep. Prog. Phys. {\bf 51}, 1105-1179 (1988).

\bibitem{Oono-paniconi}
Y. Oono  and M. Paniconi, 
Steady state thermodynamics,
Prog. Theor. Phys. Suppl. {\bf 130}, 29 (1998).

\bibitem{Sasa-Tasaki} S.-i. Sasa and  H. Tasaki,
Steady state thermodynamics, 
J. Stat. Phys. {\bf 125}, 125--224 (2006).


\bibitem{Bertin}
E. Bertin, K. Martens, O.  Dauchot, and M. Droz,
Intensive thermodynamic parameters in nonequilibrium systems,
Phys. Rev. E {\bf 75}, 031120 (2007).

\bibitem{Seifert-contact}
P. Pradhan, R. Ramsperger, and U. Seifert,
Approximate thermodynamic structure for driven lattice gases in contact,
Phys. Rev. E {\bf 84}, 041104 (2011).

\bibitem{Dickman}
R. Dickman, 
Failure of steady-state thermodynamics in nonuniform driven lattice gases,
Phys. Rev. E {\bf 90}, 062123 (2014).




  
\bibitem{Bedeaux00}
  A. R{\o}sjorde,  D. W. Fossmo, D. Bedeaux,
  S. Kjelstrup, and B. Hafskjold,
  Nonequilibrium molecular dynamics simulations
  of steady-state heat and mass transport in condensation:
  I. Local equilibrium,
  J. Coll. Int. Sci. {\bf 232}, 178-185 (2000).

\bibitem{Ogushi}
  F. Ogushi, S. Yukawa, and N. Ito,
  Asymmetric structure of gas-liquid interface,
  J. Phys. Soc. Jpn. {\bf 75}, 07301- (2006).

  






  
\bibitem{Zubarev-Morozov} 
  D. N. Zubarev and V. G. Morozov,
  Statistical mechanics of nonlinear hydrodynamic fluctuations,
  Physica {\bf 120A}, 411-467 (1983).
  
\bibitem{Morozov} 
  V. G. Morozov, On the Langevin formalism for nonlinear
  and nonequilibrium hydrodynamic fluctuations,
  Physica {\bf 126A}, 443-460 (1984).



  
\bibitem{Graham}
  R. Graham and H. Haken,
  Fluctuations and stability of stationary non-equilibrium systems
  in detailed balance, Z. Phys. {\bf 245}, 141-153 (1971).

\bibitem{Zwanzig}
  R. Zwanzig, Memory effects in irreversible thermodynamics,
  Phys. Rev. {\bf 124}, 983-992  (1961).

  
\bibitem{Itami-Sasa}
  M. Itami and S.-i. Sasa, Universal form of stochastic evolution
  for slow variables in equilibrium systems, J. Stat. Phys.
  {\bf 167}, 46-63 (2017). 


\bibitem{Nakano}
  H. Nakano and S.-i. Sasa, Statistical mechanical expressions of slip
  length, J. Stat. Phys. {\bf 176}, 312-357 (2019). 



\bibitem{Pomeau}
  Y. Pomeasu,
  Front motion, metastability and subcritical bifurcations in hydrodynamics,
  Physica 23D, 3-11 (1986). 


\end{thebibliography}
\end{document}